\documentclass[bibyear]{aa} 

\usepackage{rotating}
\usepackage{graphicx}
\usepackage{txfonts}
\usepackage{lscape}

\def\lae{\mathrel{<\kern-1.0em\lower0.9ex\hbox{$\sim$}}}
\def\gae{\mathrel{>\kern-1.0em\lower0.9ex\hbox{$\sim$}}}

\begin{document} 

   \titlerunning {Scaling relations of ETGs at $z\sim1.3$}
   \authorrunning {Saracco et al. }

   \title{Scaling relations of cluster elliptical galaxies at $z\sim1.3$:} 

   \subtitle{Distinguishing luminosity and structural evolution.}

   \author{P. Saracco\inst{1}\thanks{E-mail: 
paolo.saracco@brera.inaf.it},  
A. Casati\inst{1,2}, A. Gargiulo$^{1}$, M. Longhetti$^{1}$, I. Lonoce$^{1,3}$, 
S. Tamburri$^{1,3}$, 
D. Bettoni$^5$, M. D'Onofrio$^4$, G. Fasano$^5$, B. M. Poggianti$^5$,
K. Boutsia$^6$, M. Fumana$^7$,
E. Sani$^8$}

   \institute{INAF - Osservatorio Astronomico di Brera, Via Brera 28, 20121 Milano,
Italy \and
Universit\'a Statale di Milano, via Celoria 22, 20133 Milano, Italy 
\and Universit\'a degli Studi dell'Insubria, via Valleggio 11, 22100 Como,
Italy
\and Universit\'a degli Studi di Padova, Vicolo dell'Osservatorio 5, 35141 
Padova, Italy
\and INAF - Osservatorio Astronomico di Padova, Vicolo dell'Osservatorio 5, 35141 
Padova, Italy
\and INAF - Osservatorio Astronomico di Roma, Via Frascati 33, 00040 
Monte Porzio Catone, Italy
\and INAF - IASF, Via E. Bassini 15, 20133 Milano, Italy
\and INAF - Osservatorio Astrofisico di Arcetri, Largo Entrico Fermi 5, 50125 
Firenze, Italy }

   \date{Received 24 January 2014; accepted 29 April 2014 }

 
  \abstract
   {We studied the size-surface brightness and the size-mass relations of
a sample of 16 cluster elliptical galaxies in the mass range 
$\sim10^{10}-2\times10^{11}$ M$_\odot$ which were morphologically selected
in the cluster RDCS J0848+4453 at $z=1.27$. }
   {Our aim is to assess whether they have completed their mass growth at 
their redshift or significant mass and/or size growth can or must take 
place until $z=0$ in order to understand whether elliptical galaxies of
clusters follow the observed size evolution of passive galaxies.
   }
   {To compare our data with the local universe we considered the Kormendy relation
   derived from the early-type galaxies of a local Coma Cluster reference
   sample and the WINGS survey sample. 
The comparison with the local Kormendy relation shows that the luminosity 
evolution due to the aging of the stellar content already assembled at 
$z=1.27$ brings them on the local relation.
Moreover, this stellar content places them on the size-mass relation 
of the local cluster ellipticals.
These results imply that for a given mass, the stellar mass at $z\sim1.3$ is 
distributed within these ellipticals according to the same stellar mass 
profile of local ellipticals. We find that a pure
size evolution, even mild, is ruled out for our galaxies since it would
lead them away from both the Kormendy and the size-mass relation.
If an evolution of the effective radius takes place, this must be 
compensated by an increase in the luminosity, hence of the stellar
mass of the galaxies, to keep them on the local relations.
We show that to follow the Kormendy relation, the stellar mass
must increase as the effective radius.
However, this mass growth is not sufficient to 
keep the galaxies on the size-mass relation for the same variation 
in effective radius.
Thus, if we want to preserve the Kormendy relation, we fail
to satisfy the size-mass relation and vice versa. }
   {The combined analysis of the size-surface brightness relation with 
the size-mass relation leads to the result that these galaxies cannot 
increase solely in size and cannot significantly grow in mass.}
   {We conclude that these 16 cluster ellipticals at $z=1.27$ have, 
   for the most part, completed their stellar mass growth at the redshift they are 
   and that consequently, their evolution at $z<1.27$ will be dominated by 
   the aging of their stellar content. If this result is generalizable, 
   then it shows that elliptical galaxies in the above mass range do not 
   contribute to the observed size evolution of passive galaxies, as also 
   found by other authors.
   This evolution would be instead mainly driven by disk galaxies.
We do not find hints of differences between the properties of these 
cluster ellipticals and those of field ellipticals at comparable redshift,
even if this last comparison is still based on a low number statistics.}

   \keywords{galaxies: evolution; galaxies: elliptical and lenticular, cD;
             galaxies: formation; galaxies: high redshift
               }

   \maketitle
%

\section{Introduction}
It is now widely accepted that the mean size of 
early-type galaxies (ETGs) that is, of elliptical (E) and 
spheroidal galaxies (S0) as a population has increased with time.
This view arises many authors having found that the average 
effective radius of high-z passive and/or massive galaxies, not
necessarily early types, is smaller 
than the average effective radius of local early-type galaxies (e.g., 
Daddi et al. 2005; Trujillo et al. 2006, 2007;   
Longhetti et al. 2007;  van der wel et al. 2008; 
McGrath et al. 2008; van Dokkum et al. 2008; 
Buitrago et al. 2008; Cimatti et al. 2008; Bezanson et al. 2009;
 Damjanov et al. 2009; 
Cassata et al. 2011;  Damjanov et al. 2011; Szomoru et al. 2012;
Newman et al. 2012).
The local comparison samples of galaxies are always selected from the 
Sloan Digital Sky Survey (SDSS), often on the basis of their Sersic index $n$.
This size evolution is found to be strong even at moderate redshift:
the effective radius should increase by a factor two
since $z\sim1$ (e.g. Trujillo et al. 2011; Huertas-Company et al. 2013;
Cimatti et al. 2012; Delaye et al. 2013).
On the other hand, some studies do not confirm this result,
in particular when samples of early-type galaxies are considered 
strictly defined on the basis of their morphology 
(e.g.,  Mancini et al. 2010; Saracco et al. 2011; Stott et al. 2011; 
Jorgensen et al. 2013) and when the progenitor bias is taken into account
in the comparison between local and high-z samples (e.g., Saglia et al. 2010).

Progenitor bias represents a difficult limitation to treat, and
the more heterogeneous the sample of selected galaxies, the more
difficult it is to account for this bias.
Progenitor bias has been treated in different ways by different 
authors according to the selection of galaxies considered.
To compare the whole population of galaxies at different epochs in a meaningful 
way, some authors have selected galaxies at a constant number density at 
different redshift (e.g., van Dokkum et al. 2010). 
To compare early-type galaxies or passive galaxies 
some authors have selected  galaxies progressively older with
decreasing redshift, according to their passive aging
(e.g., Saglia et al 2010; Poggianti et al. 2013a; Carollo et al. 2013).
Other authors try to get rid of this bias by selecting galaxies at
fixed stellar velocity dispersion assuming that this quantity (defined
in the same way for disks and for spheroids) is largely unaffected by 
the merger history of the galaxies
(e.g., Belli et al. 2014).

The 
size evolution of passive/massive galaxies has been widely interpreted 
as the size evolution of the individual early types, that is to say  
elliptical galaxies would increase their size individually during their lives.
However, it was soon realized that this increase in size cannot be 
the result of a stellar mass growth since independent studies of the 
evolution of the luminosity and galaxy stellar mass function show that 
massive ETGs have already been assembled at $z\sim0.8$  and that 
they have not grown further at lower redshift both in the field 
(e.g., Pozzetti et al. 2010) and in clusters (e.g., Andreon et al. 2008).
Thus, early types should enlarge their size during their life but not grow
significantly in mass, at least from $z\sim1$.
The mechanisms proposed to increase the size of ETGs, leaving their mass
almost unchanged 
 are principally two: a pure expansion due to a significant mass
loss via quasar or stellar winds (Fan et al. 2008; 2010; Damjanov et al. 2009)
and dry minor mergers whose main effect should be adding a low stellar 
mass density envelope re-arranging the stars in the outskirts of the galaxy 
enlarging the size 
(e.g., Hopkins et al. 2009; Naab et al. 2009; Bezanson et al. 2009;
see also Nipoti et al. 2009 and 2012 for an analysis of the influence
of dry mergers on size and velocity dispersion evolution of ETGs).
Unless to only hypothesize dry minor mergers with particular orbital 
conditions (Naab et al. 2009), a pure increase in the effective radius 
of ETGs would imply a decrease in their stellar mass density within 
the same effective 
radius as the cube of the radius itself: if ETGs increase by a factor two 
since $z=1$, their effective stellar mass density should decrease by a 
factor 8, a macroscopic effect that has not yet been observed (e.g.,
 Saracco et al. 2012).

In fact, the study of the evolution of the mean size of passive and/or 
massive galaxies did not help much in constraining
the evolution of proper elliptical galaxies or of their
mass-assembly history.
This may be for two main reasons.
The first is that selecting
passive and/or massive galaxies provides samples 
with a high fraction of disk galaxies.
It is well established that at least 30-40 per cent of the passive galaxies  
at any redshift between $0.6<z<2.0$  are disk galaxies 
(e.g. Ilbert et al. 2010; van der Wel et al. 2011; Cassata et al. 2011; 
Tamburri et al. in preparation).
Moreover, the remaining fraction (60-70 per cent) of early-type galaxies
misses 
more than 20 per cent of the population of spheroidal galaxies  
(Tamburri et al.).
The selection of passive/massive galaxies therefore fails to produce 
representative samples of strictly defined ETGs, while it selects samples
of disks and spheroids.

Disk galaxies and elliptical galaxies do not share the same formation and 
evolution history.
In a hierarchical universe, elliptical galaxies are byproducts or 
descendants since their formation   
is directly linked to merger events of progenitor disk-like or irregular
galaxies 
(e.g., Khochfar and Burkert 2003; De Lucia et al. 2006; Hopkins et al. 2010).
Moreover, since merging controls the buildup and the growth 
of galaxies independently of their morphology, in a hierarchical
universe it is reasonable to expect that the average size of galaxies 
increases with time because progenitors, 
which are  smaller by definition, disappear to form the merger descendants 
which are by definition  more massive, hence larger.
In this possibly naive scheme, sampling both progenitors and 
descendants together as in a passive/massive galaxy selection to monitor the 
change of their
mean properties (such as their mean size) makes it difficult to gain 
information about the mass assembly
and the evolution either of early-type galaxies or of passive disk galaxies.
It is difficult to understand which is evolving and what kind of evolution
is taking place.
The second reason for the poor effectiveness of the study of the mean size 
of passive/massive galaxies in constraining the evolution of proper ellipticals
is that, independently of and in addition to the 
above progenitor bias, 
it is not clear whether the apparent change in the mean size of the whole 
population of passive/massive galaxies, that is disks and spheroids, 
is dominated by the size evolution of individual galaxies 
(each galaxy increases its effective radius) or rather by the 
appearance of new-born larger galaxies, by the disappearance of 
smaller ones, or by the combination of the two.
{Van der Wel et al. (2008) suggests, among the first,  the possibility 
that a combination of structural evolution of individual galaxies and the
continuous formation of early-type galaxies may account for the observed
size evolution.} 
Actually, observations suggest that most of the observed size evolution is 
due to the size evolution of the compact disk-like  
galaxies observed at $z\sim2$ (Van der Wel et al. 2011).
However, when disks and spheroids are mixed in the same sample, it is 
even more difficult to 
 distinguish between the evolution of the individual galaxies
and the evolution of the mean properties of the population.

A way to distinguish individual size evolution from other
effects could be to compare the number density of compact early-type galaxies
once selected at high and at low redshift consistently. 
Evidence of the presence   of compact 
ETGs in the local Universe similar to those observed at high-z 
has come out recently (Valentinuzzi et al. 2010a; 2010b; Poggianti et al. 2013).
Defining compact those early-types both
at low-z and at high-z
having a radius at least a factor two smaller than the radius
derived by the mean size-mass relation by Shen
et al. (2003), Saracco et al. (2010) find that the number density of compact high-z ETGs 
averaged over the interval $0.9<z<1.9$ is accounted for by their 
local counterparts showing that size evolution, if it takes place, 
cannot affect the majority of the high-z ETGs.
Recently, Carollo et al. (2013) have found no change in the number density
of compact quenched early-type galaxies with masses $<10^{11}$ M$_\odot$
at $0.2<z<1$ and a 30 per cent decrease at higher masses, suggesting
that the possible evolution of the mean size is driven mainly by 
the appearance of new-born larger early types.
Poggianti et al. (2013) find that no more than half and possibly
a smaller fraction of the compact high-z galaxies has evolved in size.

Hints that the environment may affect the size 
of early-type galaxies and hence their evolution, 
in the sense that at a given redshift cluster ETGs are larger than field 
ETGs with the same mass,  have also been recently claimed by some 
authors (e.g., Papovich et al. 2012; Delaye et al. 2013) even if not
found by others
(e.g. Damjanov et al. 2011;  Rettura et al.
2010; Raichoor et al. 2012).

The studies of the size evolution of galaxies conducted and collected so far
leave us with two basic questions about the assembly and evolution of elliptical
galaxies:
do individual elliptical galaxies grow their stellar mass and change
continuously during their life, or is their morphological modeling
the final stage of a process after which the elliptical 
evolves unperturbed in luminosity ?  
The second question is: does the destiny of an elliptical galaxy depend 
significantly on the environment?     
We are trying to assess these by adopting a systematic approach based
on the selection of early-type galaxies strictly defined morphologically
at intermediate redshift both in the field and in cluster.

In this paper we study a sample of 16 elliptical galaxies in the cluster  
RDCS J0848+4453 at $z=1.27$ to constrain their evolutionary status through
comparing their size-surface brightness 
and  size-stellar mass relations with those of a local sample of cluster
ETGs selected according to the same criteria.
In Sec. 2 we describe the data and the sample selection.
In Sec. 3 we derive the physical (stellar mass, absolute magnitude, and
age) and structural (effective radius, surface brightness) parameters
for our galaxies.
In Sec. 4 we derive the Kormendy relation at $z\sim1.3$ and
compare it with the local relation.
In Sec. 5 we derive the luminosity evolution that the stars
already formed at that redshift will experience and the consequences
of this evolution.
Then, combining the study of the size-mass relation to the size-surface
brightness relation, we constrain the evolutionary status of our 16 ellipticals
and their future evolution. 
Finally, in Sec. 6, we summarize the results and present our conclusions.
Throughout this paper we use a standard cosmology with
$H_0=70$ Km s$^{-1}$ Mpc$^{-1}$, $\Omega_m=0.3$, and $\Omega_\Lambda=0.7$.
All the magnitudes are in the Vega system, unless otherwise specified.
\begin{figure*}
\begin{center}
\includegraphics[width=8.5cm]{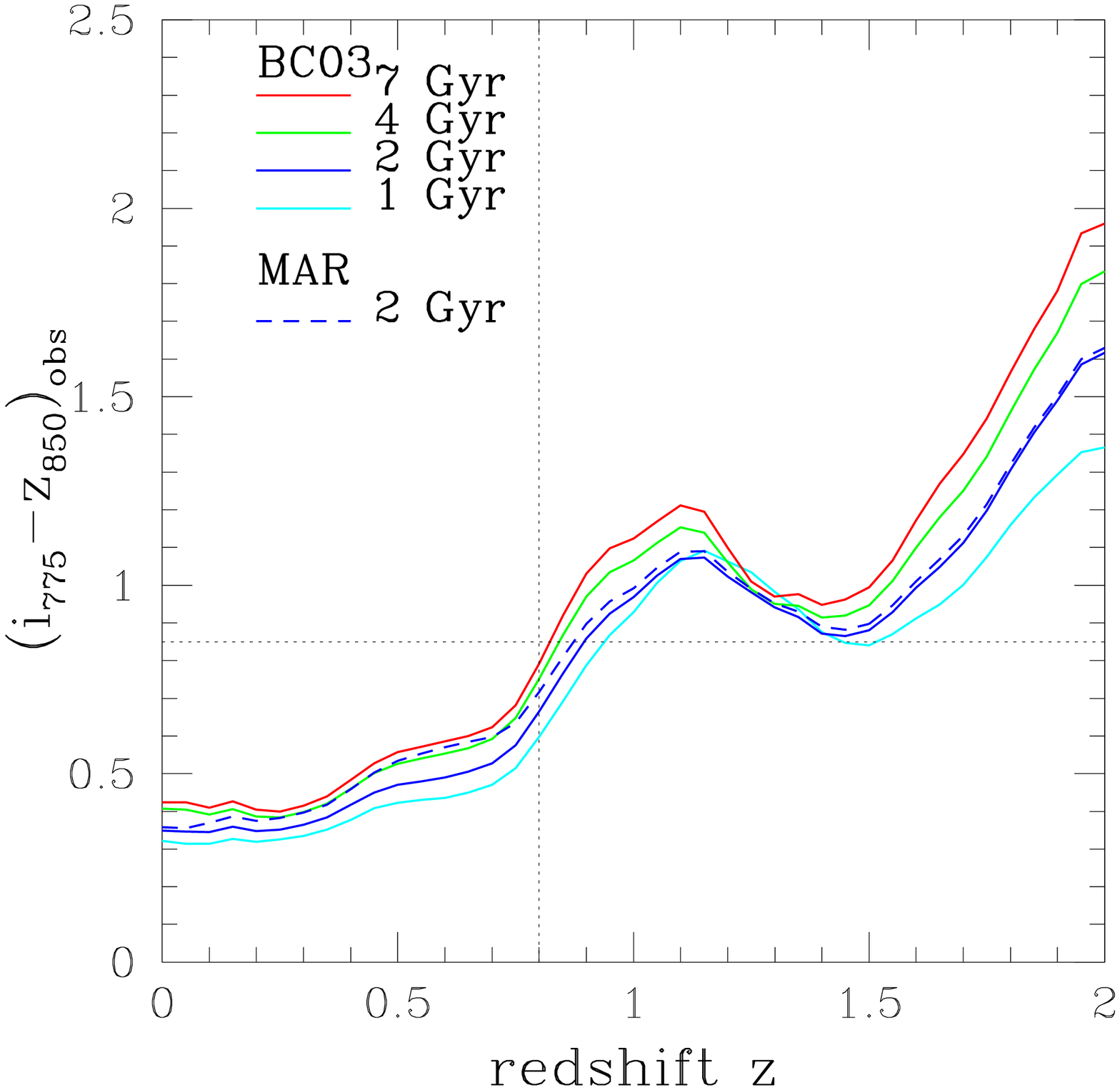}
\includegraphics[width=8.5cm]{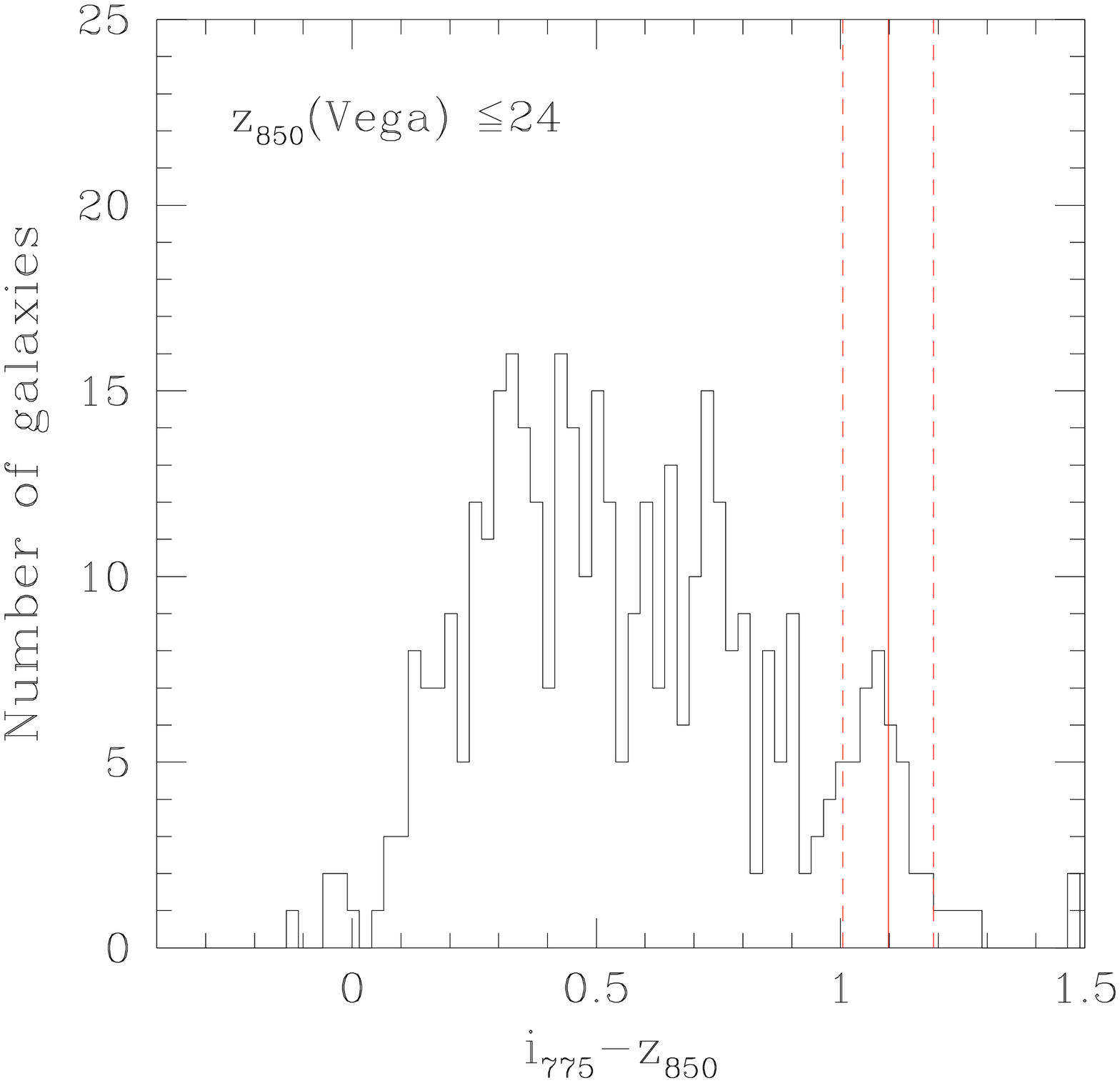}
\vskip -0.5truecm
\caption{$i_{775}-z_{850}$ color of galaxies. Left: The expected 
$i_{775}-z_{850}$ color of galaxies
is shown as a function of redshift for 4 different ages.
The different solid lines refer to (from top to bottom) 7 Gyr old (red line), 
4 Gyr old (green line), 2 Gyr old (blue line), and 1 Gyr old (cyan line).
The dashed line refers to a 2 Gyr old template obtained with the
Maraston et al. (2005) models.
The color is always $i_{775}-z_{850}<0.8$ (black horizontal line)
for redshift $z<0.8$, and it is always larger than 0.8 for $z>0.8$, 
this independently of the age of the stellar population considered.
Right: The $i_{775}-z_{850}$ color distribution of the 467 galaxies
brighter than $z_{850}<24$ falling within 1 Mpc from the cluster center 
is shown. 
The red solid line marks the mean color 
$\langle i_{775}-z_{850}\rangle=1.1\pm0.09$ of the 5 ETGs cluster members
spectroscopically identified by Stanford et al. (1997).
The dashed lines represent $\pm1\sigma$.
A second peak is evident in the distribution at the
mean color of the 5 ETGs. 
}
\end{center}
\end{figure*}


\section{Data description}
\subsection{Observations}
The analysis presented in this paper is based on Hubble Space Telescope (HST) 
and Spitzer archival data and on ground-based optical observations 
that we obtained at the Large Binocular Telescope (LBT).
The HST data retrieved from the archive are composed of optical ACS 
observations and near-IR NICMOS observations.
ACS observations (ID 9919) covering a field of about 11 arcmin$^2$
surrounding the cluster RDCS J0848+4453 were obtained in 2004 in the 
F775W (7300 s) and 
F850LP (12200 s) filters and they are described in Postman et al. (2005; 
see also Raichoor et al. 2011).
The ACS images we used have a pixel scale of 0.05 arcsec/pix and a resolution
of FWHM$_{850}\simeq0.12$ arcsec.
NICMOS observations (ID 7872) sampling a field of about 3 arcmin$^2$
centered on the same cluster were obtained with the NIC3 camera 
 in the F160W filter (11200 s) and are described in van 
Dokkum et al. (2001).
The NIC3 images have a pixels scale of 0.2 arcsec/pix and a resolution
of FWHM$_{160}\simeq0.22$ arcsec.
Spitzer data (PI S. A. Stanford) were obtained in the four IRAC band passes 
[3.6, 4.5, 5.8, 8.0]$\mu$m.
We used the fully co-added mosaics (0.6 arcsec/pix) produced by the standard 
Spitzer reduction pipeline resulting in a mean exposure time of $\sim$2200 s in 
the 3.6 $\mu$m and 5.8 $\mu$m bands and in about 1900 s in the 4.5 $\mu$m and 
8.0 $\mu$m.

Our LBT observations were carried out between 14 and 17 of
February 2013 under poor and unstable seeing conditions 
(1.0-1.8 arcsec) with the Large
Binocular Cameras (LBC\footnote{http://lbc.mporzio.astro.it/}) in the four 
Bessel U, B, V and R filters.
LBCs are two wide-field imaging cameras located at the Prime Focus stations 
of LBT.
Each LBC camera has a wide field of view equivalent to $\sim23'\times23'$,
and it provides images with a sampling of 0.23 arcsec/pixel.
The two LBC cameras are optimized  for the UV-blue wavelengths (LBC-blue, 
from 0.3 $\mu$m  to 0.5 $\mu$m) and for the red-IR bands 
(LBC-red, from 0.5 $\mu$m  1.0 $\mu$m), respectively. 
Observations have been carried out in the full binocular configuration, 
i.e., with the two LBC cameras operating simultaneously and pointing 
in the same direction of the sky.
An integration of $\sim 4$ hr in each filter has been obtained 
in 8 hr of binocular configuration time (4 hr in U at LBC-blue and
simultaneously in V at LBC-red, plus 4 hr in B (LBC-blue) and 
simultaneously in R (LBC-red)).
Observations consist of short exposures of six minutes each 
dithered by 30 arcsec in a random pattern to cover the gap between the CCDs.
The standard reduction procedure (bias and flat-field correction and cosmic 
ray removal) has been applied to the single frames before combining them 
to produce the final stacked mosaic.
 Given the large variation in the seeing conditions 
during the observations, we decided to
consider only those images taken under seeing conditions better than 
1.4 arcsec to construct final stacked images of good quality.
Thus, the final co-added mosaics have different effective exposure
times ranging from 1 hr (U and V, FWHM$\sim1.0$ arcsec) to almost 
3 hrs (B and R, FWHM$\sim1.4$ arcsec).
The final mosaic has been produced with SWarp (v.2.19.1, Bertin 2010).
\begin{table}
\caption{
Selected sample of 16 cluster elliptical galaxies.}
\centerline{
\begin{tabular}{rrrrr}
\hline
\hline
 \#ID &    RA	   &	    Dec	  &   $z_{spec}$\\ 
\hline
1    & 08:48:36.233&   44:53:55.42&  	  1.276\\
2    & 08:48:36.160&   44:54:17.24&  	  1.277\\
3    & 08:48:32.978&   44:53:46.61&  	  1.277\\
4    & 08:48:35.978&   44:53:36.12&  	  1.275\\
5    & 08:48:32.434&   44:53:34.97&  	  1.263\\
606  & 08:48:37.071&   44:53:33.99&  	  ---- \\
590  & 08:48:34.069&   44:53:32.23&  	  ---- \\
568  & 08:48:35.038&   44:53:30.83&  	  ---- \\
719  & 08:48:33.031&   44:53:39.67&  	  ---- \\
1250 & 08:48:37.341&   44:54:15.60&  	  ---- \\
1260 & 08:48:36.160&   44:54:16.16&  	  ---- \\
173  & 08:48:34.058&   44:53:02.44&  	  ---- \\
1160 & 08:48:32.768&   44:54:07.14&  	  ---- \\
657  & 08:48:32.442&   44:53:35.35&  	  ---- \\
626  & 08:48:32.390&   44:53:35.03&  	  ---- \\
471  & 08:48:29.685&   44:53:23.91&  	  ---- \\
 \hline									 
\hline									 
\end{tabular}								 
}
\tablefoot{Elliptical galaxies with $z_{850}<24$, within 1 Mpc 
radius from the cluster center and $i_{775}-z_{850}=1.1\pm0.2$.}									 
\end{table}

\subsection{Sample selection}
The sample of ellipticals used in this analysis is composed of 16 galaxies
selected to belong to the cluster RDCS J0848+4453 at $z=1.27$ 
(Stanford et al. 1997).
{Many authors have found that compact galaxies preferentially host 
older stellar populations (e.g., Gargiulo et al. 2009, 
Saracco et al. 2009; Valentinuzzi et al. 2010a; 
Poggianti et al. 2013a; Saracco et al. 2011; Taylor et al. 2010). 
Consequently, selection criteria (directly or indirectly) dealing  with the 
age could introduce a bias in favor of compact galaxies.
For instance, the median mass-size relation of galaxies is found to shift 
towards smaller radii for galaxies with older stars (Poggianti et al. 2013),
and colors bracketing the Balmer break, such as the R-K, tend to select 
an increasingly higher fraction of compact galaxies 
going toward redder colors (Saracco et al. 2011).
A selection based on the passivity as resulting from the 
specific star formation rate (sSFR) deals with the age of the stellar 
population since, at fixed stellar mass and for a fixed sSFR threshold, a 
galaxy may or not may be passive depending on the SFR of the best-fitting model, 
i.e. on its age.
Our attempt is to select elliptical galaxies by avoiding selection 
criteria based on (or related to) the age of their stellar population.}
Thus, we selected our final sample on the basis of their morphology 
without introducing any selection based on the age of their
stellar population or on their passivity.

To this end, we first
detected all the sources ($\sim2200$ up to 
a magnitude in the F850LP filter $z_{850}<27.6$) 
in the $\sim11$ arcmin$^2$ region surrounding the cluster covered
by the F850LP image.
We used SExtractor (Bertin and Arnouts 1996) both to detect sources and 
to measure their magnitude MAG\_BEST.
Magnitudes in the F775W filter ($i_{775}$ hereafter) were 
obtained by running
the procedure in double-image mode using the F850LP image as reference.
To perform a reliable and robust visual morphological classification,
we selected galaxies with magnitudes $z_{850}\le24$.
At this magnitude limit, the sample 
is 100\% complete.
From this flux-limited sample we removed stars identified by the
SExtractor stellar index CLASS\_STAR$>0.9$ and
restricted the selection to those galaxies within a projected radius
D$\le1$ Mpc ($\sim2$ arcmin) from the cluster center.
We  thus obtained a sample of 467 galaxies,
105 of which are also covered by NICMOS-F160W observations.
This sample contains the six cluster member galaxies spectroscopically 
identified by Stanford et al. (1997) within a diameter region of 
$\sim0.7$ Mpc.
One of them (galaxy \#237 in their Table 1) appears irregular
in the F850LP image.
The remaining five member galaxies are clearly ETGs, as confirmed
by their morphology 
(see below for the morphological classification).
The mean color of these five cluster member ETGs is 
$\langle i_{775}-z_{850}\rangle=1.1$
with a dispersion $\sigma_{iz}=0.09$.
At the redshift of the cluster, this color roughly corresponds 
to UV-U, so it is a measure of the slope of
the spectral energy distribution of galaxies at 
$\lambda_{rest}<4000$ \AA.

In the left hand panel of Fig. 1 the expected apparent $i_{775}-z_{850}$ color of 
galaxies with different ages is shown as a function of redshift.
It can be seen that this color clearly shows two different behaviors
depending on whether it samples the spectrum at $\lambda_{rest}>4000$ 
\AA\ or at shorter wavelengths.
In particular, the observed color is always $i_{775}-z_{850}\lae0.8$ for 
$z\lae0.8$ independently of the age of the galaxy.
At redshift $z\sim0.8-0.9$, when the region at 0.4-0.5 $\mu$m enters
the filter
F775W, the $i_{775}-z_{850}$ color  changes rapidly, thereby increasing its
value.
For $z>0.9$,  the color is always $i_{775}-z_{850}>0.8$.
The other important property of this color is its extremely
weak dependence on the age of the stellar population as clearly 
demonstrated by the small color variations ($<0.1$ mag) for
different ages.
As a result, the observed  $i_{775}-z_{850}$  trace the redshift of
the galaxies well without introducing any dependence on their age.
In the right hand panel of Fig. 1 the $i_{775}-z_{850}$ color distribution
of the 467 galaxies with $z_{850}\le24$ is shown.

The observed distribution clearly reflects the behavior shown
in the left hand panel with the bulk of the $z_{850}\le24$ galaxies 
having a color $i_{775}-z_{850}\lae 0.8$ centered on 0.3-0.4 mag 
suggesting that they are all at $z<1$.
The remaining galaxies form a second peak centered 
on $i_{775}-z_{850}\simeq 1.1$ suggesting that they are 
at $z>1$.
This second peak  
is centered on the mean color of the five elliptical cluster members
marked  in Fig.1. 
Thus, on the basis of these considerations we selected all the galaxies 
having a color $0.9<i_{775}-z_{850}<1.3$ according to the color
distribution shown in Fig. 1.
This color range corresponds to two sigmas from the mean color 
of the 5 ETGs cluster members.
For the resulting 44 galaxies we performed a morphological 
classification to identify the ellipticals.

The morphological classification is based on the
visual inspection of the galaxies carried out independently by two of us
on the ACS-F850LP image and on the fitting to their
luminosity profile described below.
In particular, we classified as ETG those galaxies having a regular 
shape with no signs of disk on the F850LP images and
no irregular or structured residuals resulting from the profile fitting.
On the basis of this morphological classification,
16 galaxies out of the 44  with 
$z_{z850}<24$ and $i_{775}-z_{850}=1.1\pm0.2$
turned out to be ellipticals.
The selected sample is summarized in Table 1.
{The F850LP images of the 16 ellipticals are shown  in Appendix
B\footnote{Published online}.}

\subsection{Multiwavelength photometry}
Magnitudes for these 16 galaxies have been measured in all the bands 
using SExtractor.
We adopted the MAG\_BEST as best estimator of the magnitude.
{Given the large difference in the PSF among the images (from $\sim$0.12 arcsec
of the ACS images to $>$2.7-3 arcsec of Spitzer images) and the relative
crowding of the field, many sources isolated in the highest resolution images
are affected by blending  with neighboring sources in the less resolved
(ground-based and Spitzer) images.
As a consequence of this, magnitudes, hence colors, measured within 
apertures scaled according to the different PSF would be affected by this 
blending in the lowest resolution images, which is difficult to quantify. 
Instead of using aperture magnitudes, we therefore preferred to measure the colors
using the MAG\_BEST estimator that, in the case of blending, tries to recover the
true flux of the target source.
We also verified that for  unblended sources, the colors derived
from MAG\_BEST are consistent with those measured within scaled apertures
by comparing them for a sample of points sources isolated in all the images.
} 
\begin{figure} 
\begin{center}
\includegraphics[width=8.5cm]{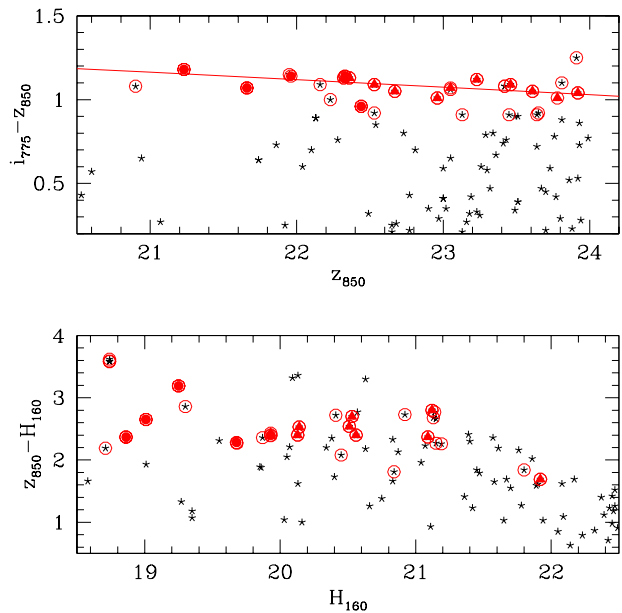}
\caption{Color-magnitude relation. The  $i_{775}-z_{850}$ (upper panel) and the
$z_{850}-H_{160}$ (lower panel) colors of the selected ETGs galaxies 
(red filled symbols) as a function of $z_{850}$ (16 galaxies)
and of $H_{160}$ (15 galaxies) are shown together with those of the 105 galaxies 
(black crosses)
covered by NICMOS observations. The five ETGs with spectroscopic redshift
are marked by red filled circles and the remaining 11 ETGs cluster members
are marked by red filled triangles.
Red open circles mark all the galaxies selected in the color range
$0.9<i_{775}-z_{850}<1.3$.
The red solid line in the upper panel is the color magnitude relation
$(i_{775}-z_{850})=2.1-0.044z_{850}$ best-fitting the
16 ellipticals.
}
\end{center}
\end{figure}

NICMOS-F160W observations cover 15 out of the 16 ETGs
selected above (and 30 out of the 44 galaxies in the color range
$0.9<i_{775}-z_{850}<1.3$), with galaxy \#471 falling outside.
In Fig. 2, the $i_{775}-z_{850}$ and the $z_{850}-H_{160}$
colors of the 16 selected ETGs are shown 
as a function of $z_{850}$ (upper panel) and of $H_{160}$ (lower panel) 
together with those of the 105 galaxies in common between NICMOS and ACS 
observations, and
galaxies selected according to the color cut  $0.9<i-z<1.3$.
It is worth noting that they span a wide color 
range, $1.7<z_{850}-H_{160}<3.6$, as visible in the lower panel of Fig. 2.
Once the morphology of these galaxies is defined, it turns out that ellipticals 
are mostly distributed within a narrower range of $z_{850}-H_{160}$ color
and that they define a red sequence, as expected.
However, there is also an exception: one of the bluest galaxy, 
$z_{850}-H_{160}=1.7$, has an elliptical morphology. 

The upper panel shows the color magnitude relation
$(i_{775}-z_{850})=2.1-0.044z_{850}$ best fitting the
16 selected ellipticals.
It is worth noting the good agreement with the relations found by Holden et al.
(2005) for a cluster at $z=1.23$ and by Mei et al. (2012) for the clusters
in the Linx structure.
For the 15 ellipticals covered by NICMOS observations, we also derived structural 
parameters in the F160W band (see \S 3.2) besides measuring 
magnitudes.
Three of them
(\#5, \#626 and \#657, see Fig. 3)  are blended in the NICMOS images 
because of the large pixel size  and the low resolution of the NIC3 camera.
For the same reasons these galaxies are not resolved either in the LBC
U-,B-,V-, and R-band images or in the Spitzer images.
For these galaxies, the magnitudes in those filters in which they are 
not resolved have been derived by redistributing the total 
flux of the resulting blended object according to the flux measured for 
each of them in the nearest filter in which they are resolved.
For instance, the $H_{160}$ magnitude for each of these three galaxies  
has been derived redistributing the total F160W flux measured for them 
blended (i.e., considered as a single object) on the basis of the 
flux measured for each of them in the  F850LP filter.

The LBC observations cover the whole sample of 16 ETGs.
As previously said, three galaxies are not resolved in the LBC images.
For them, U,B,V, and R magnitudes have been derived by redistributing 
the total U, B, V, and R fluxes on the basis of the flux measured for 
each of the galaxies in the F775W filter.

Spitzer-IRAC observations cover the whole sample of ETGs.
Magnitudes were estimated in the four IRAC bands using SExtractor
in double-image mode and adopting the 3.6 $\mu$m image as reference.
The reliability of the flux measurement was checked  by comparing the
flux measured with SExtractor for some stars in the field with the 
flux obtained using the IRAF task phot. 
For the IRAC images, besides the three galaxies above, galaxies 
\#2 and \#1260 are also not resolved due to the low resolution 
(FWHM$>2.5$ arcsec).
For them, we estimated the magnitudes in the four IRAC bands, using  
the fluxes measured in the F160W filter as
reference.
In Table 2  we report the photometry in the 11 photometric bands for the
16 ellipticals of the sample.
\begin{table*}
\caption{Photometry in the 11 photometric bands.}
\small{
\centerline{
\begin{tabular}{cccccccccccc}
\hline
\hline
 \#ID& U&  B&  V&   R& $i_{775}$&      $z_{850}$&     $H_{160}$&	   m3.6       &      m4.5	 &	m5.8	  &	m8.0	     \\ 
\hline
1    &  $>$26.8  &  $>$27.2    &  27.00  $\pm$0.15&  24.65  $\pm$0.06  &   23.40$\pm$0.03 &  22.44$\pm$0.02 &  19.25$\pm$0.01& 16.79$\pm$ 0.02 & 16.48$\pm$ 0.02  & 16.13$\pm$0.05 & 15.84$\pm$  0.07\\
2    &  $>$26.8  &  $>$27.2    &  25.88  $\pm$0.14&  23.92  $\pm$0.06  &   22.73$\pm$0.02 &  21.66$\pm$0.01 &  19.01$\pm$0.01& 16.52$\pm$ 0.10 & 16.22$\pm$ 0.10  & 16.12$\pm$0.15 & 15.68$\pm$  0.26\\
3    &  $>$26.8  &  $>$27.2    &  26.65  $\pm$0.15&  24.93  $\pm$0.07  &   23.10$\pm$0.01 &  21.96$\pm$0.02 &  19.68$\pm$0.01& 17.29$\pm$ 0.02 & 16.97$\pm$ 0.02  & 16.84$\pm$0.08 & 16.24$\pm$  0.09\\
4    &  $>$26.8  &  $>$27.2    &  25.65  $\pm$0.09&  23.76  $\pm$0.06  &   22.41$\pm$0.02 &  21.23$\pm$0.01 &  18.86$\pm$0.01& 16.03$\pm$ 0.01 & 15.70$\pm$ 0.02  & 15.70$\pm$0.06 & 15.45$\pm$  0.06\\
5    &  $>$26.8  & 28.4$\pm$0.4&  26.17  $\pm$0.10&  24.74  $\pm$0.10  &   23.47$\pm$0.01 &  22.33$\pm$0.01 &  19.93$\pm$0.04& 17.35$\pm$ 0.11 & 17.04$\pm$ 0.15  & 17.33$\pm$0.25 & 16.63$\pm$  0.25\\
606  &  $>$26.8  &  $>$27.2    &  27.25  $\pm$0.27&  25.39  $\pm$0.12  &   23.62$\pm$0.02 &  22.53$\pm$0.01 &  20.13$\pm$0.01& 17.56$\pm$ 0.02 & 17.20$\pm$ 0.02  & 17.13$\pm$0.11 & 16.41$\pm$  0.10\\
590  &  $>$26.8  &  $>$27.2    &  26.06  $\pm$0.15&  24.79  $\pm$0.06  &   23.72$\pm$0.02 &  22.67$\pm$0.01 &  20.14$\pm$0.01& 17.40$\pm$ 0.03 & 17.03$\pm$ 0.02  & 17.16$\pm$0.09 & 16.34$\pm$  0.90\\
568  &  $>$26.8  &  $>$27.2    & $>$28.10$\pm$99.0& $>$26.70$\pm$99.0  &   24.12$\pm$0.02 &  23.05$\pm$0.01 &  20.51$\pm$0.01& 18.23$\pm$ 0.02 & 17.90$\pm$ 0.03  & 18.18$\pm$0.15 & 17.96$\pm$  0.23\\
719  &  $>$26.8  &  $>$27.2    & $>$28.10$\pm$99.0&  25.26  $\pm$0.16  &   24.35$\pm$0.03 &  23.23$\pm$0.01 &  20.53$\pm$0.02& 17.68$\pm$ 0.02 & 17.33$\pm$ 0.03  & 17.48$\pm$0.13 & 16.92$\pm$  0.14\\
1250 &  $>$26.8  &  $>$27.2    &  27.87  $\pm$0.43& $>$26.70$\pm$99.0  &   24.55$\pm$0.04 &  23.46$\pm$0.02 &  21.09$\pm$0.02& 18.61$\pm$ 0.03 & 18.30$\pm$ 0.04  & 17.90$\pm$0.12 & 17.58$\pm$  0.16\\
1260 &  $>$26.8  &  $>$27.2    & $>$28.10$\pm$99.0& $>$26.70$\pm$99.0  &   24.66$\pm$0.04 &  23.61$\pm$0.02 &  21.92$\pm$0.02& 19.44$\pm$ 0.12 & 19.14$\pm$ 0.13  & 19.04$\pm$0.22 & 18.60$\pm$  0.36\\
173  &  $>$26.8  &  $>$27.2    & $>$28.10$\pm$99.0&  26.83  $\pm$0.18  &   24.96$\pm$0.03 &  23.92$\pm$0.02 &  21.12$\pm$0.03& 18.67$\pm$ 0.02 & 18.39$\pm$ 0.05  & 17.23$\pm$0.15 & 17.58$\pm$  0.23\\
1160 &  $>$26.8  &  $>$27.2    & $>$28.10$\pm$99.0& $>$26.70$\pm$99.0  &   23.97$\pm$0.03 &  22.96$\pm$0.02 &  20.56$\pm$0.02& 17.54$\pm$ 0.02 & 17.33$\pm$ 0.03  & 17.53$\pm$0.13 & 16.71$\pm$  0.11\\
657  &  $>$26.8  & 28.4$\pm$0.4&  26.17  $\pm$0.10&  24.74  $\pm$0.10  &   23.45$\pm$0.01 &  22.32$\pm$0.01 &  19.93$\pm$0.04& 17.35$\pm$ 0.14 & 17.04$\pm$ 0.15  & 17.33$\pm$0.25 & 16.63$\pm$  0.25\\
626  &  $>$26.8  & 28.4$\pm$0.4&  26.17  $\pm$0.10&  24.74  $\pm$0.10  &   23.49$\pm$0.02 &  22.36$\pm$0.01 &  19.93$\pm$0.04& 17.35$\pm$ 0.14 & 17.04$\pm$ 0.15  & 17.33$\pm$0.25 & 16.63$\pm$  0.25\\
471  &  $>$26.8  &  $>$27.2    & $>$28.10$\pm$99.0& $>$26.70$\pm$99.0  &   24.79$\pm$0.05 &  23.78$\pm$0.03 &  -----  & 19.22$\pm$ 0.05 & 19.01$\pm$ 0.06  & 18.40$\pm$0.22 & 99.00$\pm$ 99.00\\
 \hline									 
\hline									 
\end{tabular}								 
}}
\tablefoot{U, B, V and R magnitudes for each galaxy in the sample
come from the data obtained at the Large Binocular Telescope
(LBT); $i_{775}$, $z_{850}$, and  $H_{160}$ from the HST archival ACS
 and NICMOS-NIC3 data in the F775W, F850LP, and F160W filters respectively; 
 m3.6, m4.5, m5.8 and m8.0 
from the Spitzer archival images in the corresponding filters.
Photometric errors were calculated by quadratically summing to the 
Sextractor statistical errors the uncertainty in the photometric calibration 
(0.04 mag) dominant in the ground-based data and the true sky noise variations 
computed within apertures  across the images in space-based data
to account for the smoothed noise resulting from the alignment and sub pixel
shifting procedures ($\sim$0.008 mag for HST; from 0.01 mag to more 
than 0.05 mag for Spitzer.}								 
\end{table*}

\subsection{Comparison data in the local Universe}
{As comparison data in the local universe we considered two independent  
data sets relating to local cluster early-type galaxies.
The first data set comprises the well defined local Kormendy relation, 
the first comparison that we face in Sec. 4.
For this comparison, we used the Kormendy relation derived from the sample 
of 147 elliptical (E) and S0  galaxies belonging to the Coma cluster 
($z=0.024$) studied by Jorgensen et al. (1995a, 1996).
The morphological classification of this sample is based on a visual 
analysis, as for our sample.
Also, the structural parameters were derived in the Johnson B and in the 
Gunn r filters matching the F850LP and F160W filters at the redshift of our
galaxies. 
We applied a correction of -0.38 mag to pass from Gunn r to
Cousins R magnitudes (Fukugita, et al. 1995; see also
Longhetti et al. 2007).
The stellar mass range covered by this sample is approximately
$0.8\times10^{10}-3\times10^{11}$ M$_\odot$, as derived from
their stellar velocity dispersion measurements (Jorgensen et al. 1995b),
so it covers the same mass range as covered by our galaxies.
It is worth noting that this sample comprises both elliptical and S0 galaxies,
while our sample should only include elliptical galaxies.
However, the scaling relations and, in particular, the zero point and the
slope obtained using the whole sample of E and S0 galaxies do not differ 
from those obtained separately from S0 and E (Jorgensen et al. 1996).
This relation is the one most commonly used as reference 
for the local cluster galaxies (e.g., La Barbera et al. 2003, 2010;
Di Serego Alighieri et al. 2005; Saglia et al. 2010; Raichoor et al. 2012) 
and it is best suited to comparison with 
the one we derive from our galaxies at $z=1.27$, given their properties.

The other data set that we consider is 
composed of local cluster ellipticals extracted from the
Wide-field Nearby Galaxy Cluster Survey (WINGS; Fasano et al. 2006; 
Valentinuzzi et al.2010a).
The survey core is based on optical imaging of 78 nearby 
($0.04<z<0.07$) clusters in B and V, matching
the F850LP imaging of our sample. 
The morphology of the WINGS galaxies was derived on optical images 
using the automated dedicated tool MORPHOT (Fasano et al. 2012).
The morphological indicators of MORPHOT were calibrated using a control
sample of about 1000 visually classified galaxies to provide a
fine classification resembling the one  performed visually
(see Fasano et al. 2012 for a comprehensive description).
The effective radius of the WINGS galaxies were derived from optical
images using 
GASPHOT (Pignatelli et al. 2006), an automated tool that performs a 
simultaneous fit to the major and minor axis light growth curves using a 
Sersic low convoluted with the PSF.

Stellar masses are derived, as for our sample, from the BC03 stellar population 
synthesis models using the Salpeter IMF, then rescaled to Kroupa
IMF, according to the recipe in Longhetti and Saracco (2009).
We did not apply any correction to these masses since the scaling between
Kroupa and Chabrier IMF stellar masses is lower than a factor 1.1 (Longhetti
and Saracco 2009).
From the WINGS catalog, we selected  all the elliptical galaxies and the 
transition class of E/S0 galaxies, namely galaxies of morphological type 
$-5.0<T_M<-4.0$, according to the morphological classification of the 
WINGS survey (see Table 1 in Fasano et al. 2012).
We included the transition class of E/S0 since we believe that they 
cannot be distinguished in our sample at $z\sim1.3$, given 
the difficulty in distinguishing these galaxies from pure ellipticals  
in the local universe.
The resulting sample is composed of $\sim400$ ellipticals with stellar 
masses in the range $10^9-10^{12}$ M$_\odot$ and absolute magnitude 
-21$<M_B<-15.5$.
}

\section{Deriving physical and structural parameters}
\subsection{Age, stellar masses, and absolute magnitudes}
For each galaxy of the sample, we derived  the mean age of its
stellar population, the stellar mass $\mathcal{M}_*$, and
the B- and R-band absolute magnitudes, $M_B$ and $M_R$. 
These quantities were derived by fitting   the 11 available photometric 
points of the observed spectral energy distribution (SED) at the redshift 
of the cluster, $z=1.27$, with a large set of templates built 
with different models.
In particular, we considered Bruzual and Charlot models (2003, BC03), 
the later release by Charlot and Bruzual (hereafter CB07) and the models
of Maraston et al. (2005, MA05).
We considered a Salpeter initial mass function (IMF) 
for the MA05 and BC03 models  and Chabrier (Chabrier 2003) IMF
for the BC03 and CB07 models. 
In all the cases we considered four exponentially declining star formation 
histories (SFHs) with e-folding time $\tau$= [0.1, 0.3, 0.4, 0.6] Gyr and solar 
metallicity Z$_\odot$.

Extinction A$_V$ has been considered and treated as a free parameter
in the fitting. 
We adopted the extinction curve of Calzetti et al. (2000) and allowed
$A_V$ to vary in the range $0<A_V<0.6$ mag.
For 12 galaxies out of the 16 of the sample, the best-fitting template 
is defined by SFHs with $\tau=0.1$ Gyr independently of the
model and  of the IMF used. 
The remaining four galaxies of the sample 
are best fit by SFHs with $\tau=0.3$ Gyr (\#1, \#590) and $\tau=0.4$ Gyr
(\#719 and \#1160).
The stellar mass $\mathcal{M}_*$ we derived is the mass locked 
into stars at the epoch of their observation after the gas fraction returned 
to the interstellar medium.

The $M_B$ and $M_R$ absolute magnitudes have been derived using
the observed apparent magnitudes in the filters closest to the 
rest-frame B and R of the galaxies, i.e., filters F850LP and F160W sampling 
$\lambda_{rest}\sim4000$ \AA\ and $\lambda_{rest}\sim7000$, respectively,
at the redshift of the cluster.
The color k-correction term that takes the different
filters response (e.g., F850LP vs R$_{cousin}$) into account  
was derived  from the best-fitting template.

In Table 1 of Appendix A\footnote{Published online} we report  the age, the stellar mass,  
and the absolute magnitudes obtained for each galaxy with the different IMFs 
and models considered.
The mean values of the parameters are reported in the last row of the table.
It can be seen that the different models MA05, BC03, and CB07
do not provide significantly different values of age, stellar mass,
and absolute magnitudes at fixed IMF.
In contrast,  
a Salpeter IMF provides stellar masses systematically higher than a 
Chabrier IMF (see, e.g., Longhetti et al. 2009).
In particular, as to our 16 galaxies, we obtained  
$\mathcal{M}^{Sal}_*=1.7 \mathcal{M}^{Cha}_*$ on average.

In the following, we consider the values obtained
with BC03 models and Chabrier IMF summarized in Table 3.
The 16 ellipticals have stellar masses in the range 
$0.5\times10^{10}<\mathcal{M}_*^{Cha}<2\times10^{11}$ M$_\odot$
with a median value $\mathcal{M}_*\simeq6\times10^{10}$ M$_\odot$.
Their ages are in the range 0.7-4.3 Gyr with a median value of
about 1.7 Gyr.
Since the sample is magnitude-limited and all the galaxies are at the same
redshift, the minimum stellar mass for which the sample is complete
depends on the $\mathcal{M}/L$ ratio. 
According to the method used by Pozzetti et al. (2010) we estimated,
for each galaxy, the limiting mass 
$log(\mathcal{M}_{lim})=log(\mathcal{M}_*)+0.4(z-z_{lim})$ that a galaxy
would have if its magnitude was equal to the limiting magnitude of the
sample, $z_{lim}=24$ in our case.
Considering the distribution of the values of $\mathcal{M}_{lim}$ for the 
whole sample, the minimum mass $\mathcal{M}_{min}$ above which 95\% of them 
lie is $log(\mathcal{M}_{min}/{M}_\odot)\simeq 9.8$. 
A similar result is obtained if we consider only the three 
(\#173, \#471, and \#1260) faintest galaxies of the samples, which is the 
20\% faintest galaxies  for which we 
estimated $log(\mathcal{M}_{lim}/{M}_\odot)=10.52, 10.02, 9.56$.
The galaxy in our sample with the lowest mass has a stellar mass
$log(\mathcal{M}_*/{M}_odot=9.7$, so we can consider our sample nearly 
complete over the whole mass range covered.

\begin{figure}
\begin{center}
\includegraphics[width=4truecm]{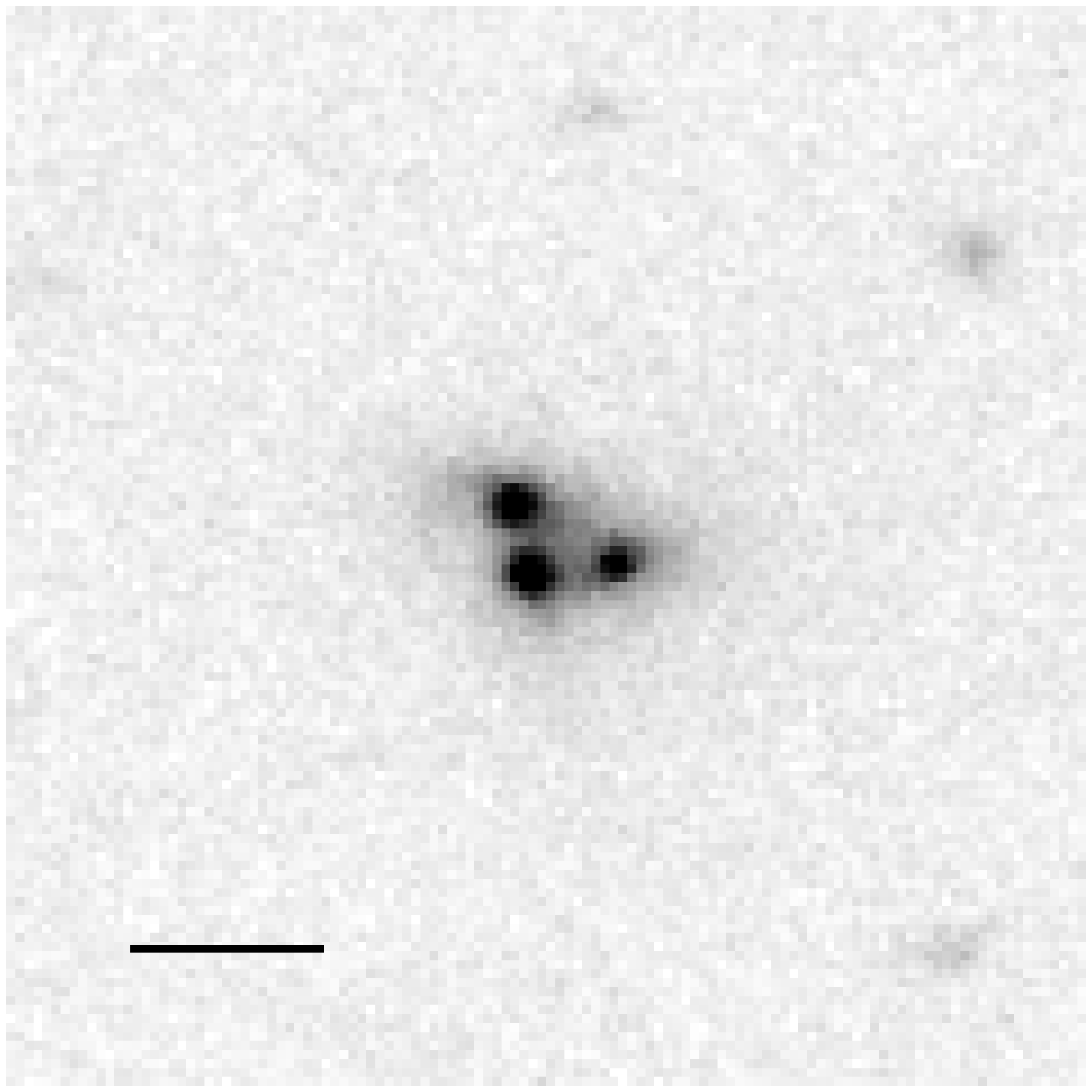}
\includegraphics[width=4truecm]{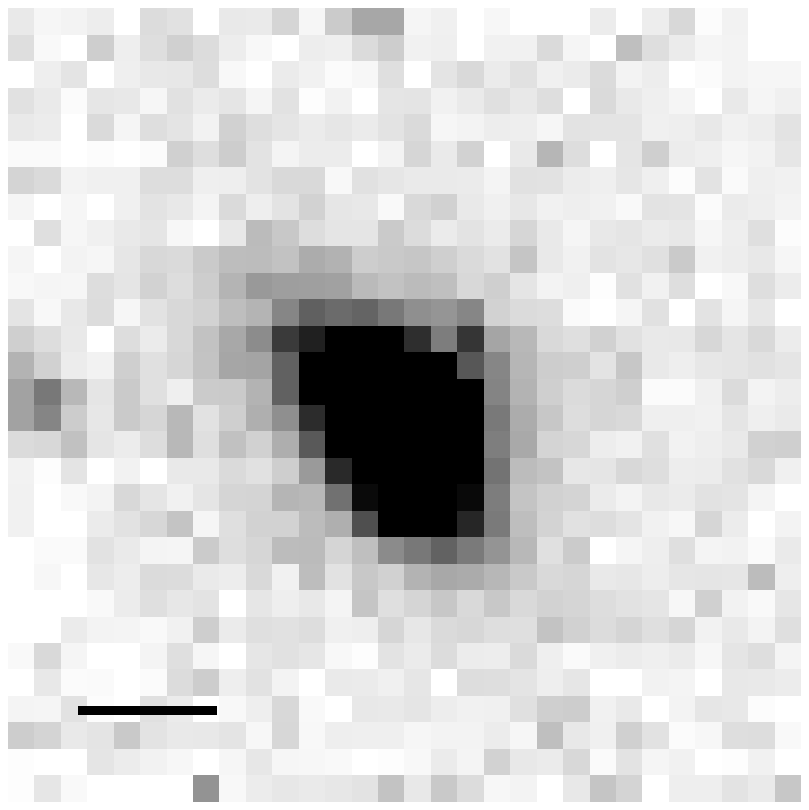}
 \caption{Three galaxies, \#5, \#626, and \#657 that are clearly resolved in
 the F850LP band image (left) thanks to both
 the resolution (FWHM$\sim0.11$ arcsec) and to the small pixel
 size (0.05 arcsec/pix). They are not resolved in the NIC3-F160W
band image (right) because of the lower resolution (FWHM$\sim0.22$ arcsec)
and pixel size (0.2 arcsec/pix).
Both the images are $6'\times6'$.
The black line represents 1 arcsec.
}
\end{center}
\end{figure}

\begin{table}
\begin{center}
\caption{Physical parameters of the sample.}
\begin{tabular}{rccccc}
\hline
\hline
   &       & \bf BC03 &\bf Cha & &   \\
\hline
 ID&age  &  logM$_*$ &M$_*(z=0)/$M$_*$& M$_B$& M$_R$ \\  
        &[Gyr]& [M$_\odot$]&     &  &	   	  \\
\hline
   1&3.75 &11.31&0.96&  -21.55& -23.39 \\  
   2&1.43 &11.16&0.93&  -22.24& -23.65 \\  
   3&2.60 &11.01&0.95&  -21.94& -23.00 \\  
   4&1.68 &11.25&0.93&  -22.65& -23.80 \\  
   5&1.43 &10.78&0.93&  -21.49& -22.71 \\  
 606&1.43 &10.70&0.93&  -21.32& -22.53 \\  
 590&2.30 &10.84&0.95&  -21.18& -22.50 \\  
 568&2.00 &10.57&0.94&  -20.81& -22.15 \\  
 719&3.50 &10.82&0.96&  -20.67& -22.10 \\  
1250&1.28 &10.24&0.92&  -20.38& -21.57 \\  
1260&0.71 & 9.70&0.89&  -20.03& -20.79 \\  
 173&4.25 &10.55&0.97&  -20.02& -21.53 \\  
1160&3.00 &10.77&0.95&  -20.89& -22.08 \\  
 657&1.43 &10.77&0.93&  -21.52& -22.73 \\  
 626&1.43 &10.76&0.93&  -21.48& -22.73 \\  
 471&1.61 &10.11&0.93&  -20.04& -21.27 \\  
\hline
\bf mean &2.10 &10.86&0.94&-21.41& -22.70\\
\hline
\hline									 
\end{tabular}								 									 
\end{center}
\end{table}

\subsection{Surface brightness profile fitting}
The effective radius R$_e$ [kpc] ($r_e$ [arcsec]) of our galaxies has 
been derived by fitting a S\'ersic profile 
\begin{equation}
I(R)=I_e exp\left[-b_n\left[\left({R\over R_e}\right)^{1/n}-1\right]\right]
\end{equation}
to the observed light profile both in the ACS-F850LP image and in the
NIC3-F160W image, i.e. in the rest-frame B and R bands of the galaxies.
The fitting was performed both assuming $n$ as a free parameter
and assuming $n=4$, i.e., a de Vaucouleour profile.
The two-dimensional fitting was performed using \texttt{Galfit} 
software (v. 3.0.4, Peng et al. 2002).
The point spread function (PSF) convolved with the S\'ersic 
profile in the F850LP image was chosen among five PSFs represented 
by four high S/N 
stars identified in the field plus a mean PSF obtained averaging the profile of 
these four stars.
The PSFs used in the NIC3 images were generated at different positions
on the array using the Tiny Tim software since the pixel size of the camera 
(0.2 arcsec/pix) does not provide a sufficient sampling of the FWHM.

We derived the effective radius $r_e=a_e\sqrt{b/a}$ 
where $a_e$ is the semi-major axis of the projected elliptical 
isophote  containing half of the total light provided by \texttt{Galfit} 
and $b/a$ is the axial ratio.
{
In all the cases considered the fit converged for all the galaxies.
In Appendix B for each galaxy we show the \texttt{Galfit} two-dimensional 
fitting model and the resulting residual.
The goodness of the fit can be seen from the residuls shown in Fig. B1
and from the one-dimensional surface brightness profiles shown in Fig. B2.
For all the galaxies the fit to the surface brightness profile extends
over more than five magnitudes and, apart from the largest galaxy (\#4), 
up to $>2$R$_e$.
}
In Table 4 we report  the best-fitting parameters $n$, R$_e$ [kpc], and
the best-fitting apparent magnitudes, $z_{850}^{fit}$ and H$_{160}^{fit}$.
The values in parentheses, were obtained by fixing $n=4$ in the fit.
The surface brightness in the B and in
the R bands was obtained from the B- and R- band absolute
magnitudes reported in Tab. 3 and corrected for 
$m_\lambda^{fit}-m_{\lambda}$, the difference between
the best fitting apparent magnitude resulting from the surface brightness
profile fitting and the observed apparent magnitude reported in Table 2.
(see Sec. 4).
The effective radii R$_e$,  as derived by fitting the S\'ersic 
profile in the ACS-F850LP images, are in the range 0.5-8 kpc,
with the exception of the dominant elliptical galaxy of the cluster 
that has an effective radius R$_e=16.7$ kpc.
The Sersic index varies in the range $2.2<n<6$ with a median value $n=4.2$.
{ 
As can be seen from Table 4 the typical uncertainty on the effective radius 
derived from the F1850LP image is about 10\%, sligthly more (15\%) in the 
F160W image.

Some authors find that the sizes derived in near-IR bands for galaxies at 
$z>1$ appear about 10\%\ to 20\% smaller than the sizes measured in optical bands
(e.g., Cassata et al. 2011; Gargiulo et al. 2012).
These works deal with field galaxies.
In fact, we do not detect this systematic in our data.
However, we have to consider that for a typical effecive radius of $\sim2$ kpc
(as for our galaxies), this systematic corresponds to 0.2-0.4 kpc, equal to the 
uncertainties (at one sigma) on our estimates of the effective radii.
Most importantly, the NIC3 camera with its pixel size (0.2 arcsec) sampling 
a physical size of $\sim1.7$ kpc at $z=1.3$ is most probably not suited to 
detecting variations on the order of a tenth of its pixel size.
The works showing differences between optical and UV rest-frame size of 
galaxies are indeed based on WFC3 data having a pixel size almost half of the 
pixel size of the NIC3 camera.
We are therefore not in the position to assess whether there is a 
trend of the size with wavelength in our data and consequently to probe
a possible dependence of this effect on the environment.
}

\begin{table*}
\caption{Morphological parameters of galaxies.}
{\small
\centerline{
\begin{tabular}{rrrrrrrcrrrr}
\hline
\hline
  ID &$n_{850}$&$b/a$& $z_{850}^{fit}$&R$_e^{F850}$ & $\langle\mu\rangle^B_e$&$Ev^B$&  &H$_{160}^{fit}$& R$_e^{F160}$ & $\langle\mu\rangle^R_e$ &$Ev^R$  \\
     &   &  &[mag]  & [kpc]                 &[mag/arcsec$^2$]     & [mag]  &  &[mag]   &    [kpc]          &   [mag/arcsec$^2$]     & mag \\
\hline
    1&  3.9& 0.7& 21.79&  7.8  (8.2)$\pm$1.2 &  20.8(20.9)$\pm$0.4 &  1.36(1.39) &  . &19.21  & 6.3(6.5) $\pm$0.2 &   19.1(19.2)$\pm$0.1 &  1.16(1.16)\\ 
    2&  6.3& 0.7& 21.08&  6.5  (3.2)$\pm$0.7 &  19.8(18.3)$\pm$0.3 &  2.35(2.21) &  . &18.51  & 9.4(4.6) $\pm$0.2 &   19.3(17.7)$\pm$0.05 & 1.90(1.86) \\ 
    3&  2.6& 0.5& 21.79&  1.3  (1.7)$\pm$0.1 &  17.0(17.6)$\pm$0.1 &  1.60(1.62) &  . &19.51  & 1.6(2.0) $\pm$0.1 &   16.5(16.9)$\pm$0.1 &  1.33(1.39)\\ 
    4&  4.4& 0.7& 20.38&  16.7(13.3)$\pm$5.6 &  21.2(20.7)$\pm$0.8 &  2.13(2.03) &  . &18.26  &10.3(8.7) $\pm$0.9 &   19.2(18.9)$\pm$0.2 &  1.73(1.75)\\ 
    5&  3.6& 0.9& 22.68&  1.7  (1.5)$\pm$0.2 &  18.6(18.4)$\pm$0.3 &  2.35(2.21) &  . &19.93  & 1.7(1.6) $\pm$0.2 &   17.0(16.8)$\pm$0.5 &  1.90(1.86) \\
  606&  4.3& 0.6& 22.22&  2.4  (2.2)$\pm$0.2 &  18.8(18.6)$\pm$0.2 &  2.35(2.21) &  . &19.69  & 3.0(2.9) $\pm$0.3 &   18.0(17.9)$\pm$0.2 &  1.90(1.86)\\ 
  590&  2.8& 0.7& 22.44&  2.4  (3.4)$\pm$0.1 &  19.0(19.8)$\pm$0.2 &  1.92(1.85) &  . &19.93  & 2.3(2.9) $\pm$0.2 &   17.7(18.2)$\pm$0.2 &  1.56(1.49)\\ 
  568&  4.2& 0.4& 22.82&  1.1  (0.8)$\pm$0.1 &  17.7(17.2)$\pm$0.2 &  1.92(1.85) &  . &20.39  & 1.2(1.2) $\pm$0.2 &   16.7(16.7)$\pm$0.3 &  1.58(1.61)\\ 
  719&  6.0& 0.7& 22.73&  0.9  (0.7)$\pm$0.1 &  17.2(16.6)$\pm$0.2 &  1.42(1.43) &  . &21.73  & 2.4(1.8) $\pm$0.6 &   19.6(19.9)$\pm$0.5 &  1.21(1.20)\\ 
 1250&  2.2& 0.8& 23.20&  2.1  (3.6)$\pm$0.3 &  19.6(20.7)$\pm$0.2 &  2.32(2.17) &  . &20.92  & 2.4(3.5) $\pm$0.2 &   18.7(19.6)$\pm$0.1 &  1.88(1.83)\\ 
 1260&  3.9& 0.7& 23.75&  2.1  (3.3)$\pm$0.4 &  20.3(21.3)$\pm$0.6 &  2.98(2.78) &  . &21.65  & 1.9(2.9) $\pm$0.2 &   18.9(19.8)$\pm$0.3 &  2.25(2.22)\\ 
  173&  3.2& 0.8& 23.63&  0.5  (0.5)$\pm$0.1 &  16.9(16.9)$\pm$0.4 &  1.21(1.31) &  . &21.23  & 0.8(0.9) $\pm$0.2 &   16.6(16.9)$\pm$0.5 &  1.03(1.09)\\ 
 1160&  4.6& 0.6& 22.54&  2.1  (1.8)$\pm$0.3 &  18.9(18.6)$\pm$0.4 &  1.58(1.52) &  . &20.19  & 2.7(2.4) $\pm$0.3 &   18.3(18.0)$\pm$0.3 &  1.31(1.30)\\ 
 657 &  2.4& 0.5& 22.12&  1.7  (2.1)$\pm$0.2 &  18.0(18.5)$\pm$0.3 &  2.35(2.21) &  . &19.93  & 1.7(2.1) $\pm$0.2 &   17.0(17.5)$\pm$0.5 &  1.90(1.86) \\
 626 &  4.2& 0.6& 21.45&  2.1  (2.5)$\pm$0.3 &  17.8(18.1)$\pm$0.3 &  2.35(2.21) &  . &19.93  & 2.2(2.5) $\pm$0.3 &   17.5(17.8)$\pm$0.5 &  1.90(1.86) \\
 471 &  4.6& 0.9& 23.33&  2.3  (1.9)$\pm$0.3 &  19.8(19.4)$\pm$0.4 &  2.13(2.03) &  . &--	  &--	     --       &   --		     &  --    \\
\hline
\end{tabular}
}}
\tablefoot{Sersic index $n$, axial ratio $b/a$, apparent magnitude and 
effective radius [kpc] as derived from the fitting to the surface brightness 
profile in the F850LP image and in the F160W image.   
The values in parenthesis have been obtained assuming $n=4$.
The terms Ev$_{B,R}$ represent the luminosity evolution that
the stellar population of each galaxy experiences in the B and R bands, 
respectively, in the $\sim8.6$ Gyr from $z=1.27$ to $z=0$, 
according to its own age at $z=1.27$ and SFH using the BC03 models 
(see text for a detailed description). The values in parenthesis have 
been obtained using the Maraston's models.}
\end{table*}

\begin{figure*}
\begin{center}
\includegraphics[width=8.5truecm]{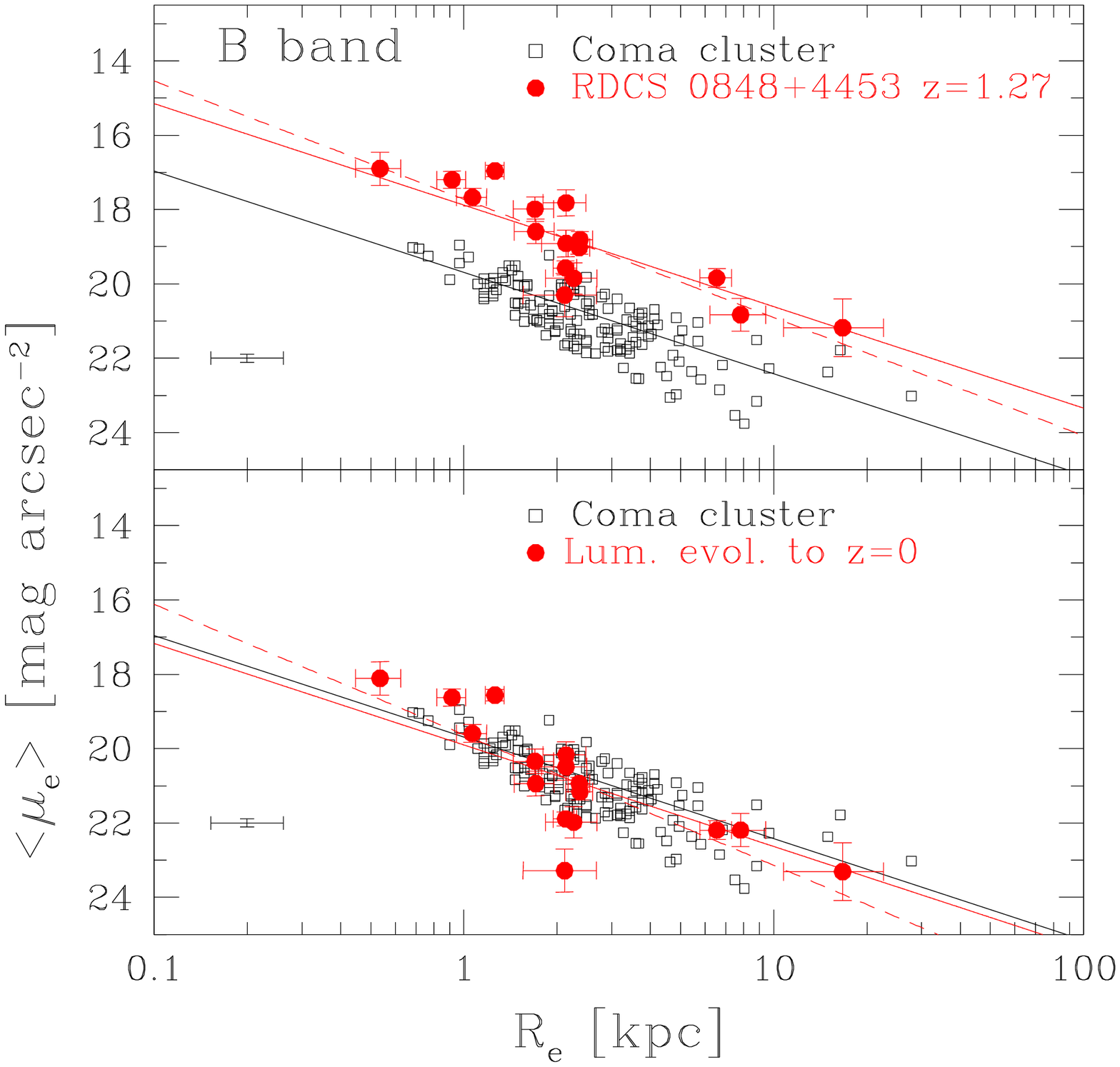}
\includegraphics[width=8.5truecm]{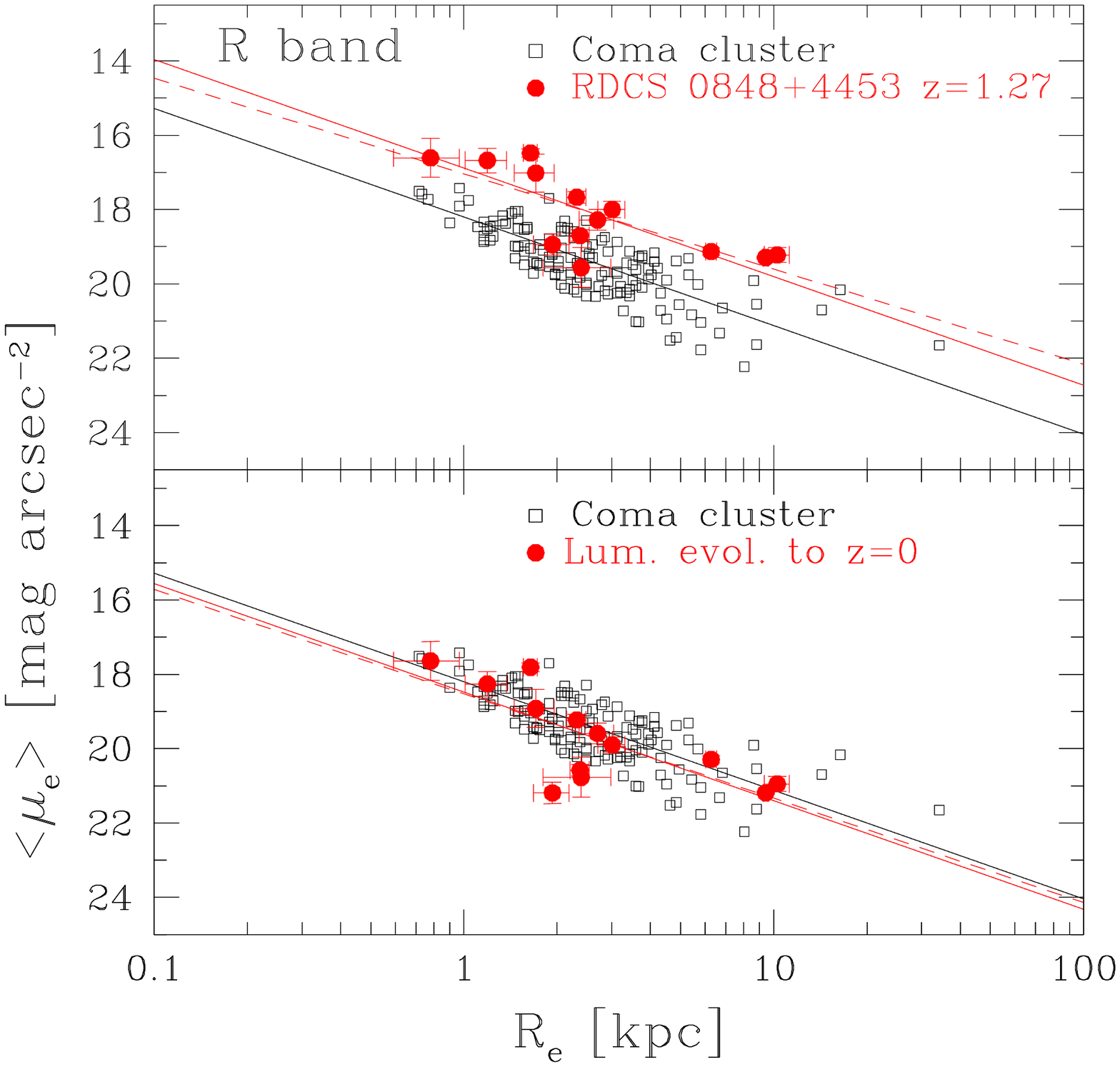}
\caption{Kormendy relation in the rest-frame B band (left panel) and
 R band (right panel). 
  Red filled symbols are our 16 cluster ellipticals at $z=1.27$ (upper panels)
  and evolved to $z=0$ (lower panels) according to the passive luminosity evolution 
  as described in Sec. 5.1 and reported in Tab. 4. 
  The red lines are the Kormendy relation reported in Tab. 5 obtained by 
  fitting eq. (4) to our 16 ellipticals assuming the slope $\beta$ at $z=0$ 
  (solid line) and leaving $\beta$ as free parameter (dashed line).
  The open squares are the sample of early-type galaxies in the Coma cluster
  studied by Jorgensen et al. (1995a). 
  The black lines represent the Kormendy relation at $z=0$ derived from
  this sample and reported in Eqs. (7) and (8).
}
\end{center}
\end{figure*}

\section{The Kormendy relation at $z=1.3$}
The mean effective surface brightness $\langle\mu\rangle_e$ [mag/arcsec$^2$]
of a galaxy 
can be computed from the apparent magnitude $m$ and the 
area included within the effective radius $r_e$ [arcsec]:
\begin{equation}
\langle\mu\rangle_e =m+2.5log(r_e^2)+2.5\log(2\pi).
\end{equation}
By substituting the observed quantities with those in the rest frame
of the galaxy, we obtain
\begin{equation}
\langle\mu\rangle_e^\lambda =M_\lambda(z)+2.5log(R_e^2)+38.57,
\end{equation}
where $M_\lambda(z)$ is the absolute magnitude of the galaxy
at the rest-frame wavelength $\lambda$ at redshift $z$, and
$R_e$ is in [kpc], after correcting for the cosmological dimming
term 10log(1+z).
This relation can be used to trace the evolution that ellipticals
undergo through time.
The mean surface brightness $\langle\mu\rangle_e$ is indeed expected
to change with redshift because of the passive luminosity evolution of the 
stellar populations that affects the absolute magnitude in Eq. (3).
Moreover, if elliptical galaxies grow with time due to merging, then their
surface brightness will change further, both due to the change in
their luminosity (proportional to the mass increase) and
to the consequent increase in their effective radius. 
The variation in R$_e$ will add up to the luminosity evolution as 
the square of its variation, significantly affecting the observed  
change of $\langle\mu\rangle_e$.
We come back to this issue in the next section.
\begin{table*}
\caption{Best-fitting $\alpha$ and $\beta$ values of eq. (4) for
the B and R band rest-frame data. The first two columns report the values
obtained by fitting the (R$_e$,$\langle\mu\rangle_e$) data derived 
with free index $n$ while the third and the fourth columns report
the values relevant to (R$_e$,$\langle\mu\rangle_e$) obtained 
assuming the index $n=4$.
The upper panel reports the values obtained by fitting
the data at $z=1.27$ while the lower panel reports the
values obtained by fitting the evolved data to $z=0$ according
to eq. (11). 
The last column in the upper panel reports the mean observed 
surface brightness evolution $\Delta\mu$ for a fixed slope. 
The values relevant to $n=4$ case coincide with those obtained 
with free index $n$.
}
\centerline{
\begin{tabular}{ccccccc}
\hline
\hline
Band&  $\alpha$	 &  $\beta$   &  $\alpha_{n=4}$ & $\beta_{n=4}$ & $\alpha(\beta_{z=0})$ & $\Delta\mu$ (mag)\\ 
\hline
$z=1.27$   &              &            &      &             &              &         \\
---------- &              &      &      &             &              &         \\ 
B& 17.7$\pm0.1$ & 3.2$\pm0.5$&	 17.9$\pm0.2$& 2.7$\pm0.6$ & 17.9$\pm0.2$ & -1.8$\pm0.2$ \\
R& 17.0$\pm0.2$&  2.6$\pm0.7$&	 17.6$\pm0.3$ & 2.6$\pm0.7$ & 16.9$\pm0.2$ &-1.3$\pm0.2$\\
 \hline	
$z=0$      &              &            &      &             &              &         \\
---------- &              &         & &             &              &         \\ 
B& 19.6$\pm0.2$ & 3.3$\pm0.6$&	 19.8$\pm0.3$& 3.0$\pm0.6$ & 19.9$\pm0.3$ & --- \\
R& 18.6$\pm0.3$&  2.8$\pm0.7$&	 19.0$\pm0.3$ & 2.6$\pm0.7$ & 18.5$\pm0.3$ & ---\\
\hline 								 
\hline									 
\end{tabular}								 
}									 
\end{table*}

The evolution of the surface brightness $\Delta\mu$ of elliptical galaxies 
is usually quantified by using the  Kormendy relation 
(KR, Kormendy 1977), a linear scaling relation 
between the logarithm of the effective radius R$_e$ [Kpc] 
and the mean surface brightness $\langle\mu\rangle_e$ 
within R$_e$:
\begin{equation}
\langle\mu\rangle_e = \alpha + \beta \log(R_{e}).
\end{equation}
Elliptical galaxies both in field and in clusters follow this tight
relation with a slope close to three out to $z\sim1$ (Hamabe and Kormendy
1987; Schade et al. 1996; Ziegler et al. 1999; La Barbera et al. 2003,
2004; Reda et al. 2004; di Serego Alighieri et al. 2005). 
On the other hand, the zero point $\alpha$ is found to vary with the 
redshift of the galaxies, reflecting 
their luminosity evolution and the possible evolution of R$_e$ 
over time. 
Since the value of $\alpha$ strictly depends on the photometric band 
and system selected to derive magnitudes and morphological parameters, 
its value needs to be transformed into that of a common rest-frame 
wavelength when comparisons at different $z$ are performed. 
Since morphological parameters have been derived in the F850LP 
and F160W bands, the comparison with the local scaling relations will 
be done considering the rest-frame B-band and R-band.   
For each galaxy of our sample, we therefore computed the mean surface brightness
in the B-band,
\begin{equation}
\langle\mu\rangle_e^B=M_B^{fit}+5log(R_e^{F850})+38.57,
\end{equation}
and in the R-band,
\begin{equation}
\langle\mu\rangle_e^R=M_R^{fit}+5log(R_e^{F160})+38.57,
\end{equation}
where $M_B^{fit}$
and 
$M_R^{fit}$
are the absolute magnitudes derived from the \texttt{Galfit} best-fitting
apparent magnitudes $z_{850}^{fit}$ and H$_{160}^{fit}$.
The surface brightnesses thus obtained are reported in Tab. 4.
In the upper panels of Fig. 4 the surface brightness of our 16
galaxies is plotted as a function of their R$_e$ 
on the  B-band [R$_e$; $\langle\mu\rangle_e$] plane (left) and on the R-band 
plane (right).  
The Kormendy relations derived from the sample
of early-type galaxies belonging to the Coma 
cluster at $z=0.024$ studied by Jorgensen et al. (1995a, 1996; see Sec. 2.4
for a description of the data) are also shown.
In particular, the KR in the B-band,
\begin{equation}
\langle\mu\rangle_e^B=19.7+2.73log(R_e) \hskip 1truecm z=0 \hskip 0.5truecm (n=4),
\end{equation}
and in the R-band,
\begin{equation}
\langle\mu\rangle_e^R=18.3+2.92log(R_e)\hskip 1truecm z=0 \hskip 0.5truecm (n=4).
\end{equation}
Figure 4 shows the KR we obtained at $z=1.27$ by 
fitting eq. (4) to our 16 ellipticals fixing the slope at $z=0$ ($\beta_{z=0}$) 
(solid line) and considering $\beta$ as free parameter (dashed line).
The resulting best fitting relation we obtained (for a free $n$ index) 
is
\begin{equation}
\langle\mu\rangle_e^B=17.7(\pm0.1)+3.2(\pm0.5)log(R_e)\hskip 1truecm z=1.27 
\end{equation}
in the B band and
\begin{equation}
\langle\mu\rangle_e^R=17.0(\pm0.2)+2.6(\pm0.7)log(R_e)\hskip 1truecm z=1.27 
\end{equation}
in the R band.
The slope $\beta$ of the KR we obtain agrees within the errors with 
the slope of the KR at $z=0$ (eqs. 7 and 8).
In Tab. 5 we report the parameters $\alpha$ and $\beta$ of the KR relation 
obtained by fitting the (R$_e$, $\langle\mu\rangle_e$) data obtained both 
for a free $n$ index and for $n=4$.
The good agreement between the slope obtained at 
$z\sim1.3$ and the local value in the case of $n=4$ is worth noting.
We also report the value of $\alpha$ obtained by fixing the 
$\beta$ slope at the value at $z=0$ and the resulting amount of evolution 
$\Delta\mu=\langle\mu\rangle_e(z=1.27)-\langle\mu\rangle_e(z=0)$
between $z=0$ and $z=1.27$. 
That the slope of the relation is not significantly 
changed in the 
past 9 Gyr means that the luminosity and the effective radius of these
elliptical galaxies scale according to the same rule seen in the local
Universe. 
In contrast, the zero point $\alpha$ of the relations at $z=1.27$ 
is significantly brighter than 
at $z=0$, 1.8 magnitudes brighter in the B-band and 1.3 magnitudes brighter 
in the R-band (see Tab. 5).
These offsets agree with those found 
by Holden et al. (2005) and by Rettura et al. (2010) for the 
cluster RDCS J1252.9-2927 at $z=1.237$ and with what was found by
Raichoor et al. (2012) on a sample of cluster and group galaxies
belonging to the Linx supercluster at $z\simeq1.27$ including
7 out of the 16 ellipticals of our sample.
These offsets account for the evolution that galaxies 
underwent and can be affected by any change that galaxies experience
caused by the passive luminosity evolution, by the possible evolution of
the effective radius, and by the possible mass accretion.
In the next section we consider how these evolutionary terms can affect
the resulting Kormendy relation.

\section{The evolution of cluster ellipticals since $z=1.3$}
In this section we discuss the  possible evolution that 
the 16 cluster ellipticals of our sample may experience since $z\sim1.3$.
The aim of this analysis is twofold. 
On one hand, we are interested in constraining the evolution that 
brings these ellipticals on the local scaling relations.
On the other hand, we want to understand whether they have completed 
their mass growth at the redshift they are observed or  significant 
structural changes (mass accretion and/or size increase) can or even must 
take place in the last 9 Gyr.
Before considering different possible evolutionary paths we have to
consider the passive luminosity evolution.
Indeed,  the stars already formed (the stellar mass already assembled) at $z=1.27$ 
will passively evolve till $z=0$ due to their aging.

\subsection{The unavoidable luminosity evolution}
The zero point $\alpha$ of the KR relation is expected to change 
because of the change in $\langle\mu\rangle_e$ with time.
The surface brightness is expected to change since 
the luminosity of a galaxy changes with time.
This is due to the aging of the stars already formed and assembled at the time 
the galaxy has been observed.
This luminosity evolution will take place and will affect the stellar
mass of the galaxy.
It is well known that the aging of a stellar population implies a 
dimming  of its luminosity with time.
The magnitude of this dimming at a given wavelength over a 
time $\Delta t$ depends primarily on the age of the stellar population at the 
beginning of this interval and it is greater for younger ages.
As shown in Tab. 3, our galaxies have different ages.
Instead of computing a mean evolution for all of them, 
we thus computed its own luminosity evolution 
Ev$^{B,R}=[M_{B,R}(z=1.27)-M_{B,R}(z=0)]$ for each of them , 
i.e., the difference between 
the absolute magnitude of the best-fitting model at $z=1.27$ and the 
absolute magnitude of the same model aged 8.6 Gyr, the time 
elapsed from $z=1.27$ to $z=0$.

The evolutionary terms $Ev^{B,R}$ thus obtained are reported in Tab. 4.
They are in the ranges 1.21 mag$<|Ev^B|<2.98$ mag for the B band 
and 1.03 mag$<|E^R|<2.25$ 
mag for the R band and, as expected, are greater for younger galaxies.
It is worth noting that these evolutionary terms are almost independent
of the IMF  and the models.
Indeed, in Tab. 4 we also report (in parenthesis) 
the evolutionary terms derived with the MA05 models and 
Salpeter IMF to be compared with those obtained with BC03 and Chabrier IMF.

For a given model, the Salpeter IMF produces differences in the 
range 0.04-0.06 mag with respect to the Chabrier IMF.
Thus, if our galaxies evolve solely in luminosity since $z=1.27$ 
according to their SFH, their surface brightness at $z=0$ would be
\begin{equation}
\langle\mu\rangle_{e,z=0}^{B,R}=\langle\mu\rangle_e^{B,R}- Ev^{B,R}
\end{equation}
In the lower panels of Fig. 4 the surface brightness evolved to $z=0$,
$\langle\mu\rangle_{e,z=0}$, of our 16
galaxies is plotted as a function of their R$_e$ 
in the  B-band (left) and in the R-band (right).
Symbols are as in the upper panels.
Our galaxies occupy the locus occupied by the Coma cluster ETGs with 
comparable mass. 
The expected Kormendy relation at $z=0$ obtained 
by fitting eq. (4) to our evolved  data is also shown.
We obtained $\alpha^B_{z=0}=19.6\pm0.3$ and 
$\alpha^R_{z=0}=18.6\pm0.4$, in agreement with the
Kormendy relation in the local universe.
Hence, the luminosity evolution  that the stars already assembled
in the 16 ellipticals will necessarily experience between $z=1.27$ to $z=0$ 
brings them on the local Kormendy relation; that is, it accounts for
the observed surface brightness evolution $\Delta\mu$ reported in Tab. 5.

It is worth noting that a similar evolution of the KR 
($\simeq1.5-2$ mag/arcsec$^2$) was also observed
by Holden et al. (2005), Raichoor et al. (2012), and Rettura et al. (2010) 
for cluster ellipticals at similar redshifts.
Analogously to our findings, Holden et al. and Raichoor et al. find that
this surface brightness evolution is consistent with the expected  
luminosity evolution due to aging and  
a similar result is also found at $z<0.9$ from the study of the 
evolution of the fundamental plane (FP) 
(Saglia et al. 2010).

This result has an important and
constraining implication:
the stellar mass underlying the luminosity of these 
ellipticals at $z=1.27$ and responsible for the observed surface brightness 
must be distributed  according to the same profile of the stellar mass 
responsible for the corresponding luminosity in ellipticals at $z=0$.
This means that
the stellar mass assembled at $z=1.27$ is not more concentrated 
than at $z=0$; otherwise, once evolved to $z=0$, the underlying stellar mass 
would result in a higher surface brightness with respect to that 
of ellipticals with the same luminosity at $z=0$. 

Indeed, comparing
the size-surface brightness relation of our 16 ellipticals evolved to 
$z=0$ with the one described by a sample of cluster ellipticals 
at $z=0$ selected in the same absolute magnitude range and
in the same stellar mass range, we obtain what it is shown in Fig. 5.
The sample of local cluster ellipticals has been extracted from the
Wide-field Nearby Galaxy Cluster Survey (WINGS; Fasano et al. 2006; 
Valentinuzzi et al.2010a) as described in Sec. 2.4.
To homogenize high-z with low-z data, we computed the effective radius 
of our 16 ellipticals again at  $z=1.27$ using GASPHOT  the same software
as was used to derive the structural parameters of WINGS galaxies.
{ For completeness, in Fig. 5 we also show our 16 galaxies in the 
case of GALFIT estimates, even if the proper comparison is the one
based on the same procedure as used to estimate the structural parameter.
However, it can be seen that the result is robust with respect to the 
software used to derive the structural parameters.}

The evolved B-band absolute magnitudes of our 16 ellipticals are in the 
range $-21<M_B<-17$, as can be derived from Tables 3 and 4 while their 
stellar masses are in the range 
$0.5\times10^{10}$ M$_\odot<\mathcal{M}_*<2\times10^{11}$ M$_\odot$.
In the left hand panel of Fig. 5 crosses represent the WINGS cluster 
ETGs selected in the same absolute magnitude range, while in the right hand panel
represent those selected in the same mass range.
In both the cases, the high-z and the low-z samples occupy the same region.
In the lower panels of Fig. 5, the distribution of the effective radius
of the selected samples at different redshifts are shown and compared.
The agreement between the distributions of the high-z and the low-z samples 
is also quantitatively confirmed by the KS test we performed, whose
probability $P$ is reported in the insets.
The stellar mass of our 16 ellipticals at $z=1.27$ is not
more concentrated than the stellar mass of local cluster ETGs with
the same luminosity and stellar mass.
The lower-right panel of Fig. 5, compares the effective 
radius of ETGs selected in the same stellar mass range and shows quantitatively
that the mean effective radius of the population of high-z cluster ETGs
does not differ from the mean effective radius of local cluster ETGs
with the same mass; that is, they follow the same size-mass relation.
This is explicitly shown in Fig. 6 where the size-mass relation described 
by our galaxies is compared with the one defined by 
the WINGS ETGs.
It is worth noting that the effect of the secular decrease in 
the galaxy stellar mass due to the stellar evolution (Poggianti et al. 2013b),
which is the gas fraction 
returned to the interstellar medium due to the evolution of the stars,
would be negligible in our case.
This is shown in Tab. 3 where we report  the correcting 
factors $\mathcal{M}_*(z=0)/\mathcal{M}_*$ for each galaxy,
namely the ratio between the stellar mass $\mathcal{M}_*(z=0)$ that 
the galaxy would have at $z=0$ owing to the evolution of the stars
and the stellar mass we estimated that they have at $z=1.27$.

In Fig. 6 the size-mass relation for a sample of field
ellipticals in the redshift range $0.9<z<1.9$ (cyan filled triangles;
Saracco et al. 2010) is also shown for comparison.
It can be seen that, as also found by Raichoor et al. (2012), field and 
cluster elliptical galaxies seem to follow the same size-mass relation 
and no appreciable differences in their effective radii are visible.
However,  we refer to a forthcoming paper for a detailed
and quantitative comparison between field and cluster ellipticals at $z>1$.  
\begin{figure*}
\begin{center}
\includegraphics[width=8.5truecm]{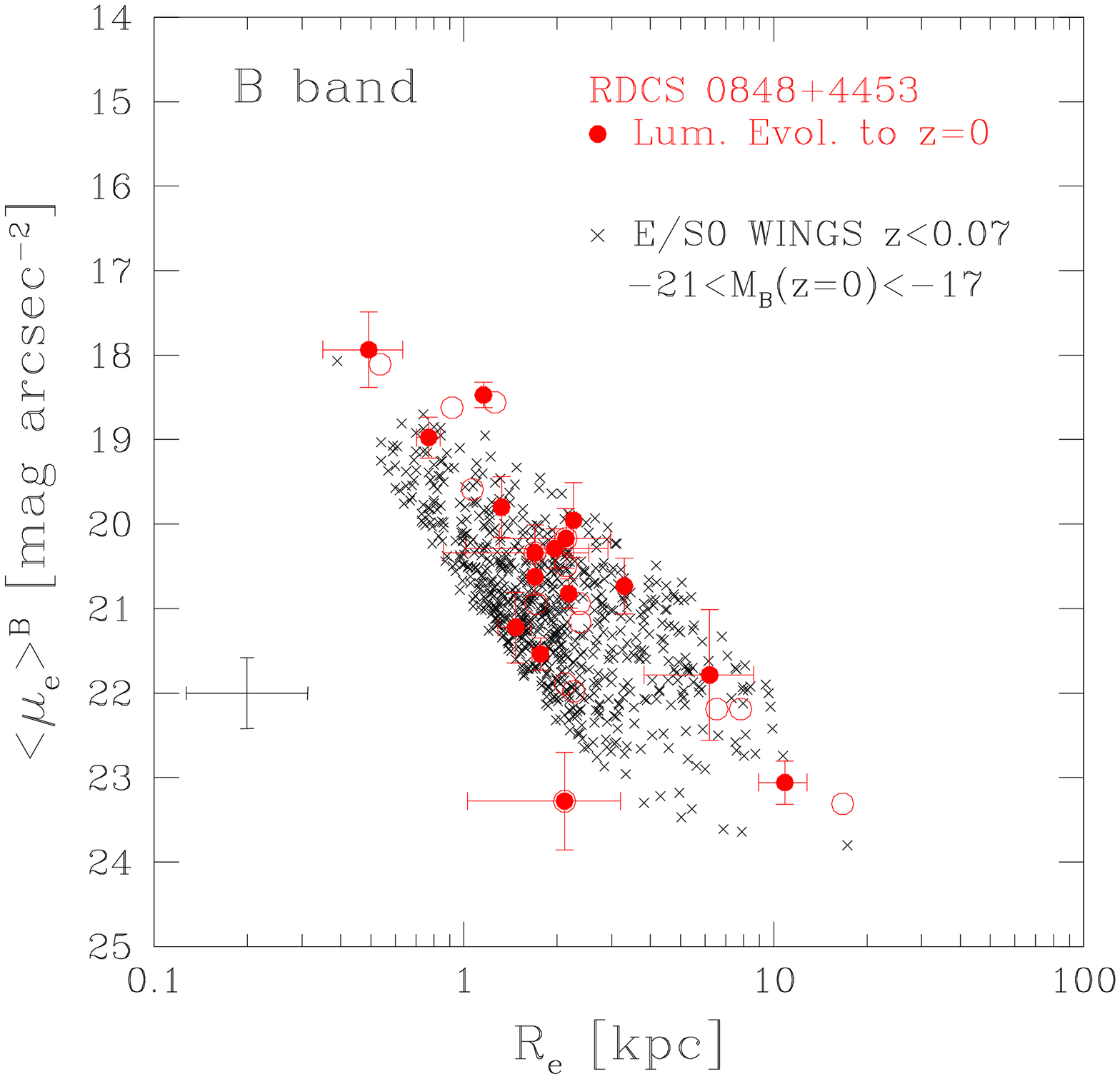}
\includegraphics[width=8.5truecm]{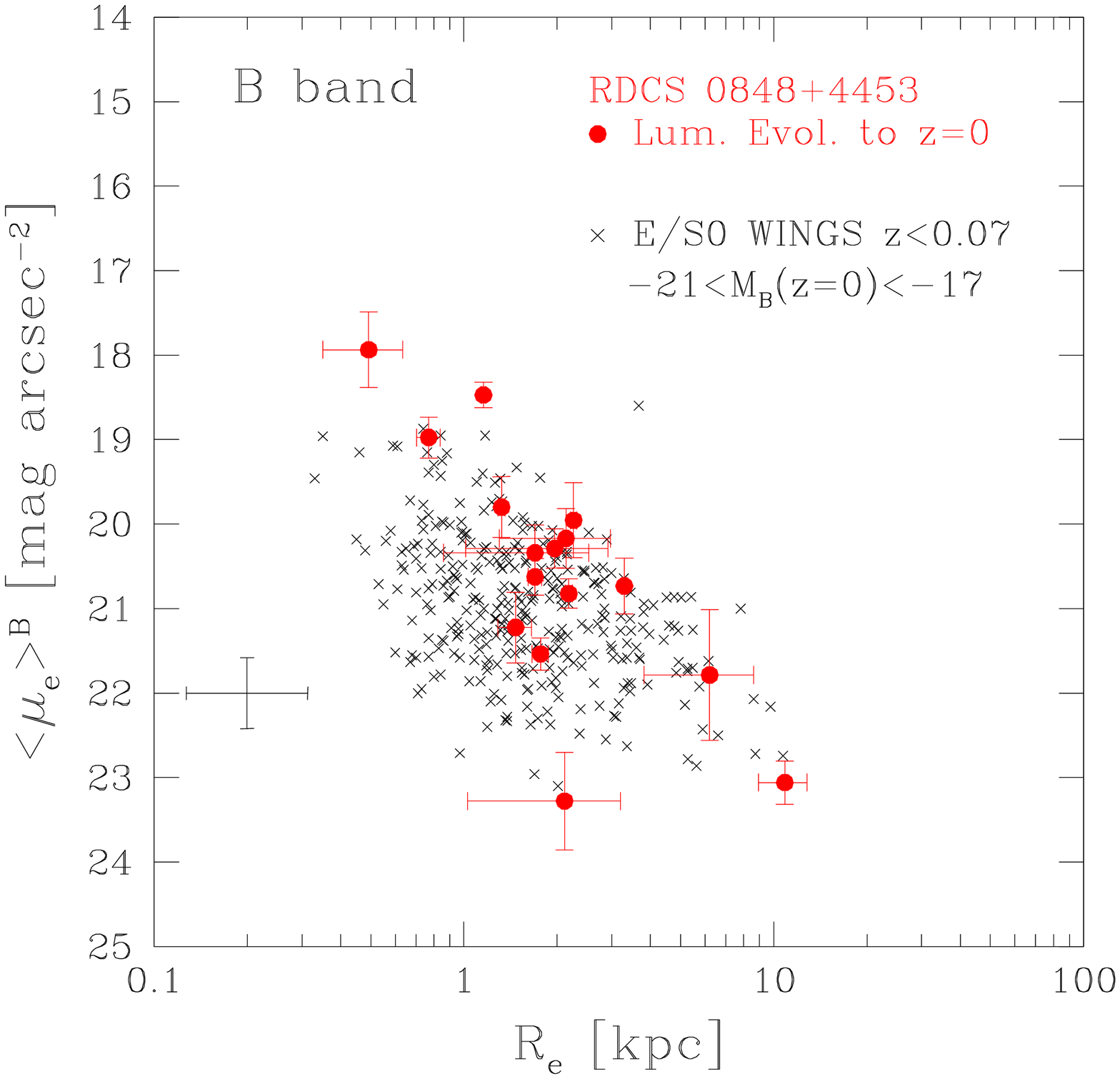}
\includegraphics[width=8.5truecm]{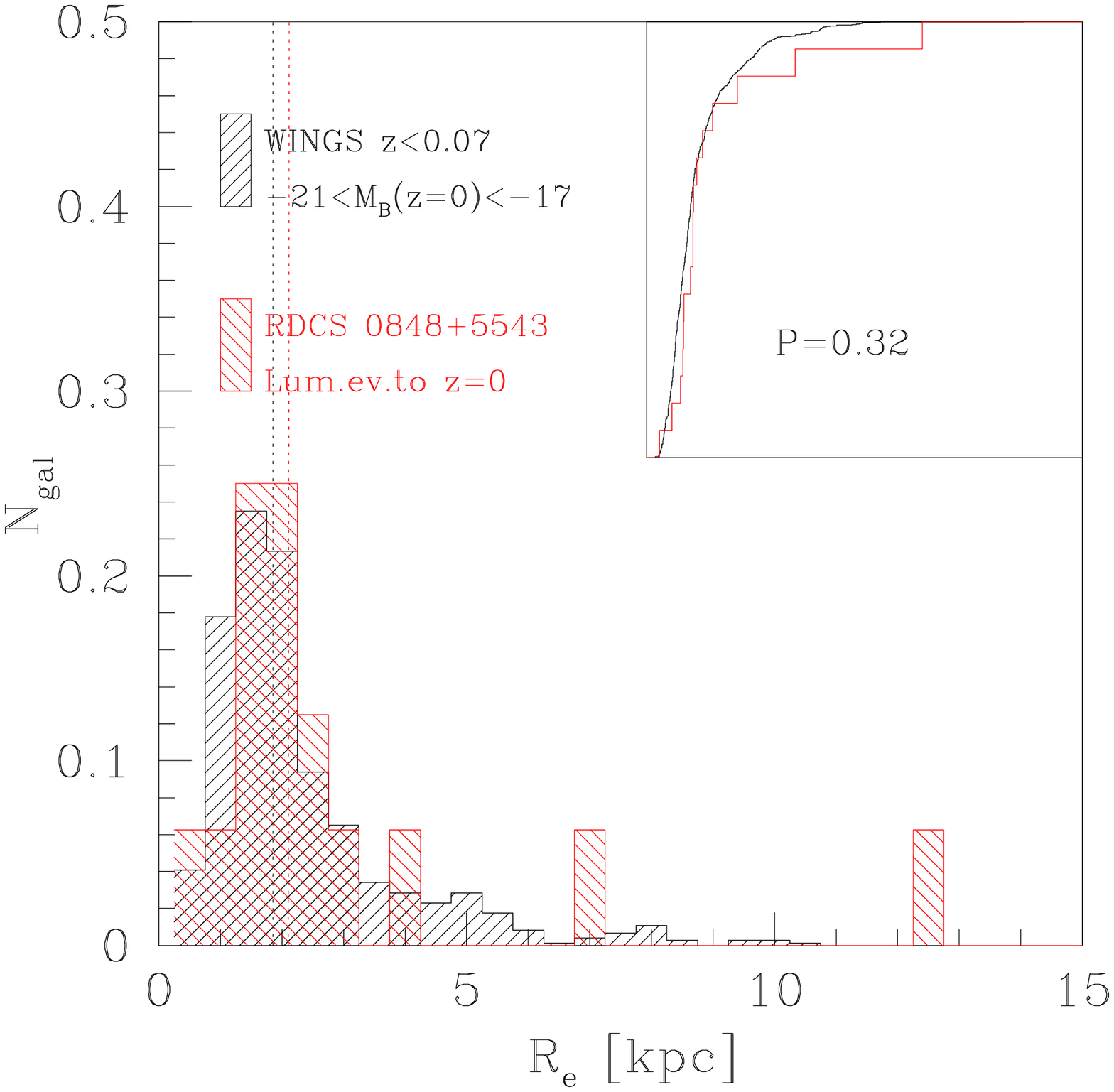}
\includegraphics[width=8.5truecm]{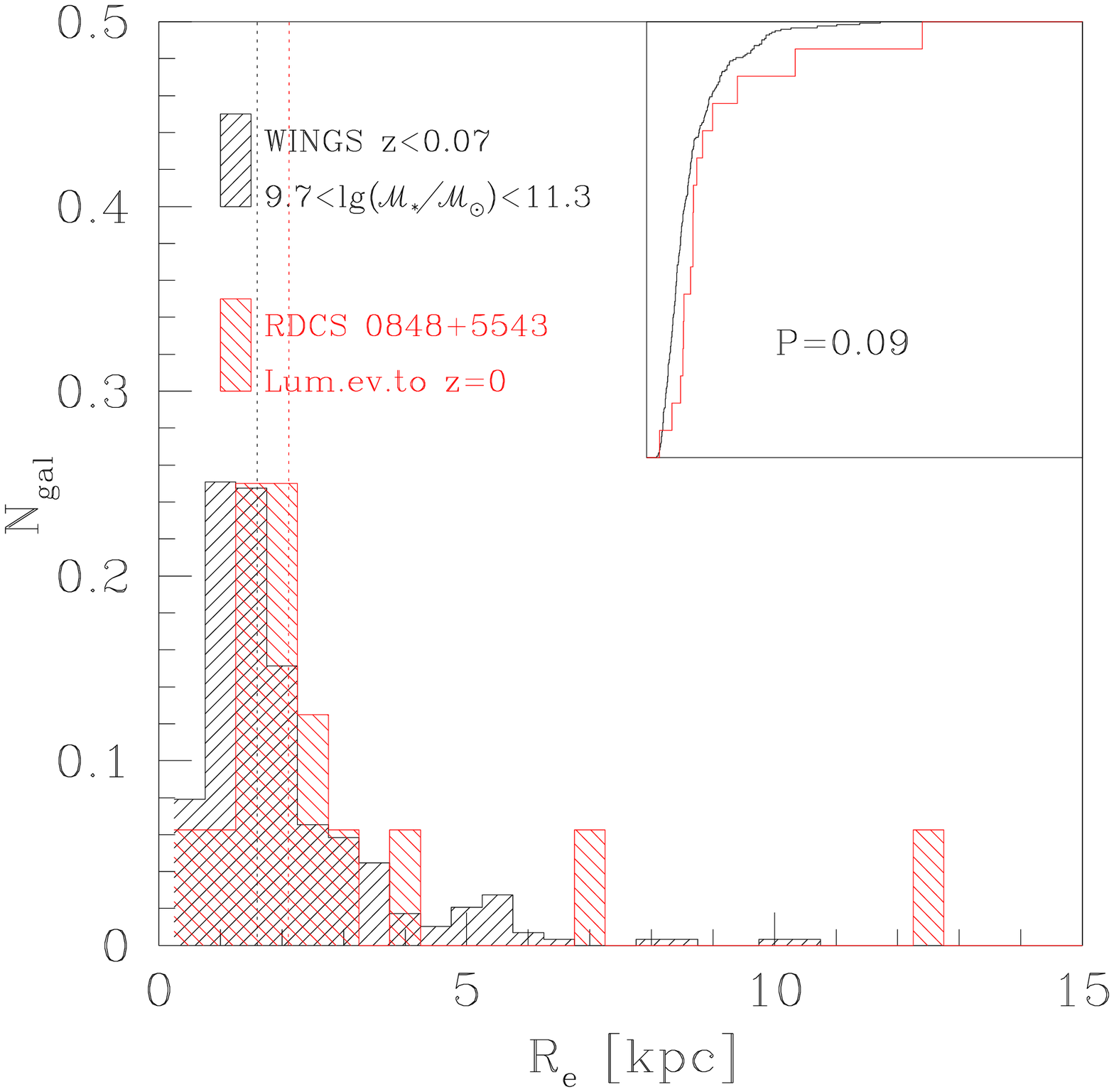}
 \caption{Upper panels - Size-surface brightness relation in the rest-frame 
 B-band. Red circles are our 16 cluster ellipticals evolved to $z=0$. 
{ Filled circles represent the values obtained using GASPHOT, consistently
with the WINGS sample, while open circles are the values derived
using GALFIT.}
Crosses are local cluster ellipticals selected from
 the WINGS sample in the same absolute magnitude range 
 ($-21<M_B<-17$; left panel) of the 16 galaxies evolved to $z=0$ 
 and in the same stellar mass range 
 ($5\times10^{10}$ M$_\odot<\mathcal{M}_*<2\times10^{11}$ M$_\odot$, 
 right panel).
 Lower panels - The distributions of the effective radius of 
 the 16 ETGs at $z=1.27$ (red histogram) and of the WINGS galaxies selected 
 according to the luminosity (left) and stellar mass (right) criteria are 
 shown and compared using the K-S test. In the small insets the cumulative
 distributions are shown together with the probability that they 
 belong to the same parent populations.
Effective radius have been computed using the same
procedures for the high-z and the low-z samples (see text).
}
\end{center}
\end{figure*}

In this section we have shown that since
the passive evolution experienced by the stars present at $z=1.27$ moves
the 16 ETGs on the Kormendy relation of local ETGs and that they follow
the same size-mass relation 
their stellar mass profile at $z=1.27$ is the same
as local cluster ellipticals  with the same stellar mass.
On the other hand, this does not imply that they cannot change their structure,
e.g., grow further at $z<1.27$ moving along the Kormendy and the
size-mass relation. 
In the next section we tackle the possible structural
evolution that they may experience.

\begin{figure}
\begin{center}
\includegraphics[width=8.5truecm]{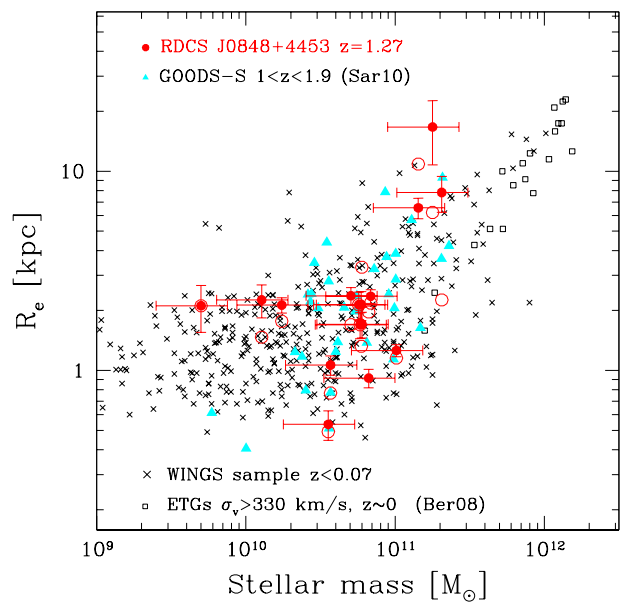}
 \caption{Size-stellar mass relation for elliptical galaxies.
 The different symbols are as follow:
 red  circles are our 16 cluster ellipticals at $z=1.27$ 
 ({ filled circles represent the R$_e$ derived using GASPHOT 
 consistently with the WINGS sample, open circles are those derived
with GALFIT)};
 blue filled triangles are field ellipticals selected from the sample
 of Saracco et al. (2010, Sar10) at $1.1<z<1.5$, 
 crosses are local cluster ellipticals selected from the WINGS 
 survey, and open squares are the local high-mass cluster ellipticals 
 with $sigma_v>330$ km/s selected by Bernardi et al. (2008, Ber08).     
}
\end{center}
\end{figure}

\subsection{Size evolution, mass accretion, and structural evolution}
We have seen that our 16 cluster ellipticals at $z\sim1.3$ share the same 
scaling relations as local cluster ellipticals; that is, they follow the
local size-mass relation and also the local Kormendy relation, 
once the aging of their stars is taken into account.
Any further evolution that may occur at $z<1.27$  
(stellar mass accretion and/or effective radius evolution), in addition
to the aging of the stellar mass already assembled,
thus must keep them  in the local scaling relations.
In practice, any variation in R$_e$ of individual galaxies must be accompanied 
by a compensating variation of their absolute magnitude and of their 
stellar mass otherwise galaxies would move away from the local 
relations.
This implies that an almost pure evolution of 
R$_e$ of individual galaxies, i.e. an expansion of the galaxy without the 
compensating variation of the luminosity/mass, is ruled out for our galaxies
since it would turn them away from the local relations.

To clearly show this effect 
we considered the mild size evolution found by Delaye et al. (2013; see also
Papovich et al. 2012)
for cluster ETGs, $R_e\propto (1+z)^b$ with $b=-0.53$.
This rate of evolution means that the effective radius of our 
galaxies at $z=1.27$ is on average 0.65 times the effective radius of the 
galaxies at $z=0$ with the same stellar mass; that is, our galaxies should 
expand by a factor 1.5 since $z=1.27$. 
We applied this evolution to each of the 16 galaxies, and the results are 
shown in Fig. 7.
The left hand panel shows the best-fitting relation to our 16 
galaxies once evolved according to the above size evolution in addition to
the passive luminosity evolution.
The offset with respect to the observed local KR relation is 
about $\Delta\mu\simeq0.8$ mag arcsec$^{-2}$. 

The middle and the right hand panels are the same plots as in Fig. 5 (lower panels),
showing the comparison between the local WINGS sample and our 16 galaxies
once expanded.
It is evident in the right hand panel of Fig. 7 the large 
discrepancy of our galaxies with respect to the local size-mass relation 
produced by the size evolution, a discrepancy even greater than the
one with the local KR shown in the middle panel. 
The above mild size evolution can be ruled out at more than 
4$\sigma$ as shown by the KS probabilities obtained.
A size evolution even stronger than this 
with rates in the range $-2<b<-1$
is claimed for field early-type galaxies 
(Damjanov et al. 2011; Huertas-Company et al. 2012; Cimatti et al. 2012).
If applied to cluster galaxies this size evolution would
imply factors 2-5 of increase in the effective radius of galaxies
since $z=1.27$.
\begin{figure*}
\begin{center}
\includegraphics[width=5.5truecm]{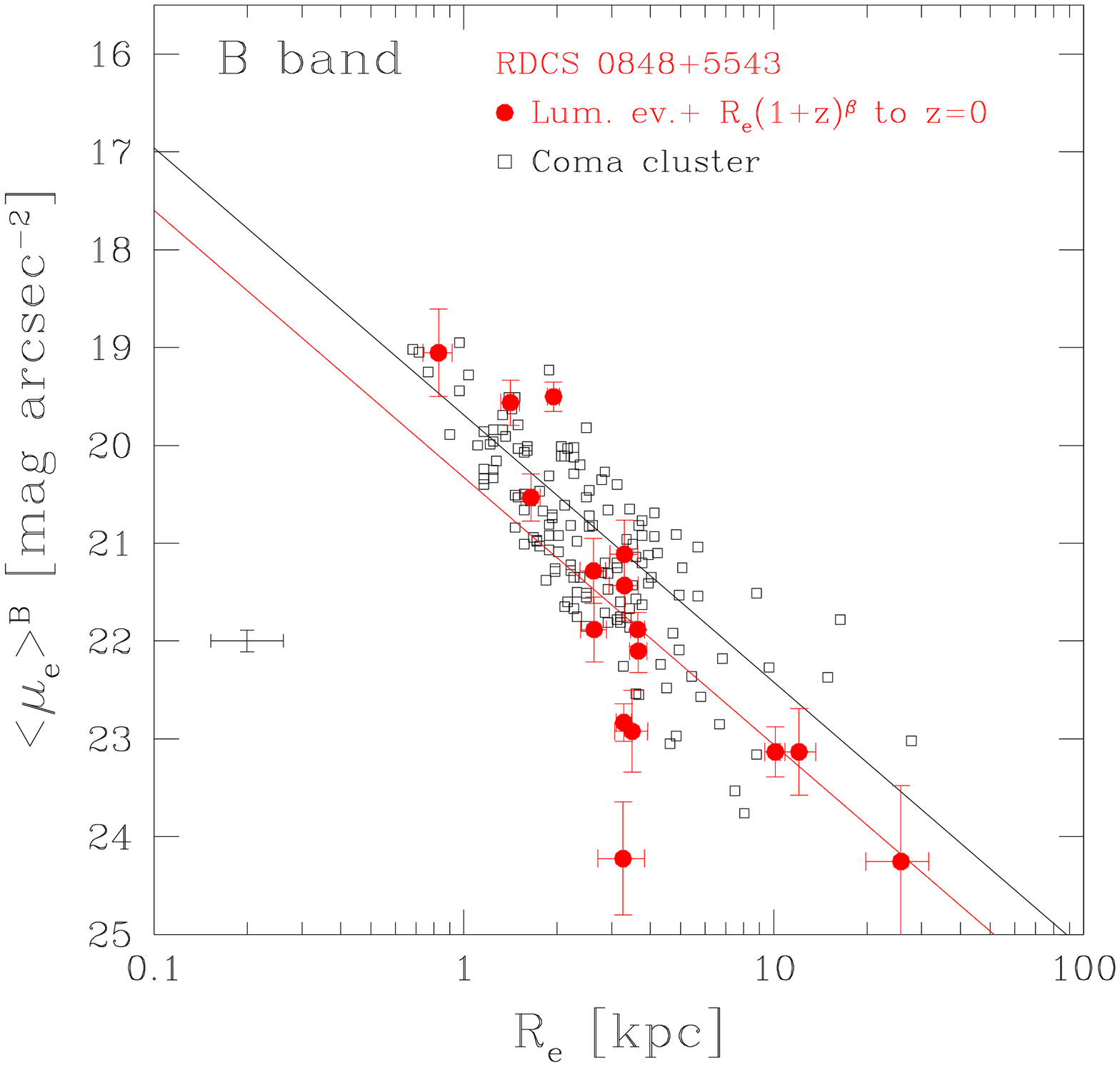}
\includegraphics[width=5.5truecm]{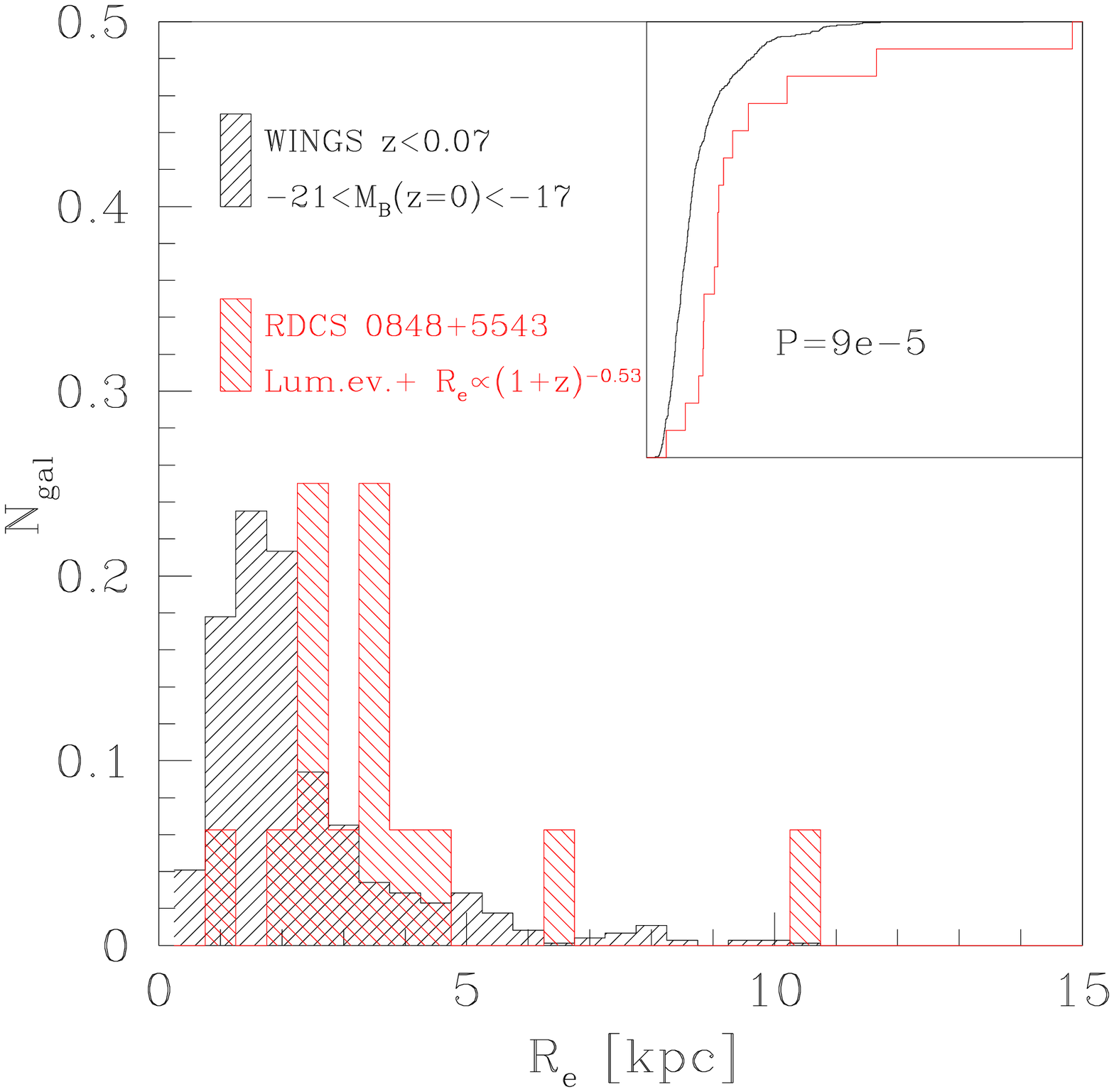}
\includegraphics[width=5.5truecm]{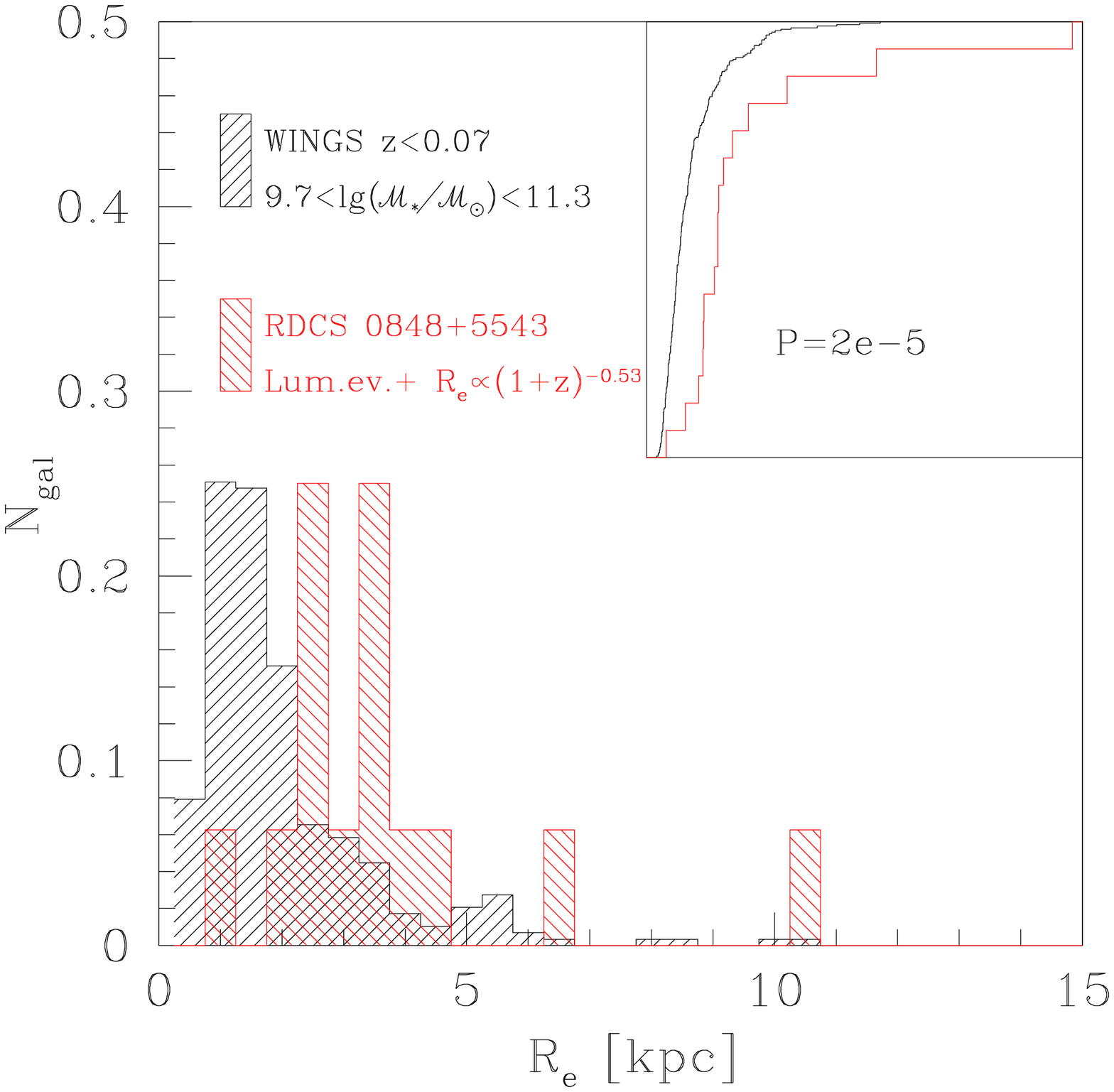}
 \caption{Effects of a pure size evolution of the form
 R$_e\propto(1+z)^b$ with $b=-0.53$ applied to the 16 cluster
ETGs at $z=1.27$.
In the left panel the resulting Kormendy relation evolved to $z=0$ (red
filled circles and red line) is compared to the local one (black line).
The middle and right panels are similar to the lower panels of Fig. 5:
they show the distributions of the evolved effective 
radius of the 16 ETGs (red histogram) and of the WINGS galaxies selected 
in the same luminosity range ($-21<M_B<-17$; left panel) and stellar mass 
range ($5\times10^{10}$ M$_\odot<\mathcal{M}_*<2\times10^{11}$ M$_\odot$;
right). The effective radii have been computed using GASPHOT as for the WINGS
sample.    
}
\end{center}
\end{figure*}

We now see what the variation in absolute magnitude
required to hold our galaxies on the 
local Kormendy relation should be in the case of a variation in R$_e$.
The relationship between the variation in the effective radius
and in the absolute magnitude (hence in the stellar mass) 
of galaxies that satisfies
this constraint can be derived from Eqs. (3) and (4) as follows.
Let $M'$ and $R'$ be the evolved absolute magnitude and effective radius of
a galaxy, that is their values at $z=0$; 
M the absolute magnitude of the stellar mass at 
$z\sim1.3$ passively evolved to $z=0$;
$\Delta M=M'-M$ the variation in the absolute magnitude 
due to the stellar mass grown at $z<1.27$; 
and $\delta_{R_e}=R'/R_e$ the variation in the effective radius.
The corresponding surface brightness $\langle\mu\rangle'$, according 
to eq. (3), can be written as
\begin{equation}
\langle\mu\rangle'=\langle\mu\rangle_e+\Delta M + 5log(\delta_{R_e}).
\end{equation}
On the other hand, $\langle\mu\rangle'$ must satisfy the KR relation,
so from eq. (4) it follows that
\begin{equation}
\langle\mu\rangle'=\langle\mu\rangle_e+\beta log(\delta_{R_e})
\end{equation}
and if the last two equations are made equal we obtain
\begin{equation}
\Delta M=(\beta -5)log(\delta_{R_e}).
\end{equation}
We can distinguish two different cases:
the case where the fraction of the accreted stellar mass at $z<1.27$ 
has a mass-to-light ratio (an age) at $z=0$ comparable to the one of the 
stellar mass already present and the case in which the mass-to-light 
ratio is significantly lower, that is the accreted component is much younger
than the bulk of the mass.
Let us consider the first case.
If the accreted stellar mass is characterized by a mass-to-light ratio 
similar to the ratio of the bulk of the mass, since
the luminosity $L$ is proportional to the stellar mass $\mathcal{M}_*$, 
it follows that 
$\Delta M=M'-M=-2.5log(\delta_{\mathcal{M}_*})$ where 
$\delta_{\mathcal{M}_*}=\mathcal{M'}_*/\mathcal{M}_*$
and eq. (14) provides the sought relationship
\begin{equation}
\delta_{\mathcal{M}_*}=\delta_{R_e}^{(2-{\beta\over{2.5}})}.
\end{equation}
The values of $\beta$ are included in the range $2.5<\beta<3$,
hence 
\begin{equation}
\delta_{\mathcal{M}_*}=\delta_{R_e}^{(0.8-1)};
\end{equation}
that is, the variation
in R$_e$ must follow a variation of the same magnitude
of the stellar mass.
Thus, the study of the evolution of the size-surface brightness relation 
 for these 16 cluster ellipticals establishes that 
{\em if their size increases, then so does their stellar mass}. 
Actually, this result is neither new nor surprising.
For instance, Jorgensen et al. (2013) on the basis of optical spectroscopy
of galaxies in clusters, find no evidence of
evolution of their velocity dispersion at a given
galaxy mass up to $z\sim0.9$.
Since effective radius and velocity dispersion are linked by the
relation
$\sigma^2_v\propto G\mathcal{M}/{R_e}$,
it follows that a simple expansion of individual galaxies would imply
a decrease in their velocity dispersion unless one hypothesizes of
a corresponding growth in their (total) mass.
In fact, a decrease in the velocity dispersion of
cluster ellipticals is not observed.
\begin{figure*}
\begin{center}
\includegraphics[width=5.5truecm]{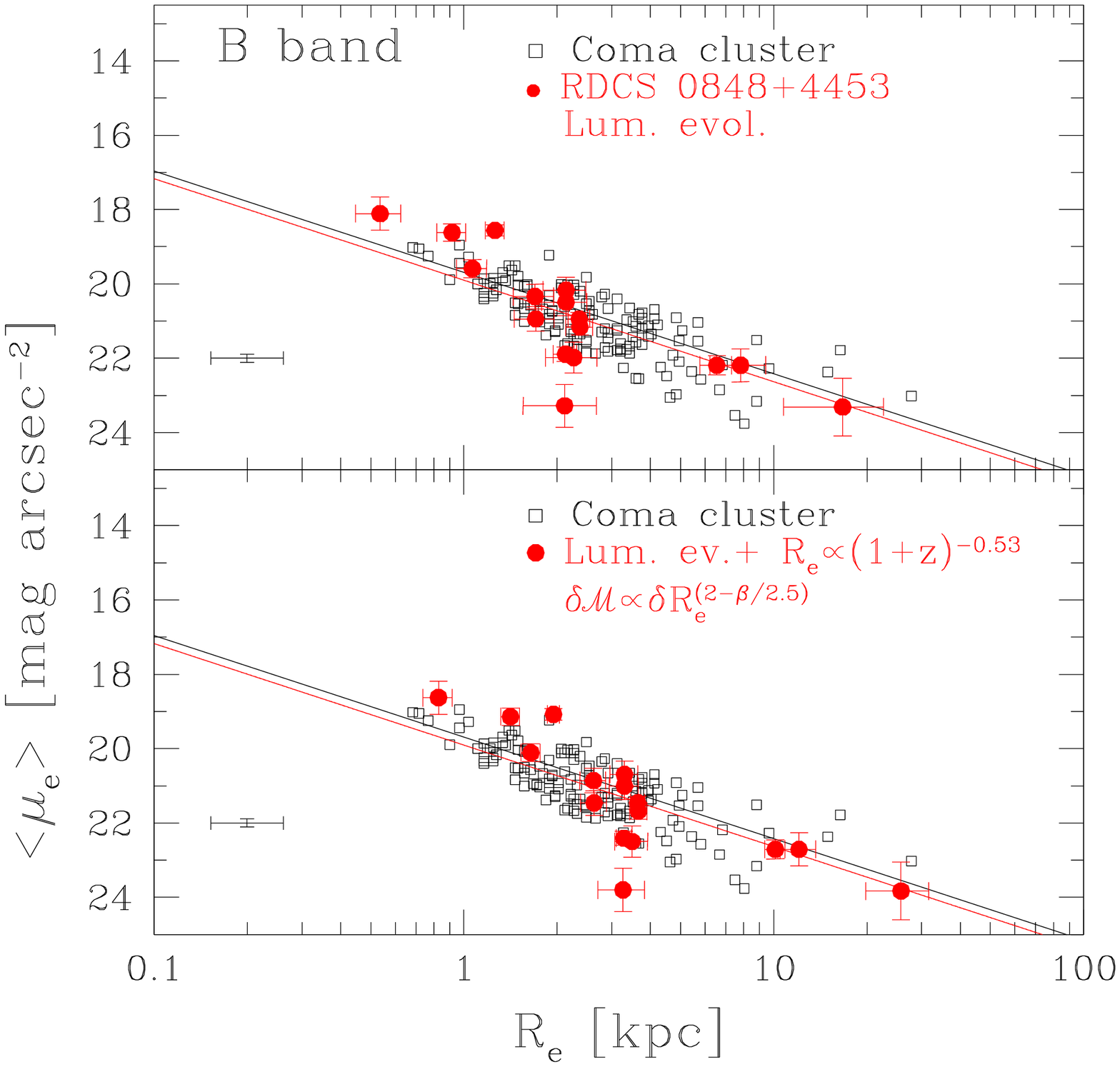}
\includegraphics[width=5.5truecm]{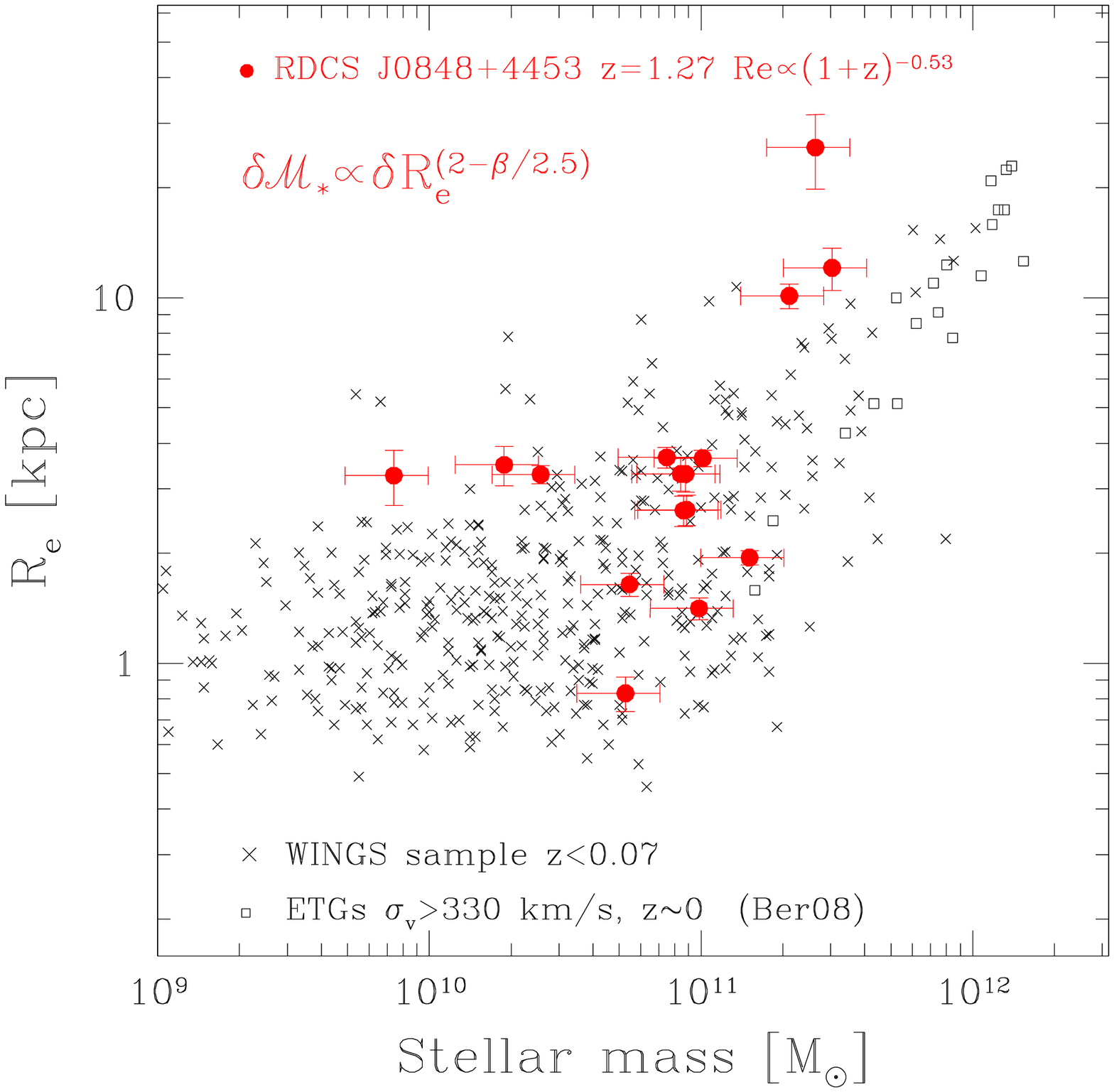}
\includegraphics[width=5.5truecm]{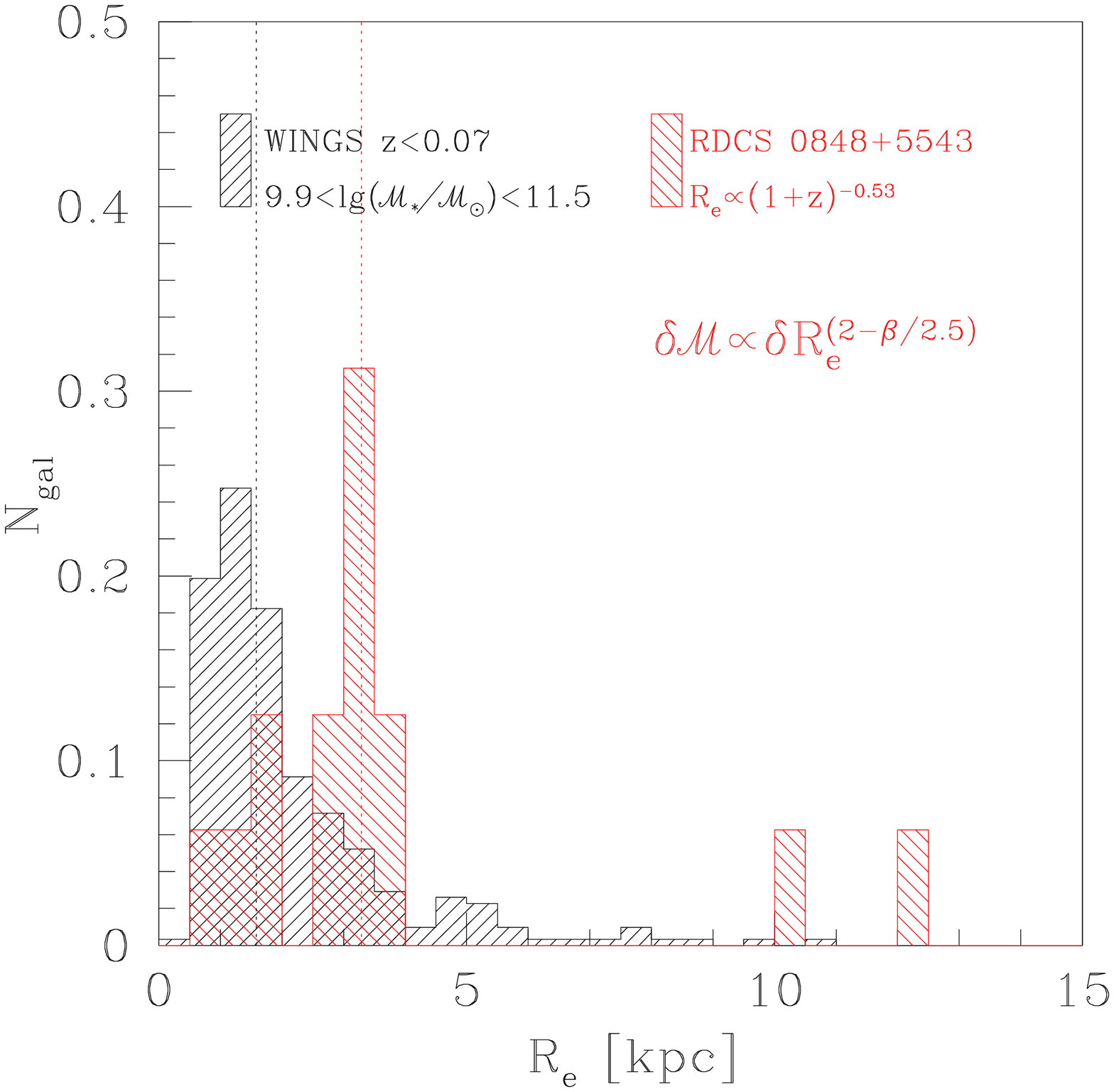}
 \caption{Left: Size-surface brightness relation described by the 16
 cluster ellipticals (filled red circles) for pure 
 luminosity evolution 
 (upper panel) and for stellar mass and effective radius evolution
 (lower panel) according to the relation 
 $\delta_{\mathcal{M}_*}=\delta R_e^{0.9}$, with 
 $R_e\propto(1+z)^{-0.53}$ (see eq. 15).
 The effect of this evolution is to move galaxies along the Kormendy relation. 
 Central: Size-stellar mass relation of our 16 cluster ETGs
 evolved to $z=0$  according to the above evolution together with 
 the relation described by the local WINGS cluster ETGs.
 Right: Effective radii of the 16 cluster ETGs evolved to $z=0$ 
 compared with the effective radii of local WINGS ETGs selected in the same
 range of evolved stellar masses.    
}
\end{center}
\end{figure*}

Returning to the earlier, if we assume that each galaxy increases its size as
 R$_e\propto(1+z)^b$ with $b=-0.5$, it follows
that its stellar mass must increase as
$\delta_{\mathcal{M}_*}=[(1+z)^b]^{(0.9)}$, having assumed 
the value $\beta=2.7$ in eq. (15).
This is the structural evolution that fulfills the constraints
imposed by the Kormendy relation; i.e., that 
leaves galaxies on the relation by moving them  
along it.
This is shown in the left hand panel of Fig. 8 where our 16  
galaxies are shown for the above 
stellar mass and size evolution 
(lower panel).
For comparison, in the upper panel, the case of pure luminosity evolution 
is shown.
In the central panel of Fig. 8 we show the effect of this structural 
evolution when we consider the size-mass relation.
In this case, the 16 ETGs still significantly deviate from the local size-mass 
relation described by the WINGS sample in spite of the increase in mass.
This is more clearly shown in the right hand panel where the distributions of
the effective radius for the two samples selected in the same stellar
mass range are compared.
The median effective radius of the two samples differs at about four sigmas:
the mass of our 16 galaxies has not grown enough to remain on the
size-mass relation.
For obvious reasons, when the accreted stellar 
component is much younger (mass-to-light ratio lower) than the bulk of the 
mass, the disagreement will be even larger than this 
(an even lower mass increase is sufficient to provide the compensating 
absolute magnitude variation).

These comparisons do not take into account that
the two samples, even if selected in the same mass range, 
could not follow the same mass distribution, which could affect
the comparison of the size distributions.
In Fig. 6 and the central panel of Fig. 8, it seems 
that the difference between the effective radius distributions
of the two samples may be due to the three lowest mass galaxies
($<2\times10^{10}$M$_\odot$ in Fig. 6) and by the two most massive galaxies
of the cluster RDCS0848.
Repeating the comparison  between the effective radius distributions 
shown in Fig. 8, considering only the 11 galaxies of the RDCS0848 sample 
in the (non-evolved) mass range $2\times10^11-1.5\times10^{11}$  M$_\odot$, 
we obtain the same result:
the two distributions differ at 97\% confidence level, as shown
in the left hand panel of Fig. 9 (at more than 99\% in the extreme case of pure
size evolution). 
However, the most proper way to perform this comparison without arbitrarily
selecting the mass range and the number of galaxies considered, is to 
extract a sample of galaxies having
the same mass distribution of the 16 ETGs at $z=1.27$ from the WINGS catalog.
We thus randomly extracted 100 samples of 48 galaxies each, following
the stellar mass distribution of the 16 RDCS0848 ETGs and compared 
their effective radius distribution using the KS test.

We considered both the case of pure size evolution
R$_e\propto(1+z)^{-0.5}$ shown in Fig. 7 and the case of size and mass growth 
 shown in Fig. 8.
As to the pure size evolution, the distribution of the effective radius
of the 16 evolved ETGs deviates at 95\% confident level from the one of the 
local WINGS early-type galaxies in 96 cases out of the 100 considered.
This confirms that a pure size evolution cannot be experienced
by these 16 ETGs since they would be significantly offset from the
local Kormendy and size-mass relations.
If the 16 ETGs increase the mass besides their size to continue to stay 
in the Kormendy relation, their effective radius distribution deviates 
from the one of the local WINGS galaxies in 87 cases out of the 100 considered.
In Fig. 9, the mass distribution (central panel)
and the size distribution (right panel) of the 16 ETGs is compared with
one of the 100 WINGS random samples as an example.
Even if at a lower significance level, this result therefore
confirms that the constraint imposed by the Kormendy relation
on the mass increase is not sufficient to keep them on the size-mass relation.

Actually, this result was expected since the size-mass relation establishes
a different relationship between effective radius and stellar mass
with respect to the one imposed by the Kormendy relation.
If we want to preserve the size-surface brightness 
relation, we fail to satisfy the size-mass relation and {\em vice versa}.
Similar conclusions about the size and luminosity/mass evolution
up to $z<0.9$ have already been reached after studying
the evolution of the FP  of cluster elliptical galaxies 
(Saglia et al. 2010).

The last result, based on eq. 15 and on the comparison with
the local scaling relations, puts a constraint on the maximum mass accretion 
and  size increase that these galaxies could experience between $z=1.27$ and
$z=0$.
It follows that they can increase their mass and their size no more than 
30\% to not depart significantly from the
size-mass distribution of local cluster ellipticals. 

Study of the size-surface brightness relation for these 16 cluster 
ellipticals thus rules out the possibility of pure size evolution.
The combined study of the size-surface brightness relation
with the size-mass relation leads to the conclusion that these 16 ETGs
have in general completed their stellar mass accretion at $z\simeq1.3$
and that, consequently, their evolution will be dominated by the 
luminosity evolution.

\begin{figure*}
\begin{center}
\includegraphics[width=5.5truecm]{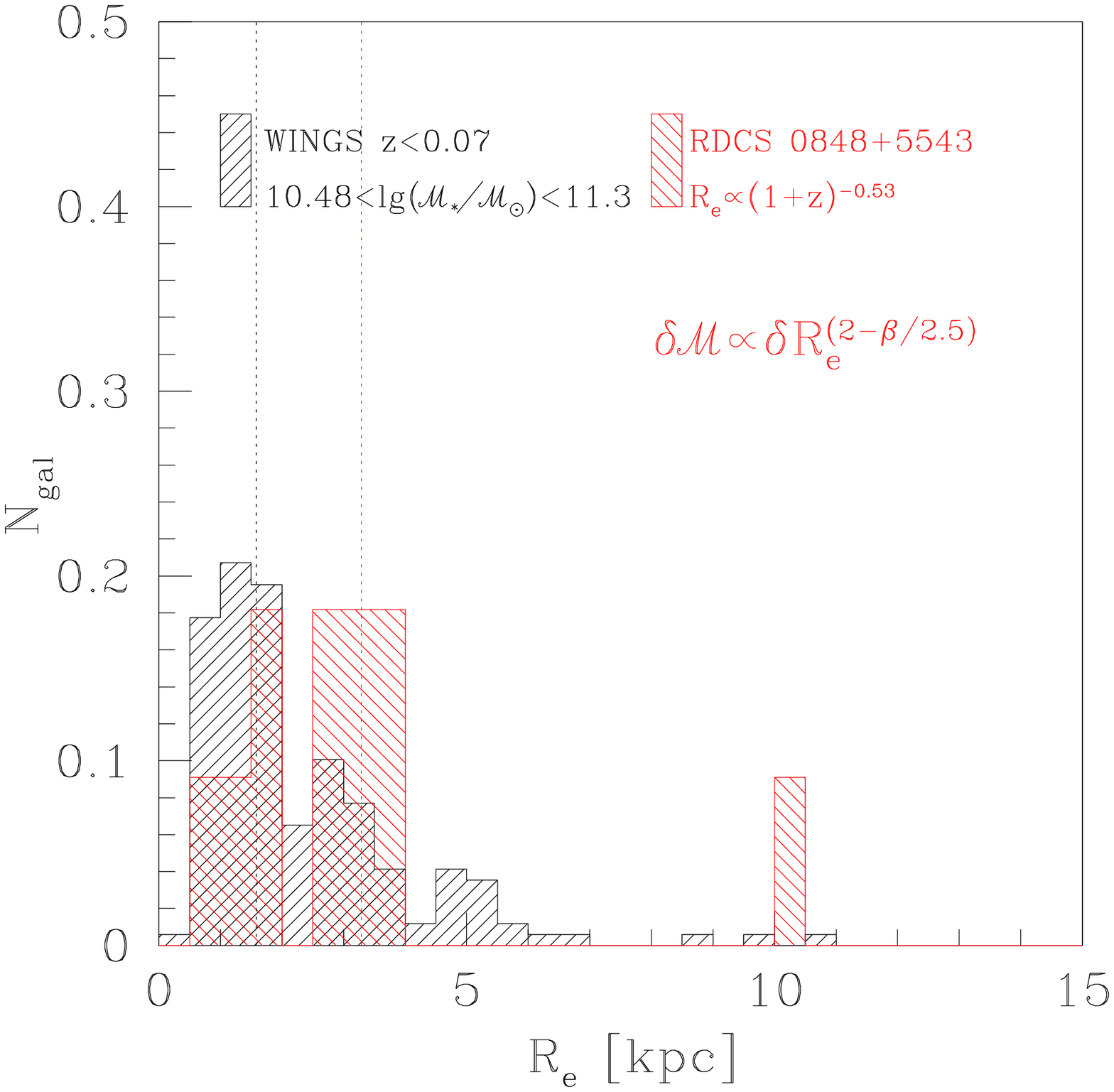}
\hskip 1.5truecm
\includegraphics[width=5.5truecm]{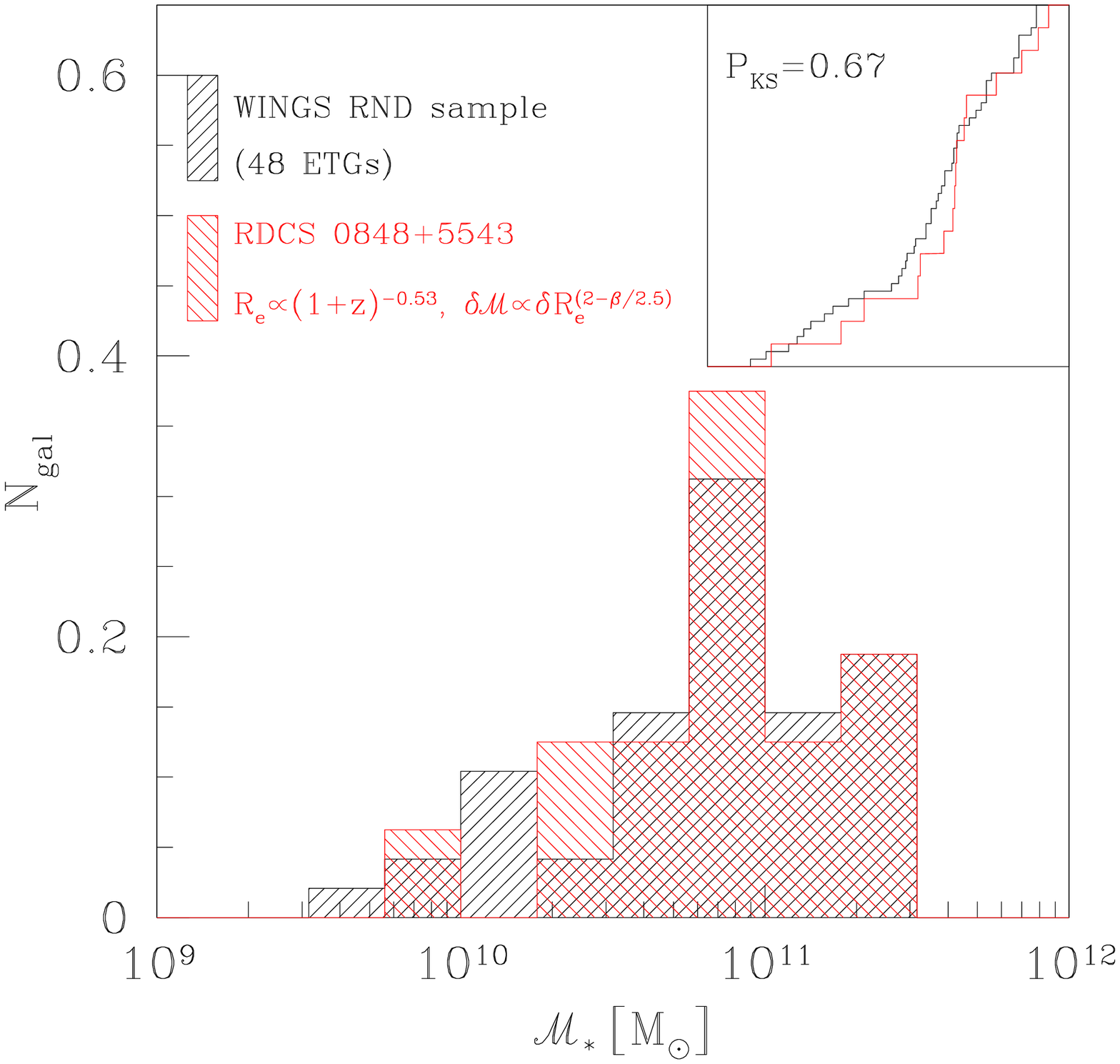}
\includegraphics[width=5.5truecm]{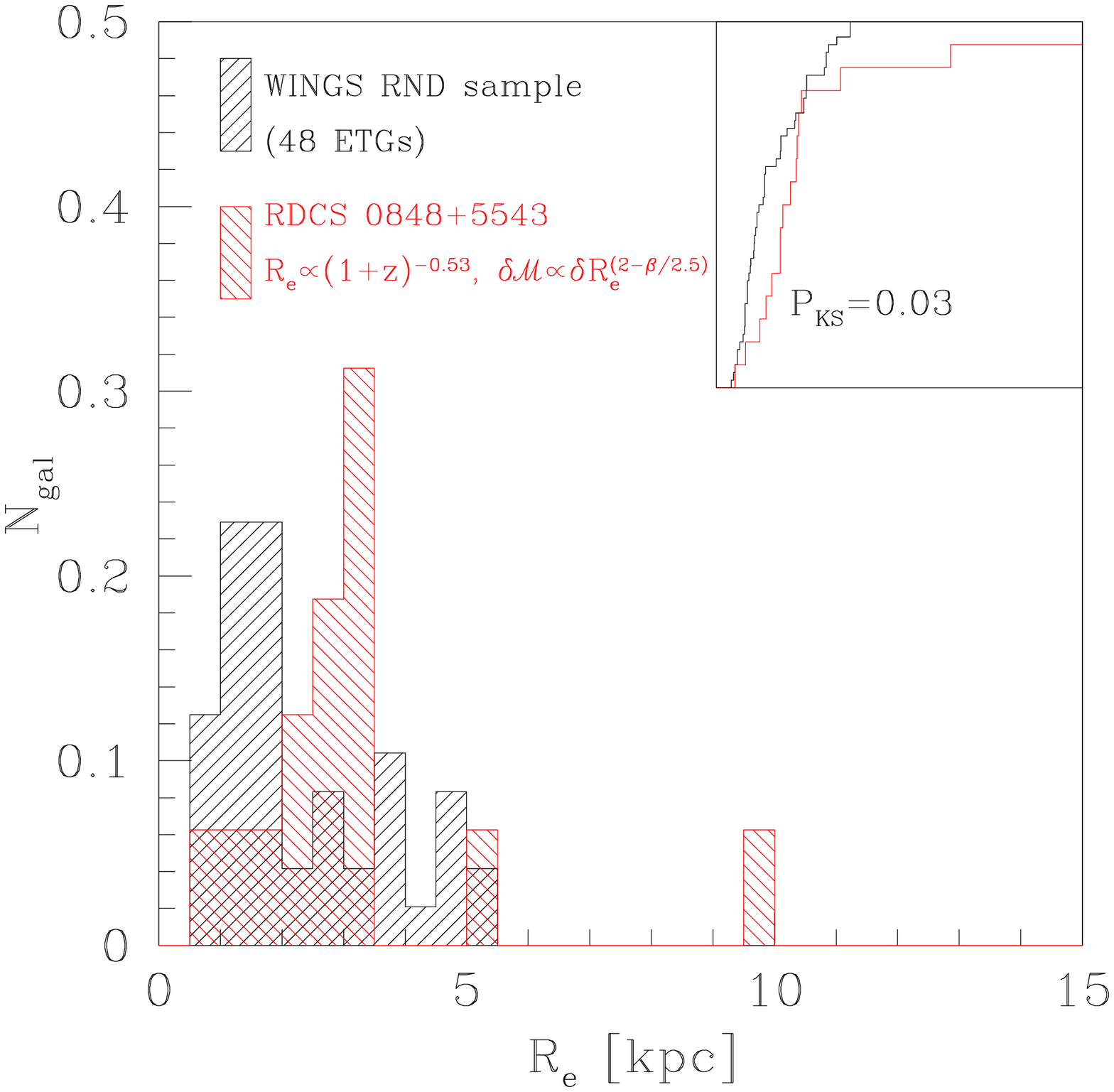}
 \caption{Left: Same as right panel of Fig. 8 but considering only
 the 11 ETGs of the RDCS0848 cluster in the (non-evolved) mass range
 $2\times10^11-1.5\times10^{11}$  M$_\odot$. 
 The distribution of the effective radius of these 11 cluster ETGs evolved 
 to $z=0$ according to mass and effective radius evolution is compared with 
 the distribution of local WINGS ETGs selected in the same range of evolved 
 stellar masses. 
 Central: The stellar mass distribution of one of the 100 random sample
 extracted from the WINGS sample is compared with distribution of
 the 16 ETGs.
 Right: The effective radius distribution of the random sample shown in
 the central panel is compared with the distribution of the 16 ETGs.    
}
\end{center}
\end{figure*}

\section{Summary and conclusions}  
We studied the relations between surface brightness, effective radius,
and stellar mass for a complete sample of 16 elliptical galaxies 
belonging to the cluster RDCS J0848+4453 at $z=1.27$. 
The aim of our analysis was to define the evolutionary status of
these galaxies by assessing whether they have completed 
their mass growth at the redshift where they are observed or they
will experience significant structural changes due to mass accretion and/or 
size growth until $z=0$.

The selection of the sample has been done on the basis of a pure
morphological criterion based on the visual inspection of their
luminosity profile in the ACS-F850LP image and of the residuals 
resulting from the profile fitting with a regular Sersic profile.
Their stellar mass and ages have been obtained through the
best fitting of their SED composed of 11 photometric points in the range
0.38-8.0 $\mu$m with different stellar population synthesis models
and IMF.
We show that the results are independent of the models used
as well as of the IMF adopted. 
Taking as reference the Chabrier IMF, 
the 16 ETGs have stellar masses in the range  
$0.5\times10^{10}-3\times10^{11}$ M$_\odot$ and ages in the range 1-4 Gyr.

We found that the size-surface brightness relation, which is the Kormendy
relation defined by these 16 cluster ellipticals at $z=1.27$, has the 
same slope as the local Kormendy relation.
This means that the luminosity 
and the effective radius of these elliptical galaxies scaled according to 
the same rule in the past 9 Gyr. 
The zero point of the Kormendy relation at $z=1.27$ 
is 1.8 magnitudes brighter in the B-band and 1.3 magnitudes 
brighter in the R-band than at $z=0$.
We found that the luminosity evolution that the stars already assembled 
at $z=1.27$ will experience up to $z=0$ brings the galaxies exactly to the
local relation accounting for these different zero-point values.
We showed that this result has important implications for the stellar mass
profile at $z\sim1.3$.
In particular, we showed that the stellar mass underlying the luminosity of these 
ellipticals was distributed according to the same stellar mass profile 
of local ellipticals having the same evolved luminosity and stellar mass.
This is confirmed by  comparison of the size-mass
relation of our galaxies with the relation described by the local
WINGS sample of elliptical galaxies: the effective radius of the galaxies
of the two samples follow the same distribution.
Actually, we did not see any differences either between cluster and field
ellipticals even if the samples used here are still too small to
firmly assess it.

We find that since the simple luminosity evolution leads our
galaxies to the local Kormendy relation, any further evolution
that may occur at $z<1.27$ must keep them on the relation.
This implies that any variation in the effective radius must be accompanied
by a compensating variation in the absolute magnitude that is in the stellar
mass.
Indeed, a pure evolution of the effective radius for these 
cluster ellipticals is ruled out.
We explicitly show this by applying to our sample 
the mild size evolution R$_e\propto(1+z)^{-0.53}$ reported in the literature
for passive cluster galaxies since $z\sim1$,
in addition to the passive luminosity evolution.
The resulting Kormendy relation differs at more than four sigmas from the
local Kormendy relation.

Thus, we are left with two possibilities.
Either the 16 ellipticals have completed their stellar mass accretion at
redshift $z\sim1.3$ or they grow in a way such that they remain in
the local Kormendy relation. 
We found that the relation satisfying this last condition is 
$\delta_{\mathcal{M}_*}=\delta_{R_e}^{(2-{\beta\over{2.5}})}$
with $\beta\sim3$ the slope of the Kormendy relation,
which is an increase in the effective radius that must be accompanied by 
an equivalent stellar mass increase. 
We applied this condition to our galaxies assuming the mild size
evolution above.
As expected, the growth of the stellar mass and of the effective radius
leads to consistency between
the galaxies and the local Kormendy relation.
On the other hand, considering the size-mass relation, we see that this mass
and size growth would lead these galaxies away  from 
the size-mass relation described by local ellipticals.
If we want to preserve the size-surface brightness 
relation, we fail to satisfy the size-mass relation and vice versa.

After combining the study of the size-surface brightness relation
with the size-mass relation, we reached the conclusion that these 16 
cluster elliptical galaxies have in general completed their stellar mass accretion 
at $z\simeq1.3$ and that, consequently, they will mainly evolve in luminosity
until $z=0$.

Our results suggest that elliptical galaxies in the mass range
$<2\times10^{11}$ M$_\odot$ do not take part in the observed size
evolution of galaxies.
They do not increase their size beyond $z\sim1.3$
either individually or as a population (due to newly added galaxies).
This result agrees with the results already found by other authors and
with the view that size evolution is mainly driven by disk galaxies.
Van der Wel et al. (2011) find that most (65\%) of the massive 
($>10^{10.8}$ M$_\odot$) compact galaxies at high-z ($z>1.5$) are disk-dominated 
with a disk scale that is substantially smaller than the disks of 
equal-mass galaxies in the present universe. 
As already noticed, Jorgensen et al. (2013) studied early-type galaxies 
in three clusters at $0.5<z<0.9$ and found no evolution of their size with respect
to $z=0$ cluster galaxies.
Stott et al. (2011) studied the brightest galaxies in clusters at $0.8<z<1.3$
finding little or no evolution with respect to counterparts at $z\sim0.2$.
Huertas-Company et al. (2013) homogeneously select and study elliptical
galaxies in the redshift range $0.3<z<1.2$ showing that  no size evolution 
takes place for ellipticals with
masses $5\times10^{10}-2\times10{11}$ M$_\odot$ in this redshift range,
as clearly shown in their figures 12 and 13.
The evolution appears at $z<0.3$ when they compare with the SDSS data.
Looking at Fig. 2 in Cimatti et al. (2012), no size
evolution of ETGs in the redshift range $0.3<z<1.6$ is visible.
Also in this case the evolution appears when the local SDSS data is
considered.

The known bias against small galaxies that affects the SDSS data and 
the effect that it has on studies of the size evolution of galaxies is 
discussed in many works (see, e.g., Gargiulo et al. 2014; Damjanov et al. 2013).
When revisiting the analysis performed by 
Belli et al. (2014), Gargiulo et al. (2014) use a sample of elliptical galaxies 
instead of passive galaxies to show that old early-type galaxies 
at high-z have local counterparts with similar structural properties, 
while the most massive and largest ones
in the local universe were not present at high-z.
Our result therefore suggests that ellipticals galaxies, at least in the
mass range probed by our sample, do not individually grow their stellar mass 
and their size continuously during their lifetimes, leading to a null contribution
to the observed size evolution of galaxies.
Whether this conclusion can be generalized to the whole population of 
ellipticals (field and cluster ETGs) at these redshifts cannot be 
assessed from these data. 
We will assess this issue in a forthcoming paper.

\section*{Acknowledgments}
This work is partially based on data collected at the European Southern 
Observatory (ESO) telescopes and with the NASA/ESA 
Hubble Space Telescope, obtained from the 
data archive at the Space Telescope Science Institute which is operated by 
the Association of Universities for Research in Astronomy. 
This work is also based on observations carried out at the Large Binocular 
Telescope (LBT). 
The LBT is an international collaboration among institutions in the 
United States, Italy, and Germany. 
We acknowledge the support from the LBT-Italian Coordination Facility for the
execution of observations, data distribution, and reduction.
We thank the anonymous referee for the useful comments and the constructive
criticism.
This paper was originally submitted to MNRAS on 13 of November 2013 and
withdrawn on 22 of January 2014 because we did not yet
receive a referee report.
This work has received financial support from Prin-INAF 1.05.09.01.05.

\begin{appendix} 
\section{Stellar masses, age and absolute magnitudes for different
models}
The table reports  the age [Gyr], the stellar mass
[log(M/M$_\odot$)], and the absolute magnitudes for each galaxy of the sample
in the B and R bands derived 
by best-fitting their spectral energy distribution with the 
stellar population synthesis models of Maraston et al. (2005, MA05), 
Bruzual and Charlot (2003, BC03), and the later release by Charlot and Bruzual (CB07). 
The last row reports the mean values.
We considered two different initial mass functions (IMFs): the Salpeter IMF
for the MA05 and BC03 models and the Chabrier IMF for the BC03 and CB07 models.
As expected, the absolute magnitudes depend neither 
on the model nor on the IMF adopted since the color k-correction 
term can vary with different models of hundredth of a magnitude. 
It can be seen that also the mean age of the stellar population is very stable 
with respect to the model used and to the IMF adopted. 
In contrast, the IMF, as is known, systematically affects the stellar mass 
with the Salpeter IMF producing masses about 1.7 times higher than the 
Chabrier IMF.

\begin{table*}
\caption{
For each galaxy of the sample we report the age [Gyr], the stellar mass
[log(M/M$_\odot$)] and the absolute magnitudes in the B and R bands derived 
through the best-fitting of their SED with the stellar population 
synthesis models of Maraston et al. (2005; MA05), Bruzual and Charlot 
(2003, BC03), Charlot and Bruzual (CB07), and with the two
Salpeter and Chabrier stellar initial mass functions.
The last row reports the mean values of the best-fitting parameters.
The typical variation in the best fitting parameters 
due to the different models and IMF is about 18\% in stellar mass 
and 25\% in age.
}
\begin{center}
\begin{tabular}{rccccccccccccccccc}
\hline
\hline
    & &MA05  &Sal  & &         &BC03 &Sal  &          \\
\hline
 ID&   age  &  logM$_*$& M$_B$& M$_R$& age  &  logM$_*$& M$_B$& M$_R$   \\  
   & [Gyr]&  [M$_\odot$]&     &        &[Gyr]&  [M$_\odot$]&     &     \\
\hline
   1&4.50 &11.59 &-21.58&-23.38& 3.50& 11.53& -21.54& -23.39\\  
   2&1.43 &11.27 &-22.22&-23.64& 1.43& 11.41& -22.23& -23.65\\  
   3&3.50 &11.27 &-21.99&-23.00& 2.60& 11.26& -21.94& -23.00\\  
   4&1.80 &11.39 &-22.65&-23.79& 1.61& 11.48& -22.64& -23.80\\  
   5&1.61 &10.98 &-21.48&-22.70& 1.43& 11.03& -21.49& -22.71\\  
 606&1.61 &10.87 &-21.32&-22.51& 1.43& 10.95& -21.32& -22.53\\  
 590&3.00 &11.06 &-21.20&-22.50& 2.30& 11.09& -21.18& -22.50\\  
 568&1.61 &10.64 &-20.79&-22.14& 1.80& 10.80& -20.80& -22.15\\  
 719&1.28 &10.71 &-20.63&-22.09& 3.50& 11.08& -20.68& -22.10\\  
1250&1.28 &10.37 &-20.35&-21.56& 1.28& 10.49& -20.38& -21.57\\  
1260&0.45 & 9.77 &-20.00&-20.76& 0.90& 10.02& -20.04& -20.79\\  
 173&3.25 &10.71 &-20.03&-21.53& 4.50& 10.81& -20.02& -21.53\\  
1160&2.30 &10.82 &-20.84&-22.07& 2.75& 10.98& -20.86& -22.08\\  
 657&1.61 &10.91 &-21.52&-22.72& 1.43& 11.02& -21.52& -22.73\\  
 626&1.61 &10.90 &-21.48&-22.72& 1.43& 11.01& -21.48& -22.73\\  
 471&3.00 &10.37 &-20.03&-21.35& 1.70& 10.36& -20.04& -21.27\\  
\hline
\bf mean &2.10& 11.03& -21.41& -22.69&  2.10& 11.10& -21.40& -22.70\\
\hline
\end{tabular}								 									 
\end{center}
\end{table*}

\begin{table*}
\begin{center}
\begin{tabular}{rcccccccc}
\hline
\hline
    &  \bf BC03 &\bf Cha &  &	     &CB07 &Cha  & \\
\hline
 ID&   age  &  logM$_*$& M$_B$& M$_R$& age  &  logM$_*$& M$_B$& M$_R$\\  
   & [Gyr]&  [M$_\odot$]&     &          \\
\hline
   1&3.75 &11.31&  -21.55& -23.39& 4.50& 11.32& -21.55& -23.40 \\  
   2&1.43 &11.16&  -22.24& -23.65& 3.25& 11.27& -22.26& -23.66 \\  
   3&2.60 &11.01&  -21.94& -23.00& 2.60& 10.94& -21.93& -23.00 \\  
   4&1.68 &11.25&  -22.65& -23.80& 2.20& 11.18& -22.64& -23.81 \\  
   5&1.43 &10.78&  -21.49& -22.71& 1.70& 10.68& -21.46& -22.71 \\  
 606&1.43 &10.70&  -21.32& -22.53& 2.75& 10.78& -21.34& -22.54 \\  
 590&2.30 &10.84&  -21.18& -22.50& 2.50& 10.72& -21.16& -22.52 \\  
 568&2.00 &10.57&  -20.81& -22.15& 2.60& 10.56& -20.81& -22.16 \\  
 719&3.50 &10.82&  -20.67& -22.10& 3.75& 10.76& -20.67& -22.11 \\  
1250&1.28 &10.24&  -20.38& -21.57& 1.02& 10.15& -20.34& -21.58 \\  
1260&0.71 & 9.70&  -20.03& -20.79& 0.77&  9.64& -20.04& -20.79 \\  
 173&4.25 &10.55&  -20.02& -21.53& 4.50& 10.50& -20.01& -21.54 \\  
1160&3.00 &10.77&  -20.89& -22.08& 1.61& 10.44& -20.86& -22.10 \\  
 657&1.43 &10.77&  -21.52& -22.73& 1.70& 10.70& -21.51& -22.73 \\  
 626&1.43 &10.76&  -21.48& -22.73& 1.70& 10.69& -21.47& -22.73 \\  
 471&1.61 &10.11&  -20.04& -21.27& 1.02&  9.86& -19.99& -21.02 \\ 								      
\hline
\bf mean &2.10 &10.86& -21.41& -22.70& 2.35& 10.83& -21.40& -22.70\\
\hline
\end{tabular}								 									 
\end{center}
\end{table*}

\section{Surface-brightness profile fitting}
The structural parameters of the galaxies were derived by fitting 
the  observed surface brightness profile in the F850LP-band image with 
a Sersic profile convoluted with the PSF using Galfit, as described 
in \SS 3.2.
In this appendix the 16 ellipticals of the sample selected according to the
criteria described in \SS 2.2 are shown (upper panels),
together with the best-fitting surface brightness model (middle panels) 
and the residuals (lower panels).
The images are $3\times3$ arcsec.
The convolution box and the fitting box ($6\times6$ arcsec) were
defined by repeatedly fitting the observed profile with increasing values 
of the box until the convergence of the best-fitting parameter values.
The goodness of the fit is shown by the lack of residuals obtained for
all the galaxies.
It is also worth noting that no structures are visible in the residuals,
showing the regularity and the symmetry of the true profile.
The good fitting is also shown in Fig. B2 where the observed surface 
brightness profile measured on the F850LP image is compared with the 
best fitting model profile.
In all cases, the profile of the galaxies were fit over
at least five magnitudes in surface brightness.
The observed surface brightness profiles of galaxies \#5, \#626, and \#657 
depart from the model profiles at $r\sim0.5$ arcsec, the mean distance 
between them. 
\begin{figure*}
\begin{center}
\includegraphics[width=18truecm]{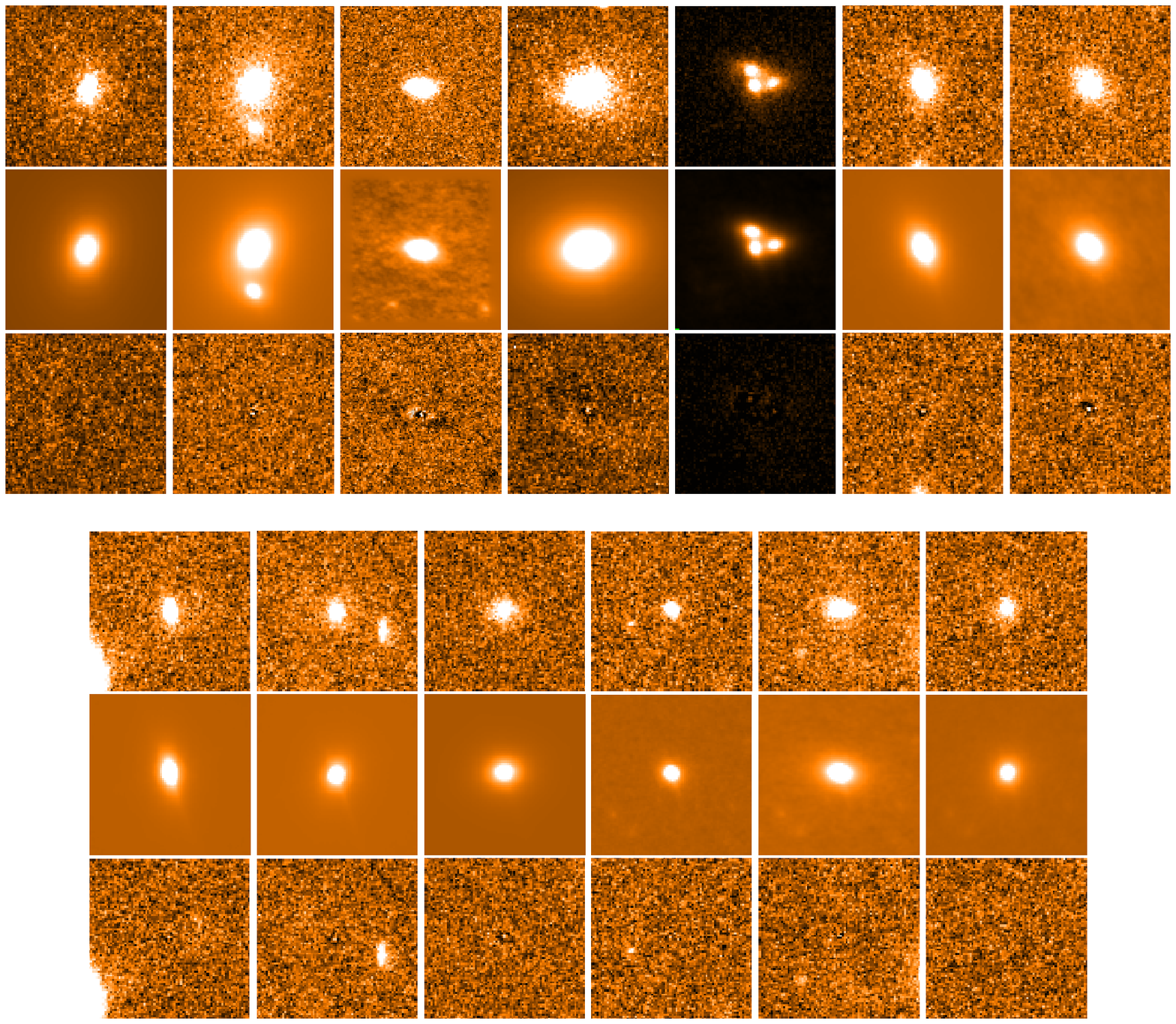}
\caption{Each column shows the GALFIT input and output for the 16 cluster 
ellipticals at $z=1.27$:
ACS-F850LP band  image of the galaxy (upper box), best-fitting Sersic
model profile (middle box) and residual image (lower box) 
obtained by subtracting the model from the image. 
Galaxies are (from top left to bottom right): (upper panel) ID. \#1, \#2, \#3, 
\#4, 
\#5 \#626 and \#657, \#606, \#590; (lower panel) \#568, \#719, \#1250, \#173, 
\#1160, \#1471. Each image is $3\times3$ arcsec.	    
}
\end{center}
\end{figure*}

\begin{figure*}
\begin{center}
\includegraphics[width=5truecm]{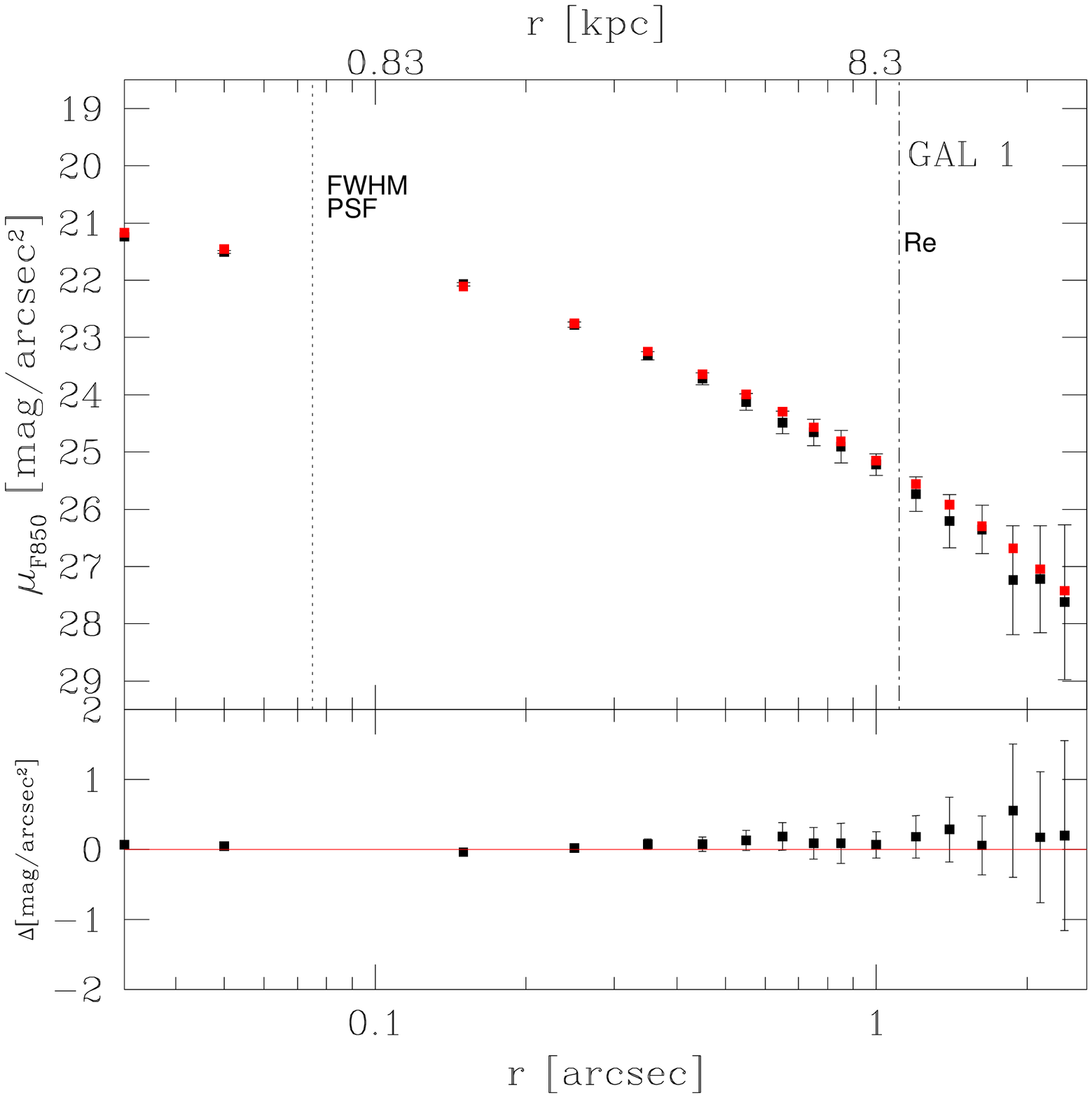}
\includegraphics[width=5truecm]{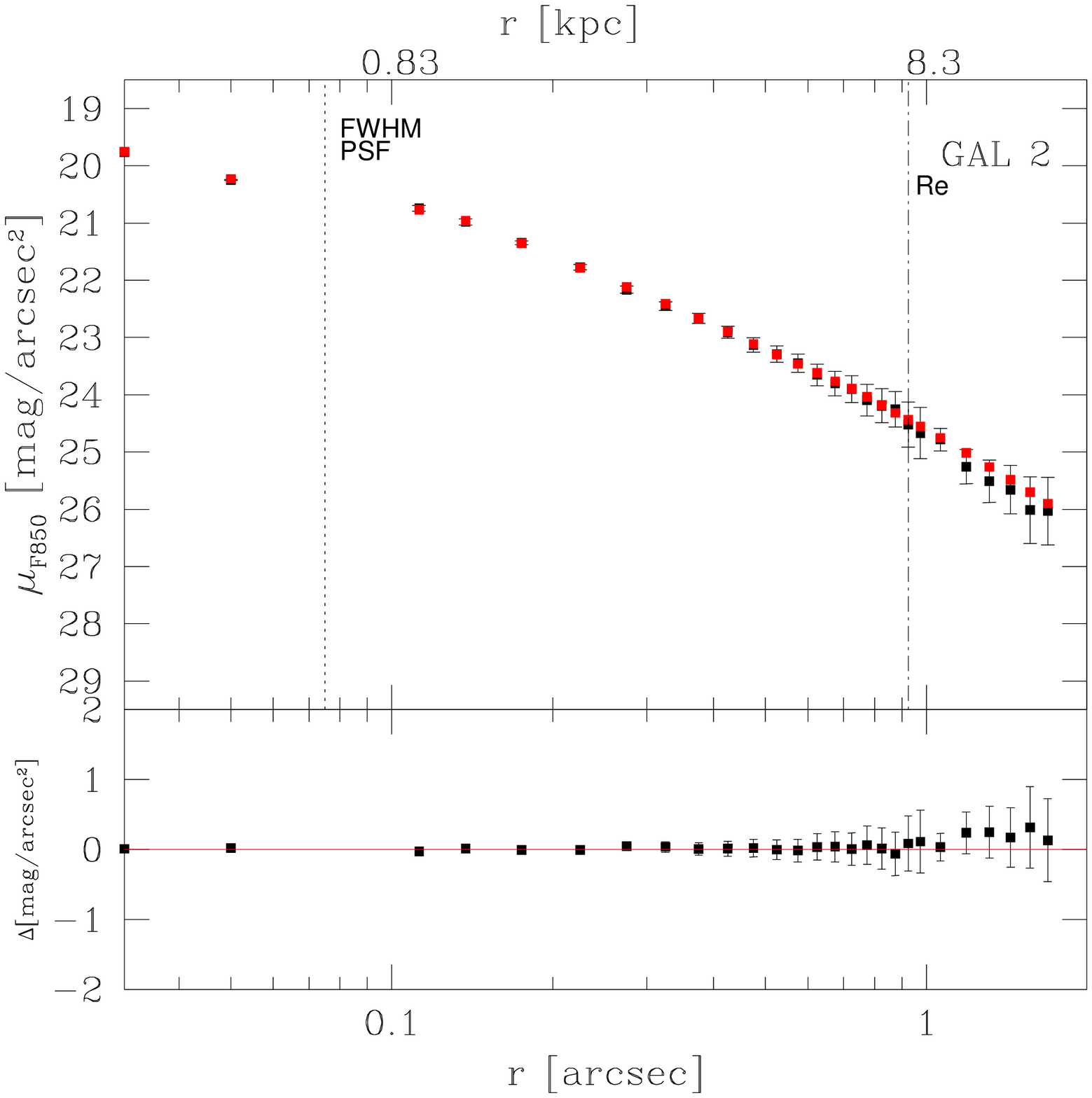}
\includegraphics[width=5truecm]{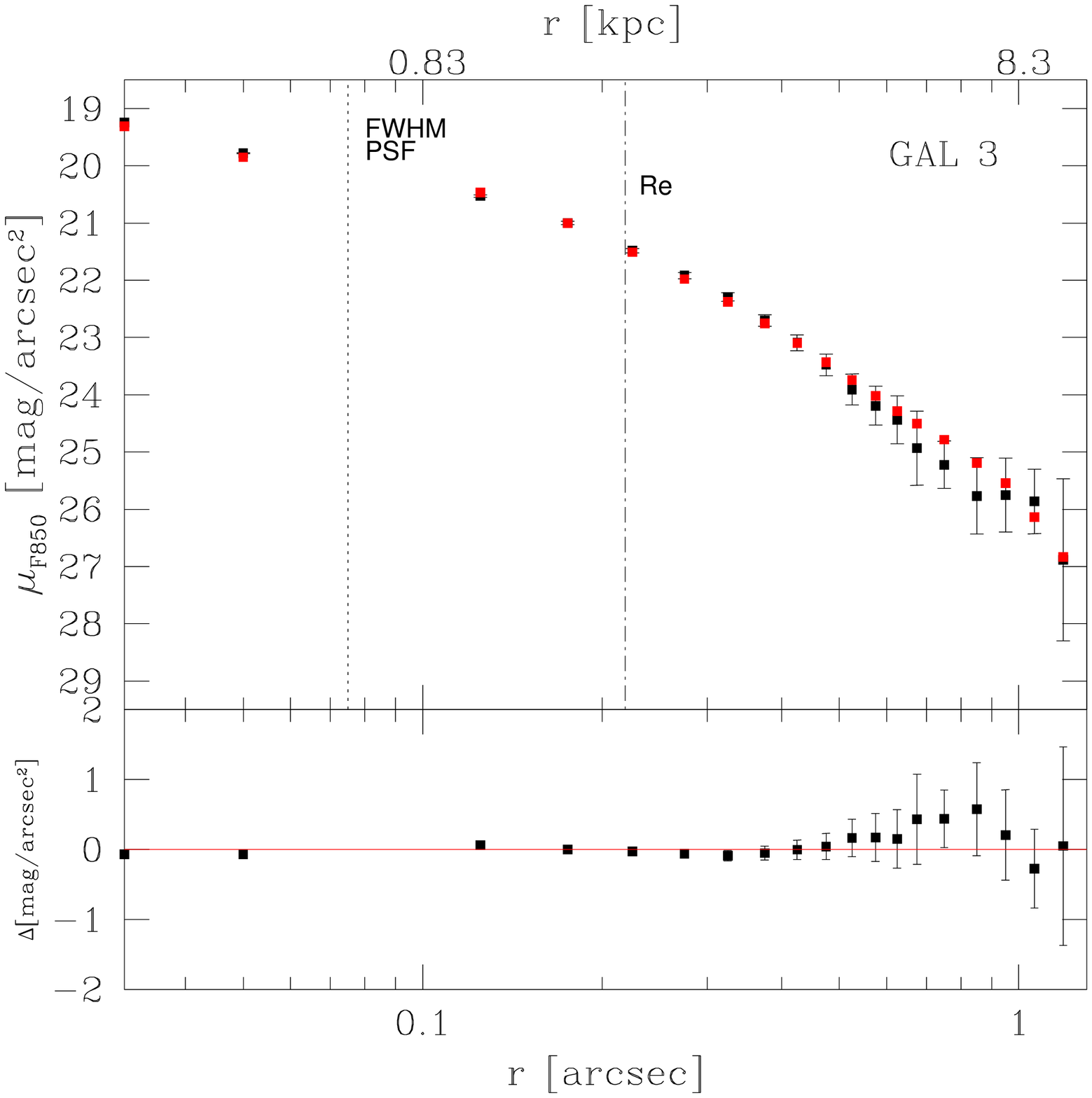}
\includegraphics[width=5truecm]{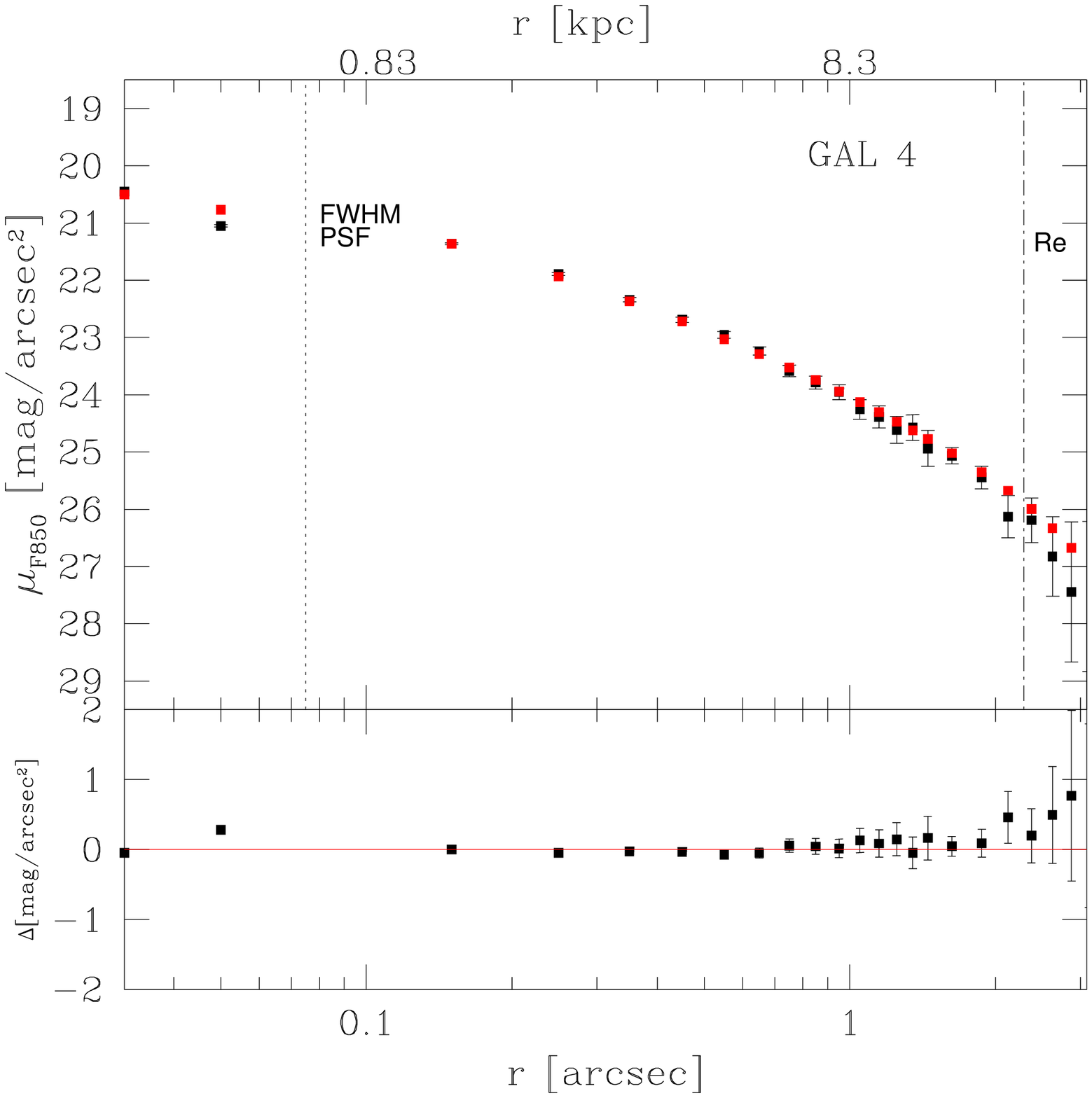}
\includegraphics[width=5truecm]{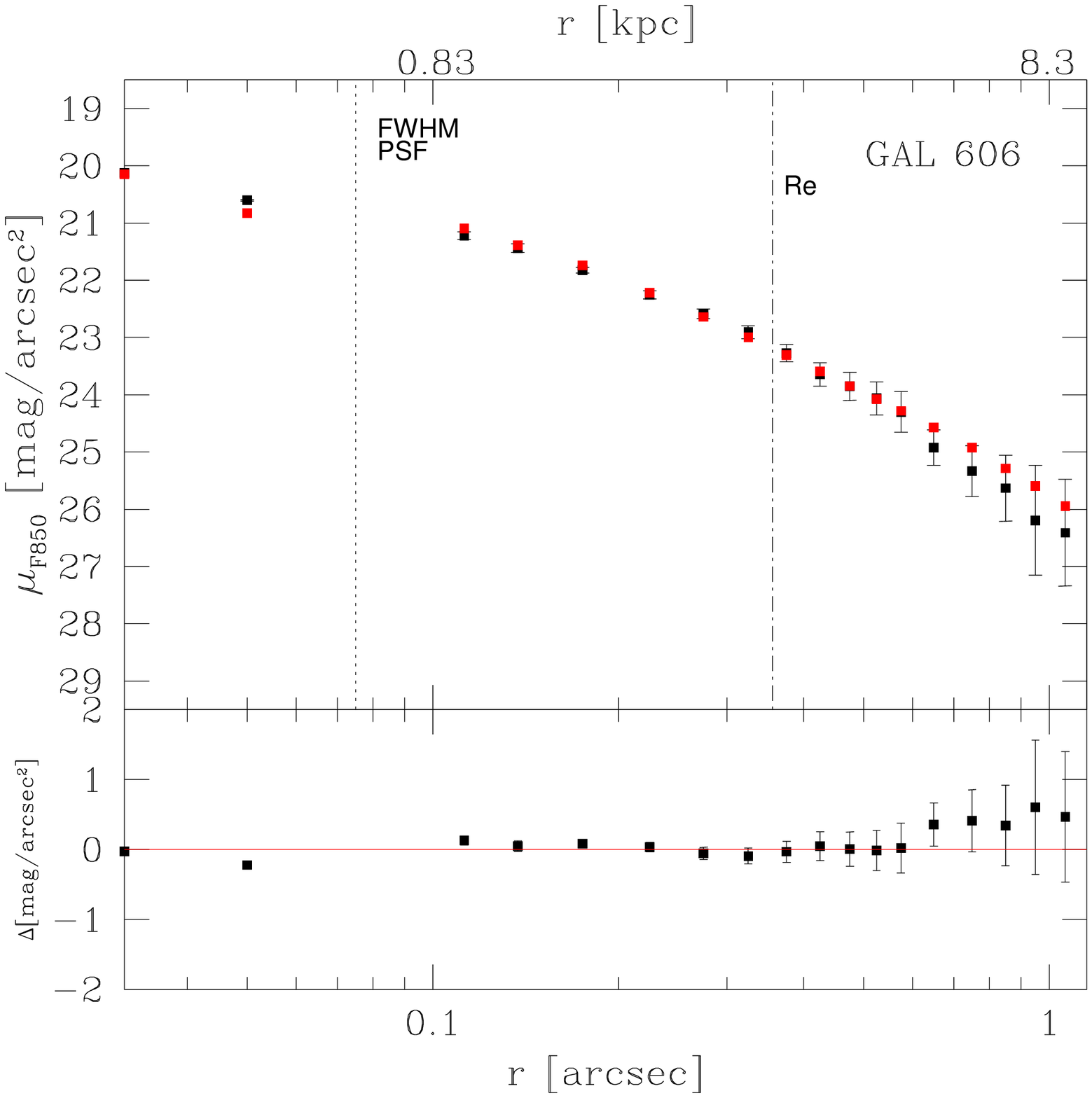}
\includegraphics[width=5truecm]{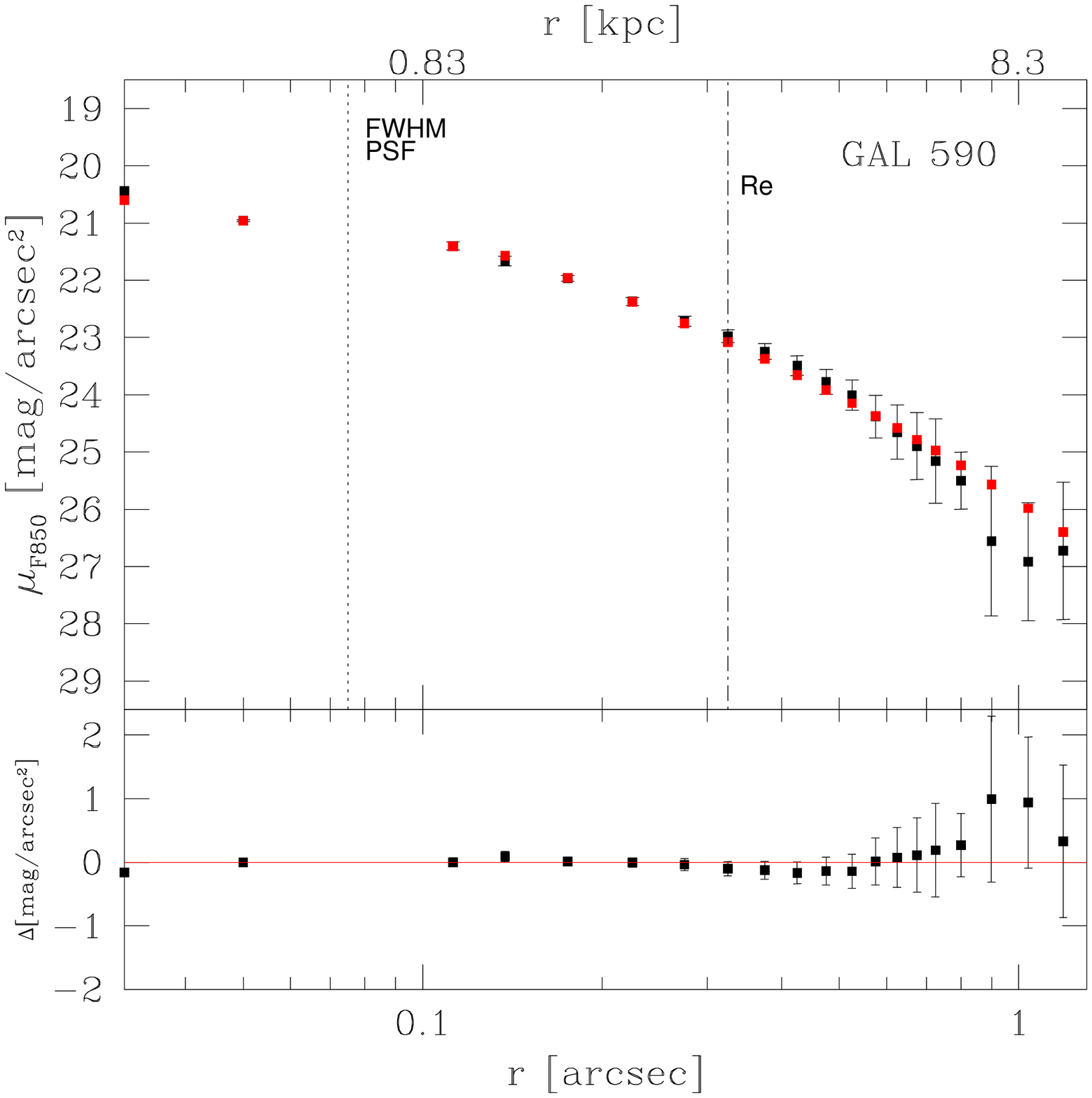}
\includegraphics[width=5truecm]{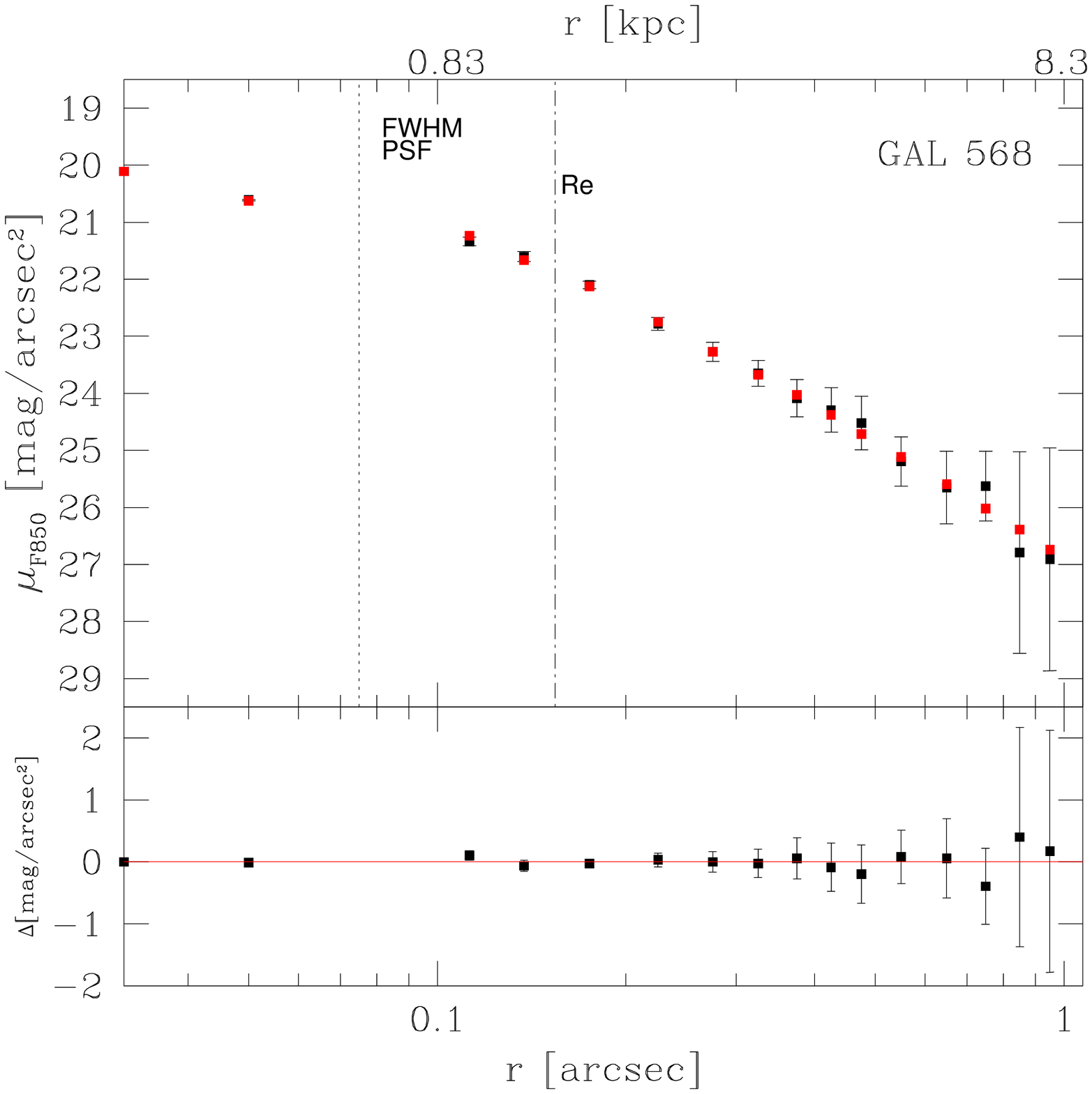}
\includegraphics[width=5truecm]{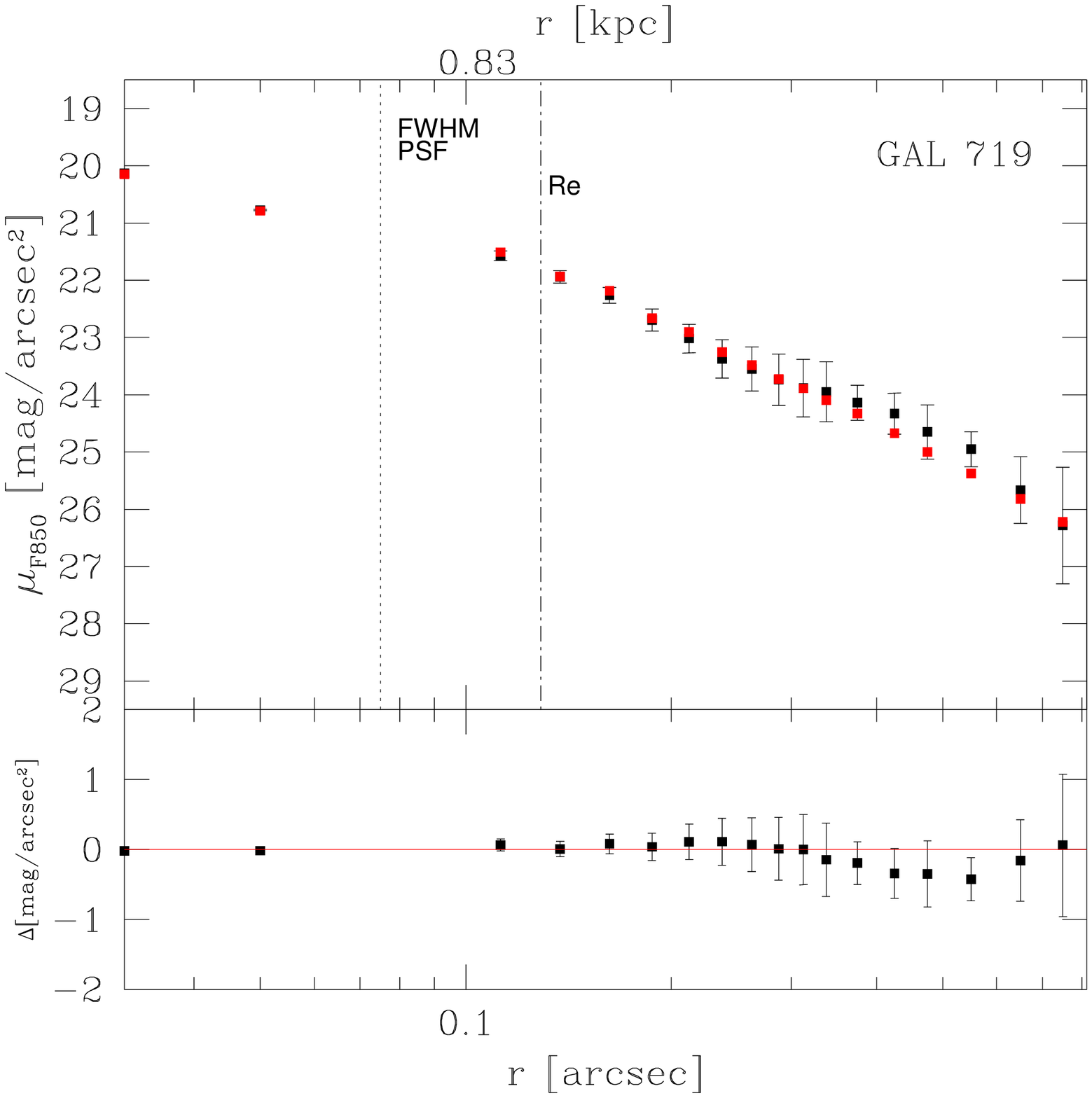}
\includegraphics[width=5truecm]{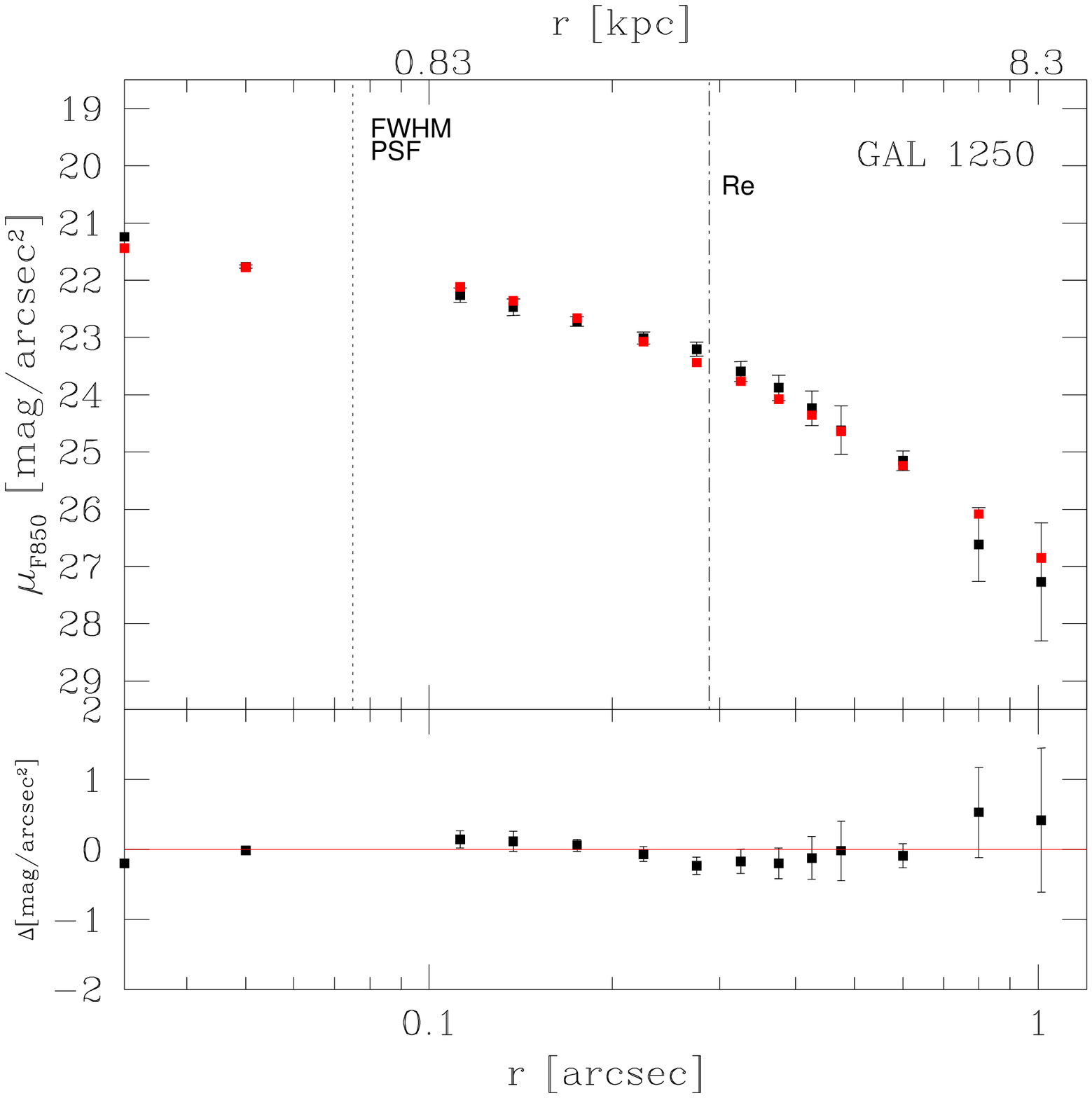}
\includegraphics[width=5truecm]{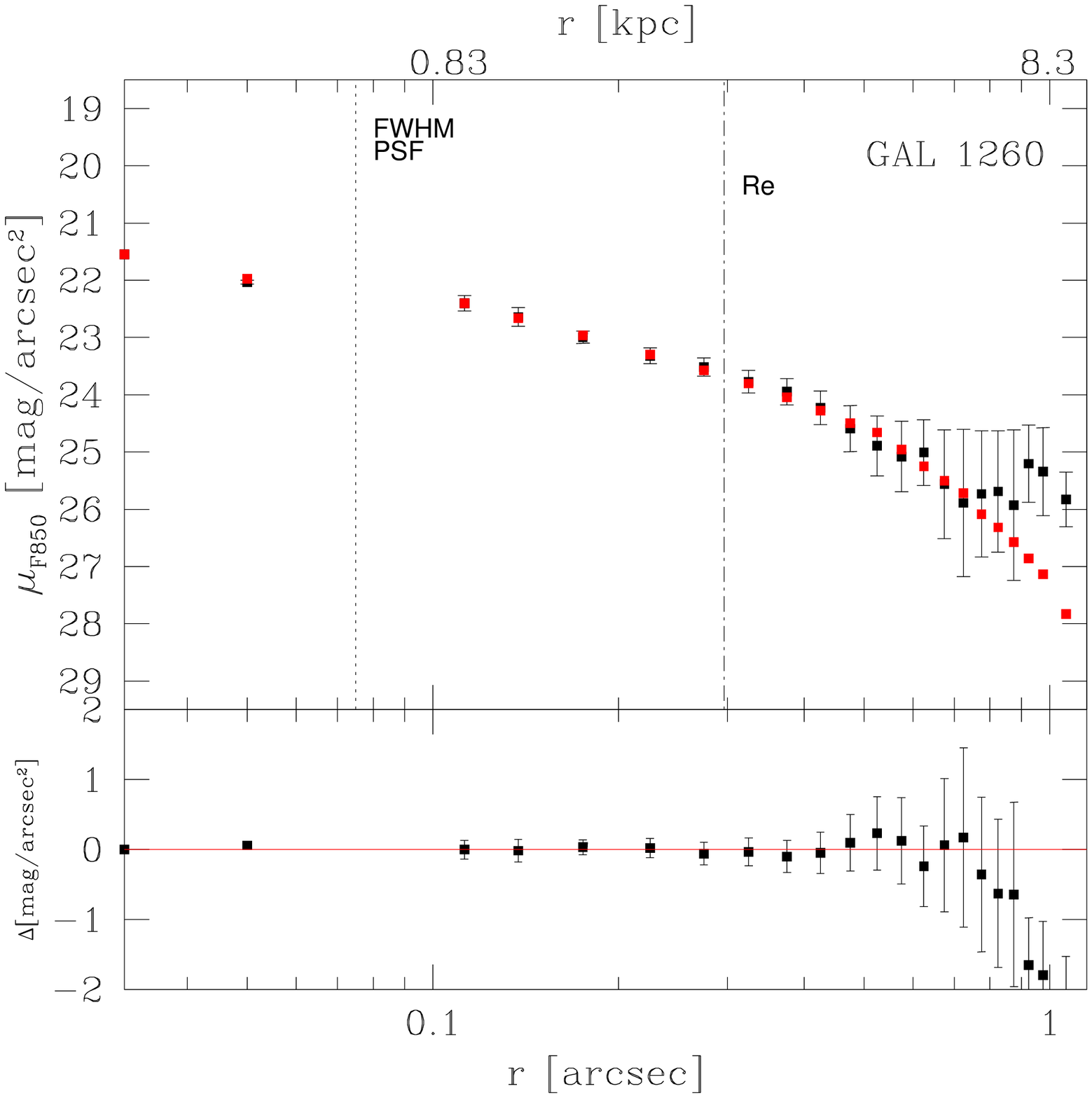}
\includegraphics[width=5truecm]{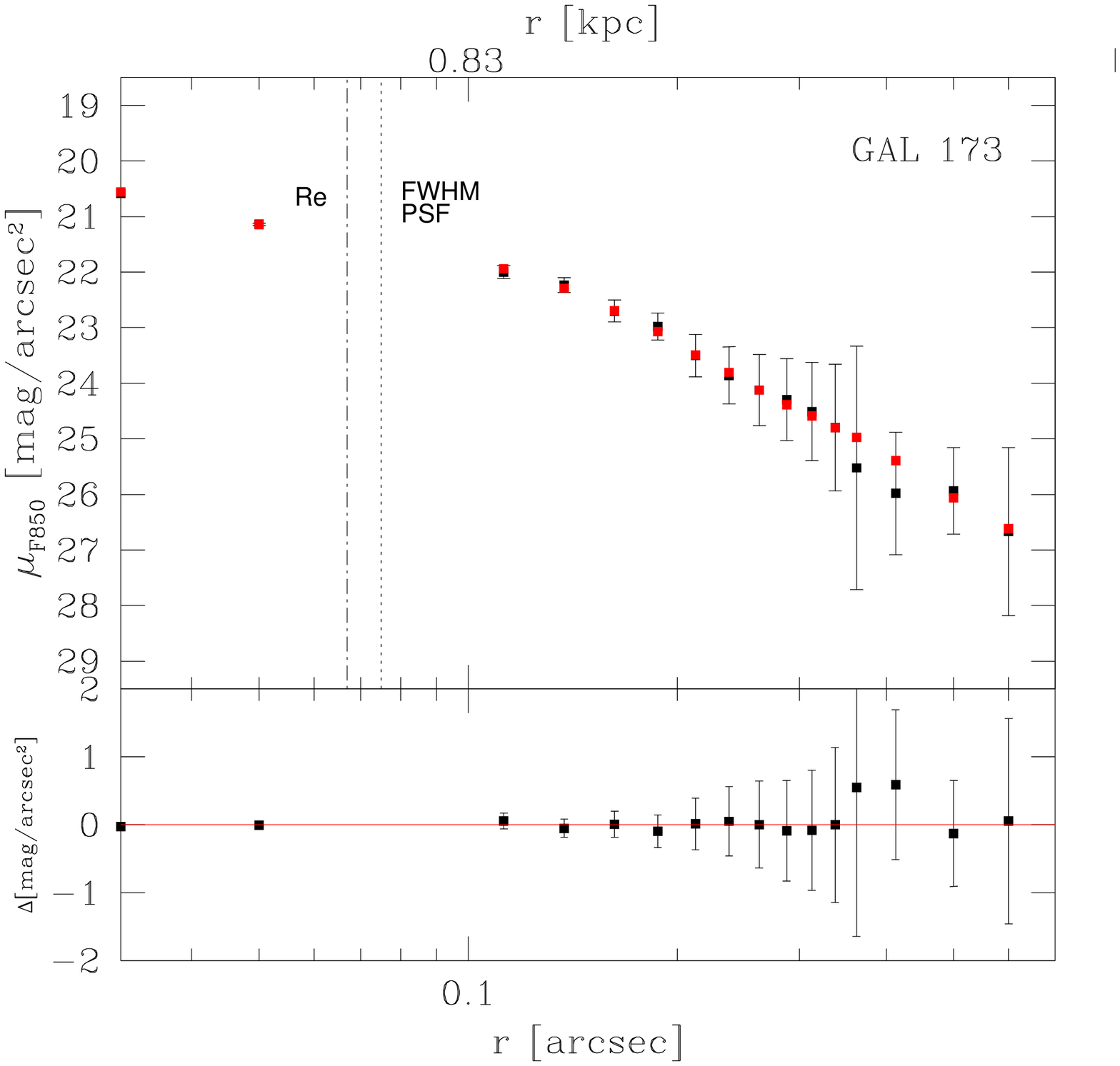}
\includegraphics[width=5truecm]{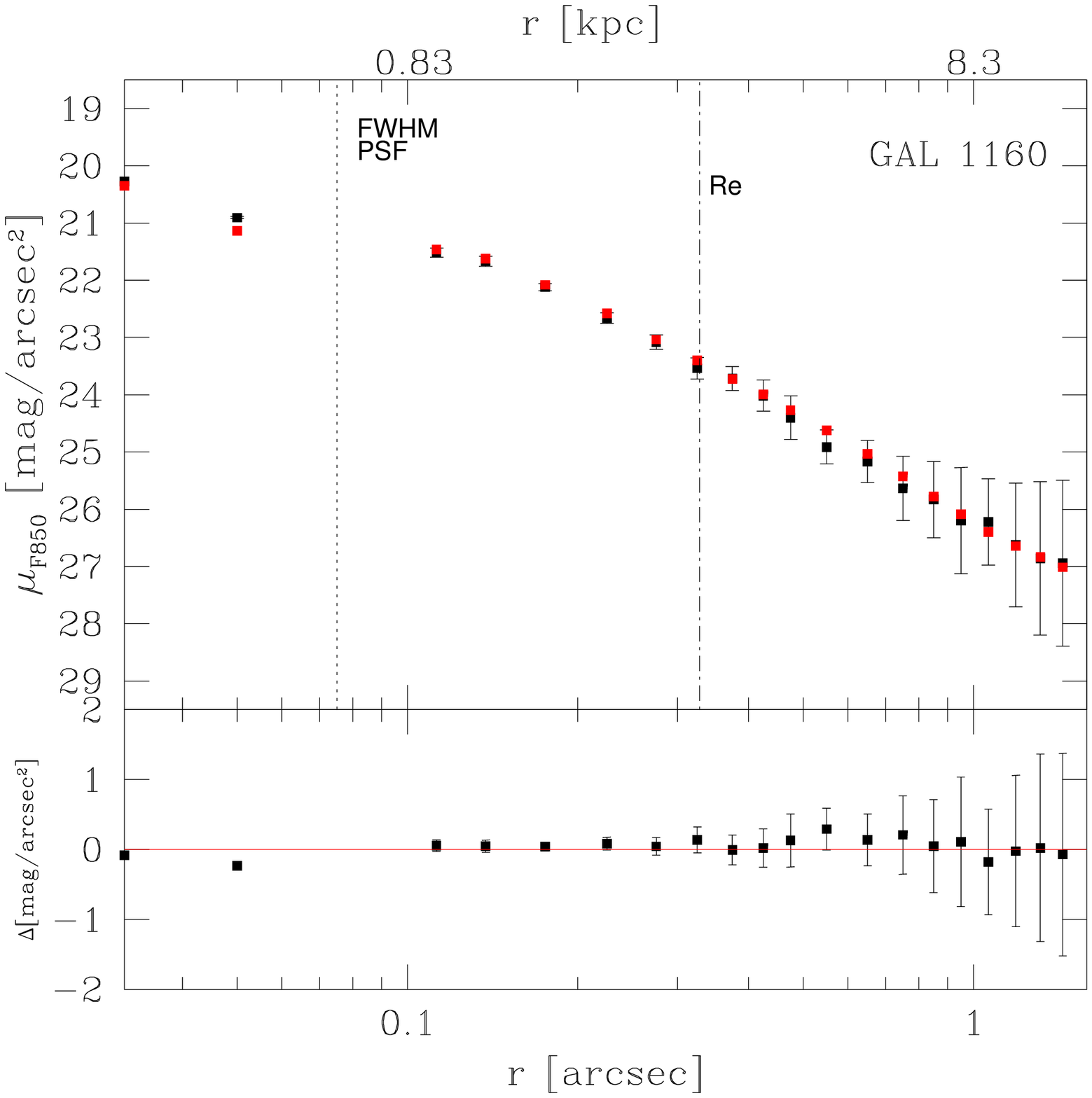}
\includegraphics[width=4truecm]{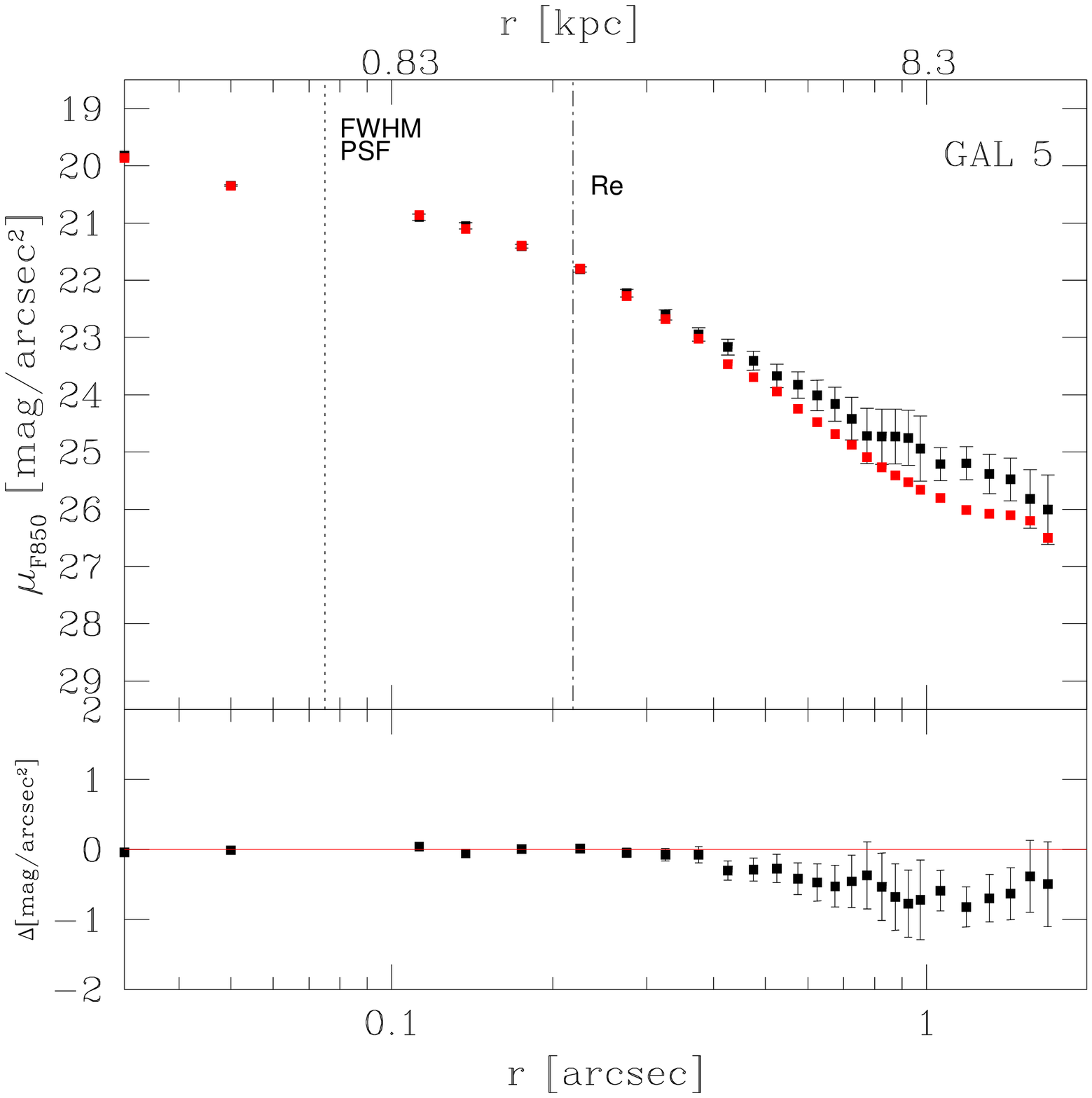}
\includegraphics[width=4truecm]{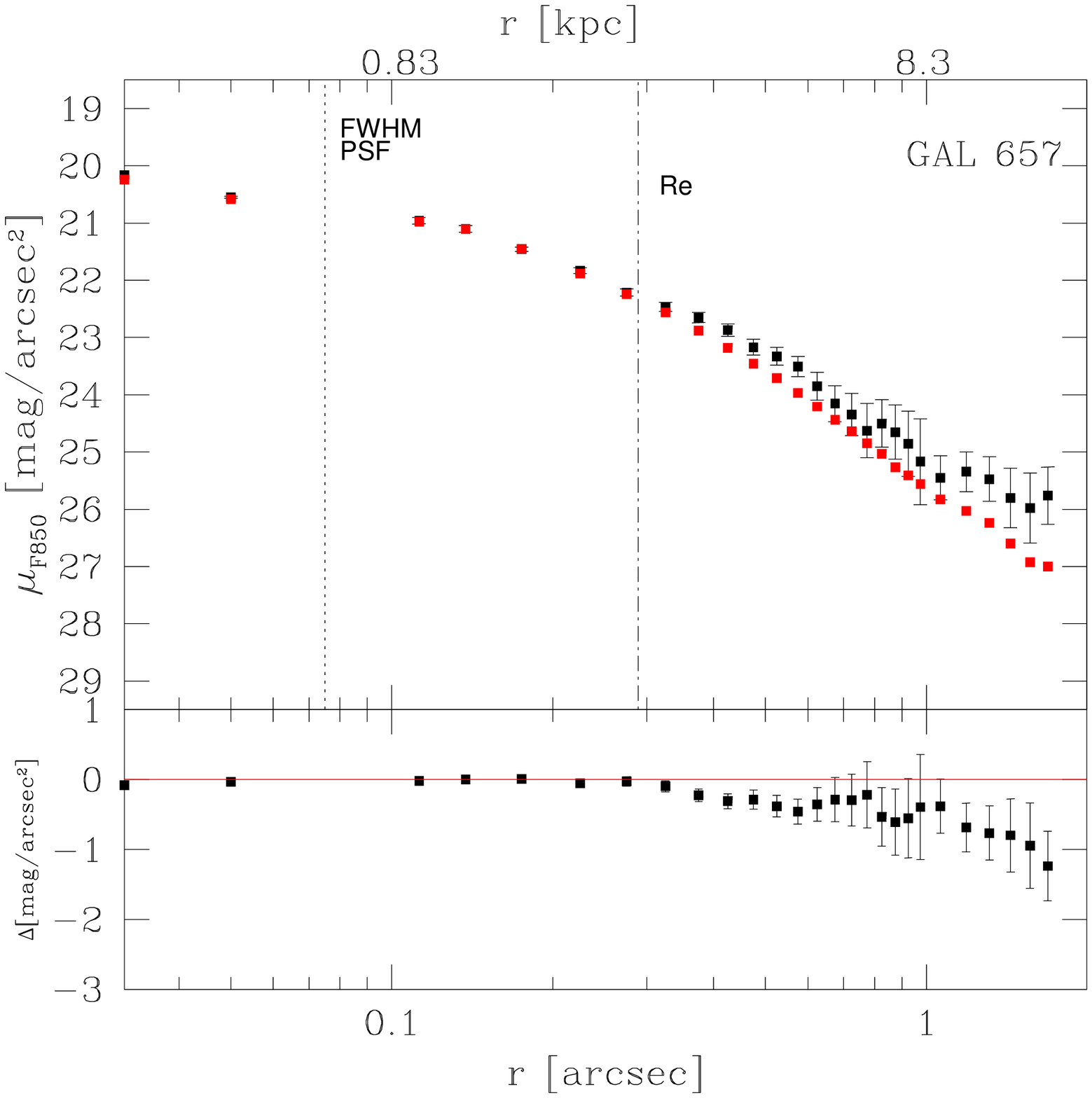}
\includegraphics[width=4truecm]{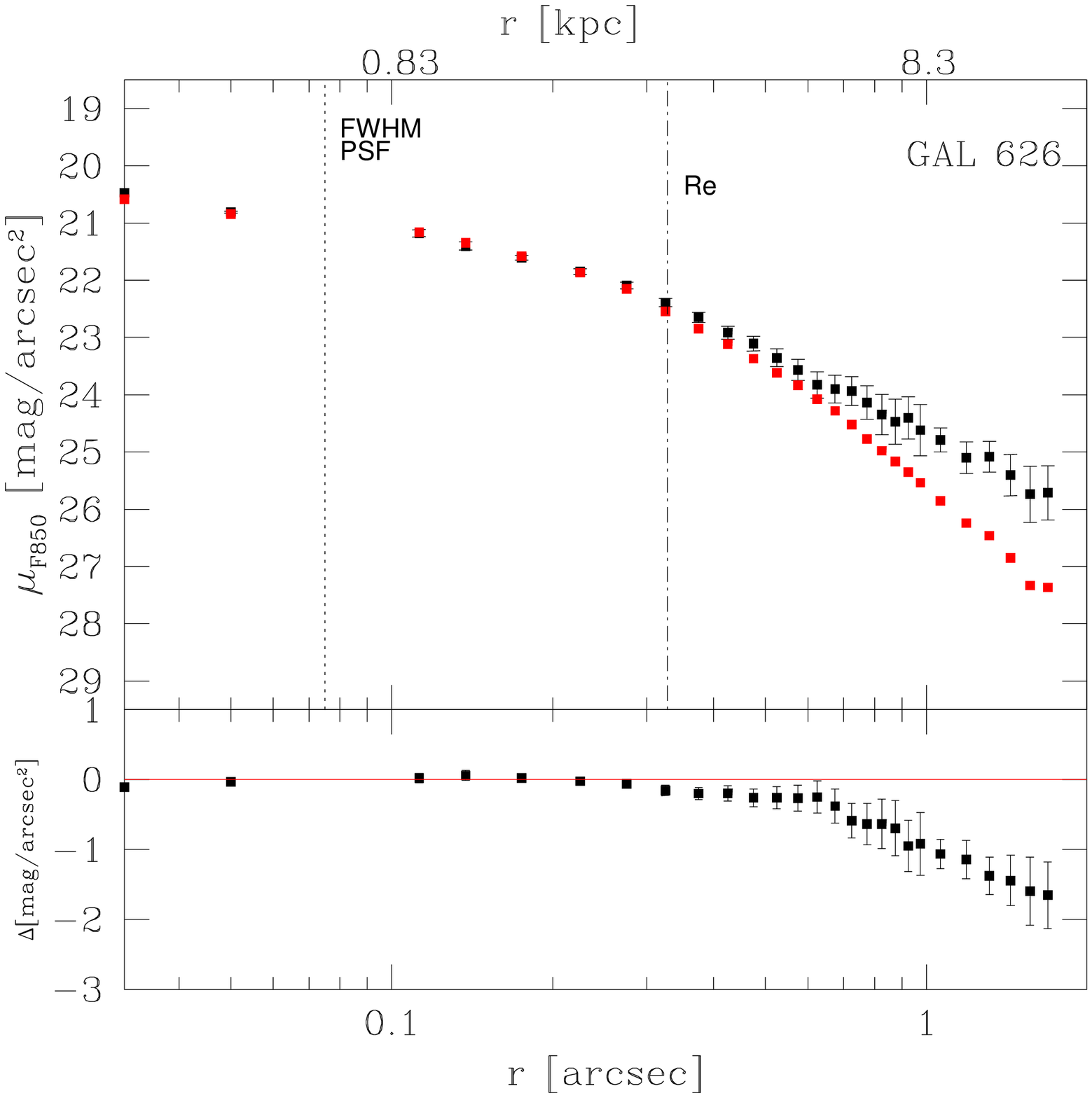}
\includegraphics[width=4truecm]{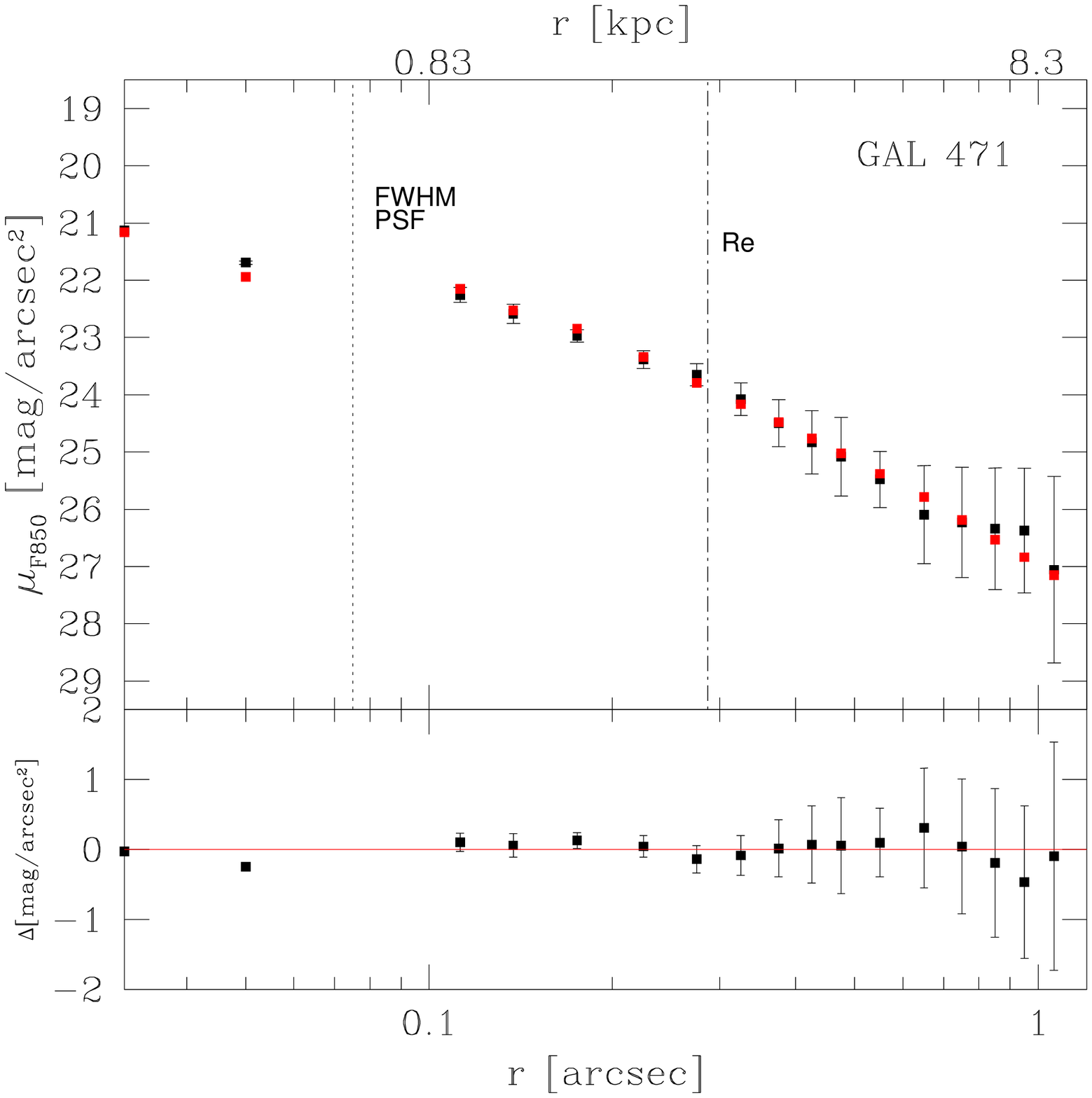}
\caption{Surface brightness in the 
F850LP band measured within circular annulus centered on  
each galaxy (black points), compared with the surface brightness
resulting from the best fitting S\'ersic model profile (red points). 
The dotted line marks the radius of the FWHM while the dotted-dashed line
marks the effective radius.
In the lower panels the residuals of the fitting obtained as the difference 
between the data points and the models are shown. 
}
\end{center}
\end{figure*}

\end{appendix}


\begin{thebibliography}{}
\bibitem[]{} Andreon S.,  Puddu E., de Propris R., Cuillandre J.-C. 2008,
MNRAS, 385, 979
\bibitem[]{} Belli S., Newman A. B., Ellis R. S., 2014, ApJ, 783, 117
\bibitem[]{} Bezanson R., van Dokkum P. G., Tal T., Marchesini D., Kriek M.,
Franx M., Coppi P. 2009, ApJ, 697, 1290
\bibitem[]{} Bruzual A.,G. \& Charlot S. 2003, MNRAS 344, 1000
\bibitem[]{} Buitrago F., Trujillo I., Conselice C. J., Bouwens R. J., Dickinson
M., Yan H. 2008, ApJ, 687, L61
\bibitem[]{} Calzetti D., Armus L., Bohlin R. C., Kinney A. L., Koorneef J.,
Storchi-Bergmann T. 2000, ApJ, 533, 682
\bibitem[]{} Carollo C., Bschorr T. J., Renzini A., et al. 2013, ApJ, 773, 112
\bibitem[]{} Cassata P.,  Giavalisco M., Guo Y., al. 2011, ApJ, 743, 96
\bibitem[]{} Chabrier G. 2003, PASP, 115, 763
\bibitem[]{} Cimatti A.,  Cassata P., Pozzetti L., et al. 2008, A\&A, 482, 21
\bibitem[]{} Cimatti A., Nipoti C., Cassata P., 2012, MNRAS, 422, L62
\bibitem[]{} Daddi E., Renzini A., Pirzkal N., et al. 2005, ApJ, 626, 680
\bibitem[]{} Damjanov I., McCarthy P. J., Abraham R. G., et al. 2009, ApJ, 
695, 101
\bibitem[]{} Damjanov I., Abraham R. G., Glazebrook K., et al. 2011, ApJ, 739, L44
\bibitem[]{} Damjanov I., Chilingarian I., Hwang H. S., Geller M. J., 2013, ApJ,
L48
\bibitem[]{} Delaye L., Huertas-Company M., Mei S., et al. 2014, MNRAS, 441, 203
\bibitem[]{} De Lucia G., Springel V., White S. D. M., Croton D., 
Kauffmann G. 2006, MNRAS, 366, 499
\bibitem[]{} di Serego Alighieri S., Vernet J., Cimatti A., et al. 2005, A\&A, 442, 125
\bibitem[]{} Fasano G., Marmo C., Varela J., et al. 2006, A\&A, 445, 805
\bibitem[]{} Fasano G., Vanzella E., Wings Team, 2007, ASP Conf. Ser., 374, 495
\bibitem[]{} Fan, L., Lapi, A., De Zotti, G., Danese, L., 2008, ApJ, 689, L101
\bibitem[]{} Fan, L., Lapi, A., Bressan A., Bernardi M., De Zotti G., Danese L.,
2010, ApJ, 718, 1460
\bibitem[]{} Gargiulo A., Haines C. P., Merluzzi P., et al., 2009, MNRAS, 397, 75
\bibitem[]{} Gargiulo A., Saracco P., Longhetti M. La Barbera F., Tamburri S.,
2012, MNRAS, 425, 2698
\bibitem[]{} Gargiulo A., Saracco P., Longhetti M., Tamburri S., Lonoce I.,
Ciocca F., 2014, A\&A, submitted
\bibitem[]{} Hamabe M., Kormendy J. 1987, IAUS, 127, 379
\bibitem[]{} Holden B. P., Blakeslee J. P., Postman M., et al. 2005, ApJ, 626, 809
\bibitem[]{} Hopkins, P., Bundy, K., Murray N., Quataert E., Lauer T. R.,
Ma C.-P. 2009, MNRAS, 398, 898
\bibitem[]{} Hopkins, P. F., Bundy K., Croton D., al. 2010, ApJ, 715, 202
\bibitem[]{} Huertas-Company M., Mei S., Shankar F., et al. 2013, MNRAS, 428, 1715
\bibitem[]{} Ilbert O., Salvato M., Le Floc'h E., et al. 2010, ApJ, 709, 644
\bibitem[]{} Jorgensen I., Franx M., Kjaergaard P. 1995a, MNRAS, 273, 1097
\bibitem[]{} Jorgensen I., Franx M., Kjaergaard P. 1995b, MNRAS, 276, 1341
\bibitem[]{} Jorgensen I., Franx M., Kjaergaard P. 1996, MNRAS, 280, 167
\bibitem[]{} Jorgensen I., Chiboucas K. 2013, AJ, 145, 77
\bibitem[]{} Khochfar S., Burkert A., 2003, ApJ, 597, L11
\bibitem[]{} Kormendy J. 1977, ApJ 218, 333
\bibitem[]{} La Barbera F., Busarello G., Massarotti M., Merluzzi P., Mercurio
A. 2003, A\&A, 409, 21
\bibitem[]{} La Barbera F.,Merluzzi P., Busarello G., Massarotti M., Mercurio A.
2004, A\&A, 425, 797
\bibitem[]{} La Barbera F., de Carvalho R. R., de la Rosa I. G., Lopes P. A. A.,
2010, MNRAS, 408, 1335
\bibitem[]{} Longhetti M., Saracco P. 2009, MNRAS, 394, 774
\bibitem[]{} Longhetti M., Saracco P., Severgnini P., et al., 2007, MNRAS, 374, 614
\bibitem[]{} Mancini, C., Daddi E., Renzini A., et al. 2010, MNRAS, 401, 933
\bibitem[]{} Maraston C. 2005, MNRAS, 362, 799
\bibitem[]{} McGrath E., Stockton A., Canalizo G., Iye M., Maihara T. 2008,
ApJ, 682, 303
\bibitem[]{} Mei S., Stanford S. A., Holden B. P., et al. 2012, ApJ, 754, 141
\bibitem[]{} Naab T., Johansson P. H., Ostriker J. P. 2009, ApJ, 699, L178
\bibitem[]{} Newman A. B., Ellis R. S., Bundy K., Treu T., 2012, ApJ, 746, 162
\bibitem[]{} Nipoti C., Treu T., Auger M. W., Bolton A. S. 2009, ApJ, 706, L86
\bibitem[]{} Nipoti C., Treu T., A. Leauthaud, K. Bundy, A. B. Newman, M. W.
Auger, 2012, MNRAS, 422, 1714
\bibitem[]{} Papovich C., Bassett R., Lotz J. M., et al. 2012, ApJ, 750, 93
\bibitem[]{} Peng C.Y., Ho L.C., Impey C.D. \& Rix H-W 2002, AJ 124, 266
\bibitem[]{} Pignatelli E., Fasano G., Cassata P., 2006, A\&A, 446, 373
\bibitem[]{} Poggianti B. M., Calvi R., Bindoni D., et al. 2013a, ApJ, 762, 77
\bibitem[]{} Poggianti B. M., Moretti A., Calvi R., D'Onofrio M., Valentinuzzi
T., Fritz J., Renzini A. 2013, ApJ, 777, 125
\bibitem[]{} Postamn M., Franx M., Cross N. J. G., et al. 2005, ApJ, 623, 721
\bibitem[]{} Pozzetti L., Bolzonella M., Zucca E., et al. 2010, A\&A 523, 13 
\bibitem[]{} Raichoor A., Mei S., Nakata F., et al. 2011, ApJ, 732, 12
\bibitem[]{} Raichoor A., Mei S., Stanford S. A., et al. 2012, ApJ, 745, 130
\bibitem[]{} Reda F. M., Forbes D. A., Beasley M. A., O'Sullivan E. J.,
Goudfrooij P. 2004, MNRAS, 354, 851
\bibitem[]{} Rettura A., Rosati P., Nonino M., et al., 2010, ApJ, 709, 512
\bibitem[]{} Saglia R. P., Sa\'nchez-Bla\'zquez P., Bender R., et al. 2010, A\&A, 524, 6
\bibitem[]{} Salpeter E. E. 1955, ApJ, 121, 161
\bibitem[]{} Saracco P., Longhetti M., Andreon S., 2009, MNRAS, 392, 718
\bibitem[]{} Saracco P., Longhetti M., Gargiulo A. 2010, MNRAS, 408, L21
\bibitem[]{} Saracco P., Longhetti M., Gargiulo A. 2011, MNRAS, 412, 2707
\bibitem[]{} Saracco P.,  Gargiulo A., Longhetti M. 2012, MNRAS, 422, 3107
\bibitem[]{} Schade D., Carlberg R. G., Yee H. K. C., Lopez-Cruz O.,
Ellingson E. 1996, ApJ, 464, L63
\bibitem[]{} Shen S., Mo H. J., White Simon D. M., et al., 2003, MNRAS, 343, 978
\bibitem[]{} Stanford S. A., Elston R., Eisenhardt P. R., Spinrad H., Stern D.,
Dey A. 1997, AJ, 114, 2232
\bibitem[]{} Stott J. P., Collins C. A., Burke C., Hamilton-Morris V., Smith G.
P., 2011, MNRAS, 414, 445
\bibitem[]{} Szomoru D., Franx, M., van Dokkum P. G., 2012, ApJ, 749, 121
ApJ, 621, 673
\bibitem[]{} Taylor E. N., Franx M., Blazebrook K., Brinchmann J., van der Wel
A., van Dokkum P. G., 2010, ApJ, 720, 741
\bibitem[]{} Trujillo I., Feulner G., Goranova Y., et al. 2006, MNRAS, 373, L36
\\bibitem[]{} Trujillo I., Conselice C. J., Bundy K., Cooper M. C., Eisenhardt P,
Ellis R. S., 2007, MNRAS, 382, 109
\bibitem[]{} Trujillo I., Ferreras I., de La Rosa I. G. 2011, MNRAS, 415, 3903
\bibitem[]{} Valentinuzzi P.,  Fritz J., Poggianti B. M., et al. 2010a, ApJ, 712, 226
\bibitem[]{} Valentinuzzi P., Poggianti B. M., Saglia R. P., et al. 2010b, ApJ, 721, L19
\bibitem[]{} van der Wel A., Holden B. P., Zirm A. W., Franx, M., Rettura, A.,
Illingworth G. D., Ford H. C., 2008, ApJ, 688, 48
\bibitem[]{} van der Wel A., Rix H.-W., Wuyts S., al. 2011, ApJ 730, 38
\bibitem[]{} van Dokkum P. G., Stanford S. A., Holden B. P., Eisenhardt P. R.,
Dickinson M., Elston R., 2001, ApJ, 552, L101
\bibitem[]{} van Dokkum P. G., Franx M., Kriek M., et al. 2008, ApJ, 677, L5
\bibitem[]{} van Dokkum P. G., Whitaker K. E., Brammer G., et al. 2010, ApJ, 709, 1018
\bibitem[]{} Ziegler B. L., Saglia R. P., Bender R., Belloni P., Greggio L.,
Seitz S. 1999, A\&A, 346, 13
\end{thebibliography}
\end{document}